\documentclass[pdflatex,sn-mathphys-num, iicol, referee]{sn-jnl}

\usepackage{graphicx}%
\usepackage{multirow}%
\usepackage{amsmath,amssymb,amsfonts}%
\usepackage{amsthm}%
\usepackage{mathrsfs}%
\usepackage[title]{appendix}%
\usepackage{xcolor}%
\usepackage{textcomp}%
\usepackage{manyfoot}%
\usepackage{booktabs}%
\usepackage{algorithm}%
\usepackage{algorithmicx}%
\usepackage{algpseudocode}%
\usepackage{listings}%
\usepackage{anyfontsize}

\usepackage{nicefrac}       
\usepackage{microtype}      
\usepackage{comment}

\usepackage[english]{babel}
\usepackage{multirow}
\usepackage{amsthm}
\usepackage{bbm}
\usepackage{float}
\usepackage{subcaption}
\usepackage{multirow}
\usepackage{makecell}
\usepackage{mathtools}
\usepackage{bm}
\usepackage{physics}
\usepackage{appendix}
\usepackage{enumitem}
\usepackage{csquotes}
\usepackage{cancel}
\allowdisplaybreaks

\renewcommand{\v}[1]{\ensuremath{\bm{#1}}}

\newcommand{\bfI}{\boldsymbol{\mathbbm{I}}}
\newcommand{\bfD}{\boldsymbol{D}}
\newcommand{\bfmu}{\boldsymbol{\mu}}
\newcommand{\bfSigma}{\boldsymbol{\Sigma}}

\newcommand{\bfalpha}{\boldsymbol{\alpha}}
\newcommand{\bfdelta}{\boldsymbol{\delta}}

\newcommand{\bfy}{\boldsymbol{y}}
\DeclareMathOperator{\argmax}{argmax}
\DeclareMathOperator{\diag}{diag}

\begin{document}

\title[Logit unfolding models]{Logit unfolding choice models for binary data}

\author*[1]{\fnm{Rayleigh} \sur{Lei}}\email{rlei13@uw.edu}

\author[1]{\fnm{Abel} \sur{Rodriguez}}\email{abelrow@uw.edu}

\affil*[1]{\orgdiv{Department of Statistics}, \orgname{University of Washington}, \orgaddress{\street{Padelford Hall}, \city{Seattle}, \postcode{98915}, \state{WA}, \country{United States}}}


\abstract{Discrete choice models with non-monotonic response functions are important in many areas of application, especially political sciences and marketing.  This paper describes a novel unfolding model for binary data that allows for heavy-tailed shocks to the underlying utilities.  One of our key contributions is a Markov chain Monte Carlo algorithm that requires little or no parameter tuning, fully explores the support of the posterior distribution, and can be used to fit various extensions of our core model that involve (Bayesian) hypothesis testing on the latent construct.  Our empirical evaluations of the model and the associated algorithm suggest that they provide better complexity-adjusted fit to voting data from the United States House of Representatives.}


\maketitle

\section{Introduction}\label{sec:intro}
Discrete choice models \citep{McFaddenModelingChoiceResidential1978} have been widely used to  understand decision making in a wide range of fields, including resource selection in ecology \citep{CooperMillspaughApplicationDiscreteChoice1999, McCrackenEtAlUseDiscreteChoiceModels1998}, consumer choice \citep{ZwerinaDiscreteChoiceExperiments1997, AndersonEtAlDiscreteChoiceTheory1992}, travel demand and transportation policy \citep{BrownstoneDiscreteChoiceModeling2001, ChandraR.BhatRecentMethodologicalAdvances1997}, and energy saving measures \citep{JridiEtAlHouseholdPreferencesEnergy2015}.  In political science, spatial voting models, which have been extensively used to model the latent ideological preferences of political actors based on their votes, 
can be described as discrete choice models \citep{davis1970expository,enelow1984spatial,poole1985spatial,JackmanMultidimensionalAnalysisRoll2001}.

Discrete choice models assume that individuals have a utility over a discrete set of outcomes that is subject to random shocks, and that they make decisions consistent  with the highest \textit{realized} utility. A common approach to building the utility functions is to assume that the preferences of the individuals live in a low-dimensional latent (unobserved) space, and that the utility is a function of the distance between their latent position and decision-specific positions associated with each of the outcomes under consideration, plus an additive independent random noise. In settings where choices are binary and the latent space is Euclidean and unidimensional, this type of formulation usually results in monotonic response functions, i.e., the probability of choosing each option increases or decreases as a function of the latent position.  Such behavior, while very natural in many applications, is often inappropriate in political science ones where members of deliberate bodies that are on the extremes of the political spectrum sometimes vote together against those in the middle \citep{Lewis2019a,Lewis2019b,yu2021spatial,duck2022ends,LeiRodriguezNovelClassUnfolding2023}. In these kinds of situations, traditional models that only allow for monotonic response functions often inaccurately estimate the legislators having centrist preferences despite the legislators being considered extreme according to most political analysts.


The literature on models that can accommodate non-monotonic response functions is limited.  One contribution in this area is the work of \citet{roberts2000general}, which proposed the generalized graded unfolding model (GGUM)
. GGUMs rely on an item response theory formulation of the response function and cannot be easily linked to a set of underlying utility functions. 
More recently, \citet{duck2022ends} introduced a Bayesian version of the GGUM (which we refer to as the BGGUM) to the political science literature and showed that the model is able to correctly capture the ideological preferences of politicians on the ``ends'' of the political spectrum. Meanwhile,  \citet{LeiRodriguezNovelClassUnfolding2023} developed a novel class of unfolding models for binary data that are built within the framework of discrete choice models and allow for non-monotonic responses by allowing outcomes to be associated with more than one latent position.  Their experiments suggests that this novel class of unfolding model leads to more accurate estimates of political actors' ideologies than those of BGGUM.

While the methodology introduced in \cite{LeiRodriguezNovelClassUnfolding2023} is defined for a broad class of probability distributions for the utility shocks, their paper focused on \textit{probit} unfolding models where the shocks follow Gaussian distributions.  This choice simplified computation by enabling the design of a Markov chain Monte Carlo (MCMC) algorithm that relied on Gibbs sampling steps.  This is in contrast to BGGUM, which relies on a Metropolis coupled Markov chain Monte Carlo (MC3) implemented through random walk Metropolis-Hastings steps that require careful tuning.  This paper extends the work of \cite{LeiRodriguezNovelClassUnfolding2023} by developing a similar computational algorithm based on Gibbs sampling steps for the logistic version of the model in which the random shocks follow (standard) Gumbel distributions.  The use of a logit link is appealing because it is robust to certain types of outliers which, in the context of our unfolding model, helps avoid overfitting.  Indeed, in ``end-against-the-middle'' votes, it is legitimate to ask whether legislator's behavior really reflects a different set of voting principles or is merely an artifact related to sporadic, large random shocks to the underlying utility functions that Gaussian shocks cannot accommodate.

The algorithm we present in this paper builds on the work of \citet{Fruhwirth-SchnatterFruhwirthAuxiliaryMixtureSampling2007, Fruhwirth-SchnatterFruhwirthBayesianInferenceMultinomial2012}.  More specifically, we approximate the Gumbel distribution associated with the utility shocks with a finite mixture of normal distributions. Such an approximation is pre-computed by minimizing the Kullback–Leibler divergence between the standard Gumbel density and that of the finite mixture.  The development of such specialized algorithm is important because the structure of our model is such that black-box algorithms, such as \texttt{Stan} \citep{Rstan}, fail to converge to the right posterior distribution.  Furthermore, such an algorithm enables the development of extensions to hierarchical settings, particularly those that involve model comparisons or other discrete random variables (e.g., see \citealp{rodriguez2015measuring}, \citealp{lofland2017assessing} and \citealp{MoserEtAlMultipleIdealPoints2021}), for which computation using off-the-self software is unfeasible. 

The rest of the paper is organized as following. In Section \ref{sec:model_formulation}, we describe the logit unfolding model and discuss prior specification. In Section \ref{sec:computation}, we discuss our computatinal algorithm, including the derivation of the approximation to the Gumbel distribution based on a mixture of normal distributions that we exploit to derive our Gibbs sampler. Section \ref{sec:empirical} discusses  our motivating application to estimating legislator's preferences in the U.S.\ House of Representatives, and illustrates the performance of the model on four datasets corresponding to the 103\textsuperscript{rd} (1993-1994), 107\textsuperscript{rd} (2001-2002), 110\textsuperscript{th} (2007-2008), and 116\textsuperscript{th} (2019-2020) Houses. Finally, Section \ref{sec:discussion} presents a brief discussion and future directions of research.

\section{Model Formulation}\label{sec:model_formulation}

To be concrete, we discuss the construction of the model in the context of our motivating application:  the estimation of preferences of legislators from roll call vote data in deliberative bodies.  Let there be $I$ legislators and $J$ issues on which they vote. Set $y_{i, j}$ to be legislator $i$'s vote on issue $j$ for $i = 1, 2,\ldots, I$ and $j = 1, 2,\ldots, J$, where $y_{i, j}$ is a binary outcome variable such that $y_{i, j} = 1$ indicates an affirmative (``Aye'') vote whereas $y_{i, j} = 0$ denotes a negative (``Nay'') vote. 

We assume that each legislator has a preferred position in a latent, one-dimensional Euclidean space, i.e. \textit{the policy space}. Denote an individual $i$'s preferred position as $\beta_i$ for $i = 1, 2, \ldots, I$. Each decision $j$ has associated with it three positions in the same policy space, $\psi_{j, 1}$, $\psi_{j, 2}$, and $\psi_{j, 3}$, for $j = 1, 2, \ldots, J$. In our model, $\psi_{j, 2}$ corresponds to the position associated with an affirmative vote whereas $\psi_{j, 1}$ and $\psi_{j, 3}$ are two distinct positions that both correspond to a negative vote. We make the further assumption that either $\psi_{j, 1} < \psi_{j, 2} < \psi_{j, 3}$ or $\psi_{j, 3} < \psi_{j, 2} < \psi_{j, 1}$.  Under this order constraint, one of the negative positions (say, $\psi_{j, 1}$) can be interpreted as the \textit{status quo}, the positive position $\psi_{j,2}$ as the \textit{proposed policy}, and $\psi_{j, 3}$ as a \textit{more extreme alternative policy} that some legislators might find preferable to the proposed one (and to the status quo).  This interpretation reflects a voting pattern sometimes observed in deliberative bodies where the most extreme members of the majority party vote against their own party's proposed policy because it is not extreme enough. 

We further assume that each individual selects an option based on quadratic utility functions. This utility function depends on the distances between their ideal point and the various policy positions:
\begin{equation}
\begin{aligned}
    U_{N^{-}}(\beta_i, \psi_{j, 1}) &= -\left(\beta_i - \psi_{j, 1}\right)^2 + \epsilon_{i, j, 1}, \\
    U_{Y}(\beta_i, \psi_{j, 2}) &= -\left(\beta_i - \psi_{j, 2}\right)^2 + \epsilon_{i, j, 2}, \\
    U_{N^{+}}(\beta_i, \psi_{j, 3}) &= -\left(\beta_i - \psi_{j, 3}\right)^2 + \epsilon_{i, j, 3}.
\end{aligned}    
\label{eq:utility_def}
\end{equation}

In this paper, $\epsilon_{1, i,j}, \epsilon_{i, j, 2}, \epsilon_{i, j, 3}$ are assumed to be independently and identically distributed according to a standard Gumbel distribution. A positive vote occurs if and only if $U_{Y}(\beta_i, \psi_{j, 2}) > U_{N^{-}}(\beta_i, \psi_{j, 1})$ and $U_{Y}(\beta_i, \psi_{j, 2}) > U_{N^{+}}(\beta_i, \psi_{j, 3})$. It is then easy to show (e.g., see \citealp{TrainDiscreteChoiceMethods2009}) that the implied response function reduces to :
\begin{multline*}
    p(y_{i, j} = 1 \mid \beta_i, \psi_{j, 1}, \psi_{j, 2}, \psi_{j,3}) 
    =\\  
    \frac{\exp\left\{-\left(\beta_i - \psi_{j, 2}\right)^2\right\}}{\sum_{k = 1}^3 \exp\left\{-\left(\beta_i - \psi_{j, k}\right)^2\right\}}.
\end{multline*}

If we divide the numerator and denominator of the response function by $\exp\left\{-\left(\beta_i - \psi_{j, 2}\right)^2\right\}$ and perform some standard algebraic manipulations, the response function can be rewritten as:
\begin{multline}\label{eq:logisticBGUM}
     p(y_{i, j} = 1 \mid \beta_i, \alpha_{j,1}, \delta_{j,1}, \alpha_{j,2}, \delta_{j,2}) =\\  \frac{1}{1 + \exp\left\{-\alpha_{j,1}(\beta_i - \delta_{j,1})\right\} + \exp\left\{-\alpha_{j,2}(\beta_i - \delta_{j,2})\right\}} ,
\end{multline}
where $\alpha_{j,1} = 2(\psi_{j,2} - \psi_{j,1})$, $\alpha_{j,2} = 2(\psi_{j,2} - \psi_{j,3})$, $\delta_{j,1} = (\psi_{j,1} + \psi_{j,2})/2$ and $\delta_{j,2} = (\psi_{j,3} + \psi_{j,2})/2$. Notice that, as $\psi_{j, 3} \rightarrow \infty$, we recover the standard logistic item response theory model (e.g., see \cite{fox2010bayesian}):
\begin{multline*}
    p(y_{i, j} = 1 \mid \beta_i, \alpha_{j,1}, \delta_{j,1}, \alpha_{j,2}, \delta_{j,2}) =\\ \frac{1}{1 + \exp\left\{-\alpha_{j,1}(\beta_i - \delta_{j,1})\right\}}. 
\end{multline*}
A similar result holds if $\psi_{j,1} \to -\infty$ instead.

The model is completed by eliciting priors on the unknown parameters $\beta_1, \beta_2, \ldots, \beta_I$, $\v{\alpha}_1, \v{\alpha}_2, \ldots, \v{\alpha}_J$, and $\v{\delta}_1, \v{\delta}_2, \ldots, \v{\delta}_J$. In this paper, we consider a standard normal distribution as the prior for $\beta_i$, a common choice throughout the related literature. By fixing the mean and variance the prior on the $\beta_i$'s, the model becomes weakly identifiable to translations and rescalings of the latent policy space. On the other hand, lack of identifiability to reflections of the latent space is addressed by fixing the sign of one carefully selected legislator. For further discussion on model identifiability, please see Section 1 of the Supplementary Materials.

On the other hand, the priors on $\v{\alpha}_1, \v{\alpha}_2, \ldots, \v{\alpha}_J$, and $\v{\delta}_1, \v{\delta}_2, \ldots, \v{\delta}_J$ are chosen carefully because of the need to account for the constraints, $\psi_{j,1} < \psi_{j,2} < \psi_{j,3}$ or $\psi_{j,3} < \psi_{j,2} < \psi_{j,1}$. Note that a sufficient condition for these order constraints is that $\alpha_{j, 1} < 0 < \alpha_{j, 2}$ or $\alpha_{j, 1} > 0 > \alpha_{j, 2}$. To enforce this constraint, we propose a mixture of two truncated multivariate Gaussian distributions with density
\begin{multline}\label{eq:alphadeltaprior}
p \left( \bfalpha_j, \bfdelta_j \right)
      =\\
      \frac{1}{32 \pi^2 \omega^2 \kappa^2} \exp\left\{ -\frac{1}{2} \left(\frac{1}{\omega^2 }\bfalpha_j'\bfalpha_j  + \right.\right.\\
      \left.\left. \frac{1}{\kappa^2}(\bfdelta_j - \v{\vartheta})'(\bfdelta_j - \v{\vartheta})\right)\right\}\\ \mathbbm{1}(\alpha_{j,1}>0, \alpha_{j,2}<0) + \\
      \frac{1}{32 \pi^2 \omega^2 \kappa^2} \exp\left\{ -\frac{1}{2} \left(\frac{1}{\omega^2 }\bfalpha_j'\bfalpha_j  + \right.\right.\\
      \left.\left. \frac{1}{\kappa^2}(\bfdelta_j + \v{\vartheta})'(\bfdelta_j + \v{\vartheta})\right)\right\}\\ \mathbbm{1}(\alpha_{j,1}<0, \alpha_{j,2}>0).
\end{multline}
for some $\v{\vartheta} \in \mathbbm{R}^2$ and $\omega, \kappa \in \mathbbm{R}^+$.  Note that the first component of the mixture corresponds to the case $\psi_{j,1} < \psi_{j,2} < \psi_{j,3}$, while the second corresponds to $\psi_{j,3} < \psi_{j,2} < \psi_{j,1}$.  Furthermore, note that the marginal distributions for both $\alpha_{j,1}$ and $\alpha_{j,2}$ are univariate Gaussians with mean zero and variance $\omega^2$.  This property is important in order to allow for both increasing and decreasing response functions within the same dataset.  We discuss specific choices for the hyperparameters $\omega^2$, $\v{\vartheta}$ and $\kappa^2$ in Section \ref{sec:empirical} in the context of our application.

It is worthwhile emphasizing that we place our priors on $(\alpha_{j, 1},\alpha_{j, 2},\delta_{j, 1},\delta_{j, 2})$ rather than the original parameters $(\psi_{j, 1},\psi_{j, 2},\psi_{j, 3})$. There are two reasons for this choice. The first is computational. It is not clear what form the conditionally conjugate priors for $(\psi_{j, 1},\psi_{j, 2},\psi_{j, 3})$ would take or even whether they exist.  Without them, sampling $(\psi_{j, 1},\psi_{j, 2},\psi_{j, 3})$ would require some sort of Metropolis-Hastings step that would potentially need a fair bit of tuning. Doing so partially defeats the purpose of the algorithm being proposed. The second reason is model flexibility. As  pointed out by the reviewers, if we had started by placing priors directly on $( \psi_{1,j},\psi_{2,j},\psi_{3,j})$, the implied prior on $(\alpha_{j, 1}, \alpha_{j, 2}, \delta_{j, 1}, \delta_{j, 2})$ would be constrained to a three dimensional manifold. Our initial simulations suggest that working with the four dimensional space associated with $(\alpha_{j, 1}, \alpha_{j, 2}, \delta_{j, 1}, \delta_{j, 2})$ and ensuring some level of compatibility between the three dimensional and four dimensional spaces through the sign constraints on the $\alpha_j$'s leads to improved model fit while preserving interpretability.

\section{Computation}\label{sec:computation}

\subsection{Computational challenges}\label{sec:compchallenges}

\begin{figure*}[!htb]
    \centering
    \begin{subfigure}[t]{0.4\textwidth}
        \includegraphics[width = \textwidth]{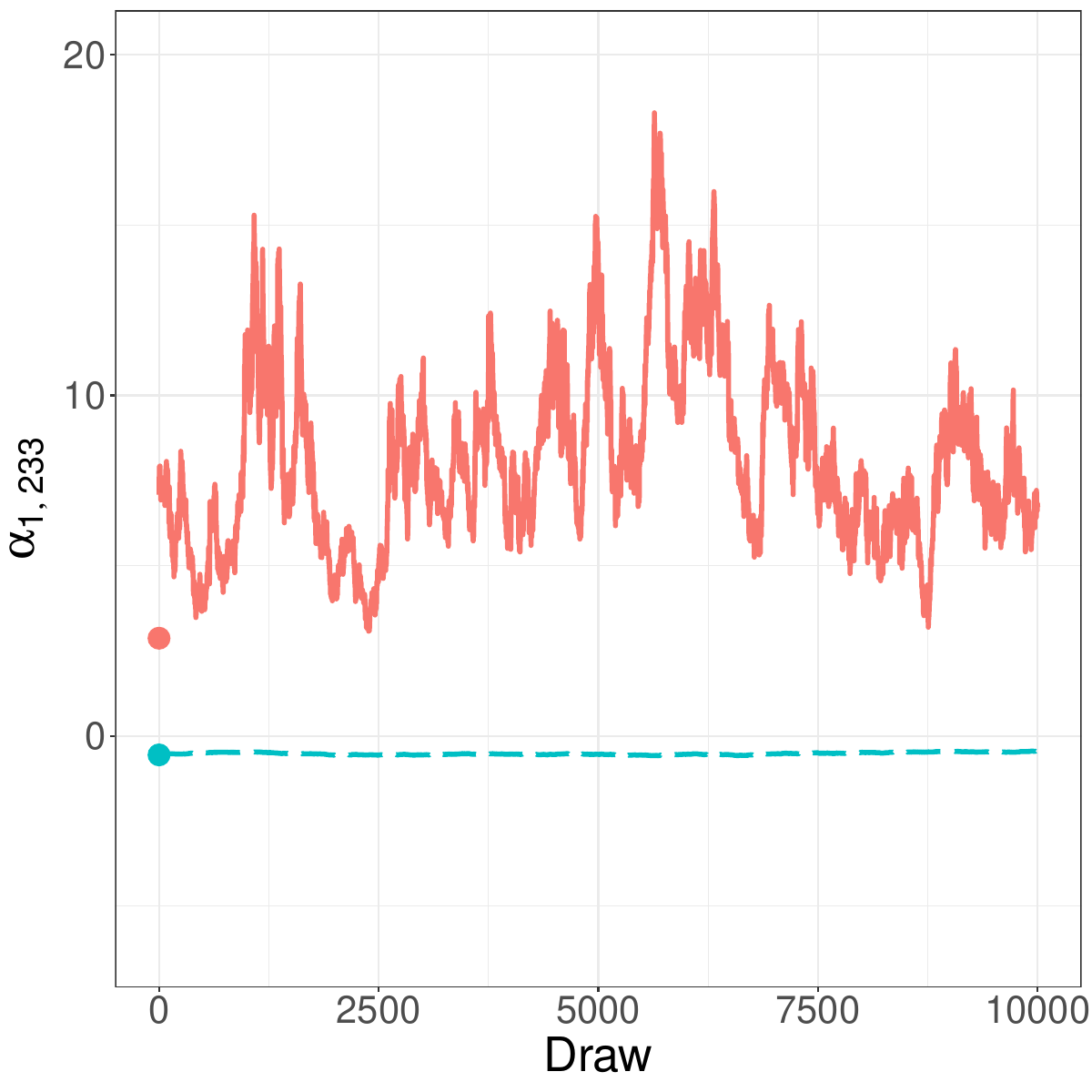}
        \caption{}\label{fig:stan_traceplot}
    \end{subfigure}
    \begin{subfigure}[t]{0.4\textwidth}
        \includegraphics[width = \textwidth]{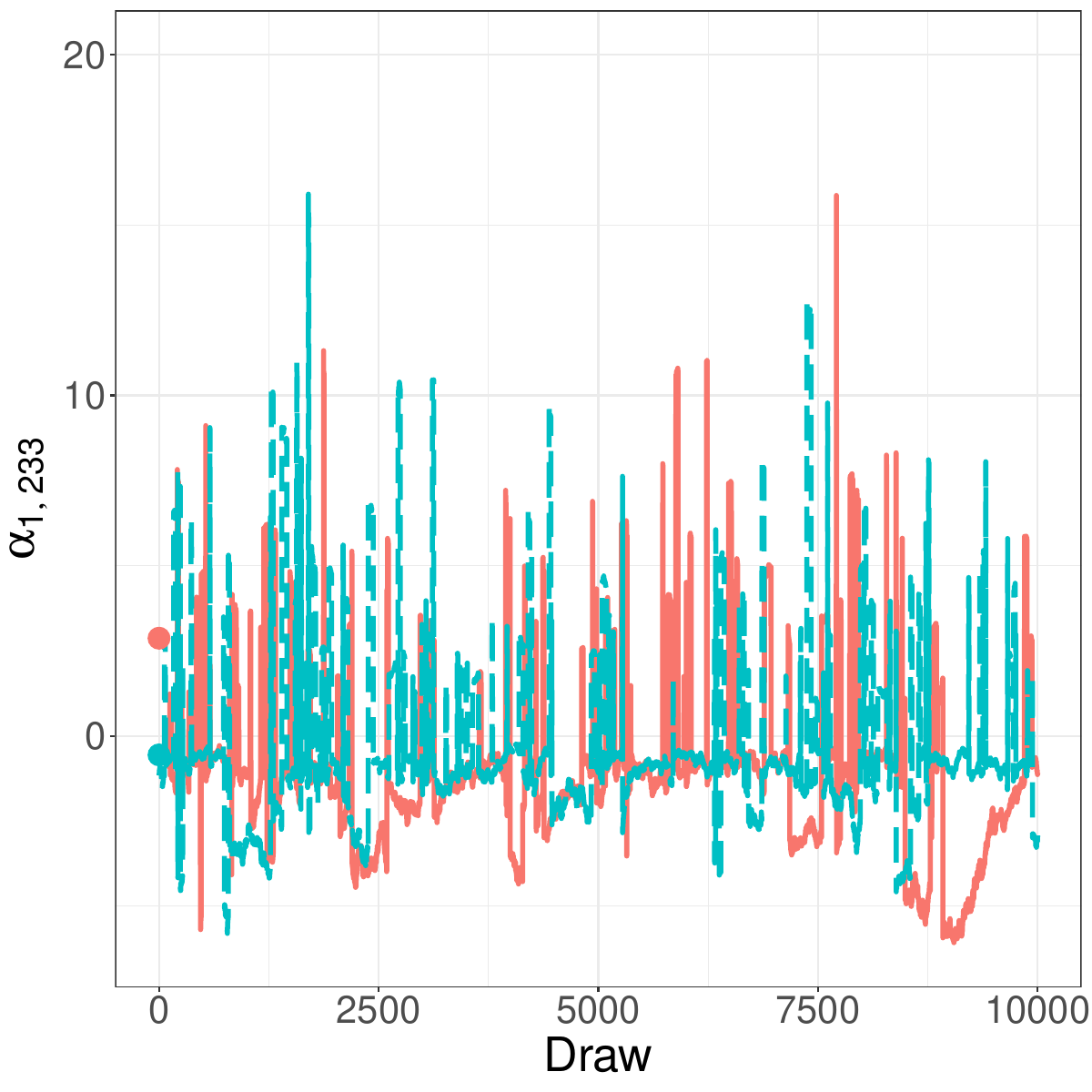}
        \caption{}\label{fig:gibbs_traceplot}
    \end{subfigure}
    \centering
    \begin{subfigure}[t]{0.4\textwidth}
        \includegraphics[width = \textwidth]{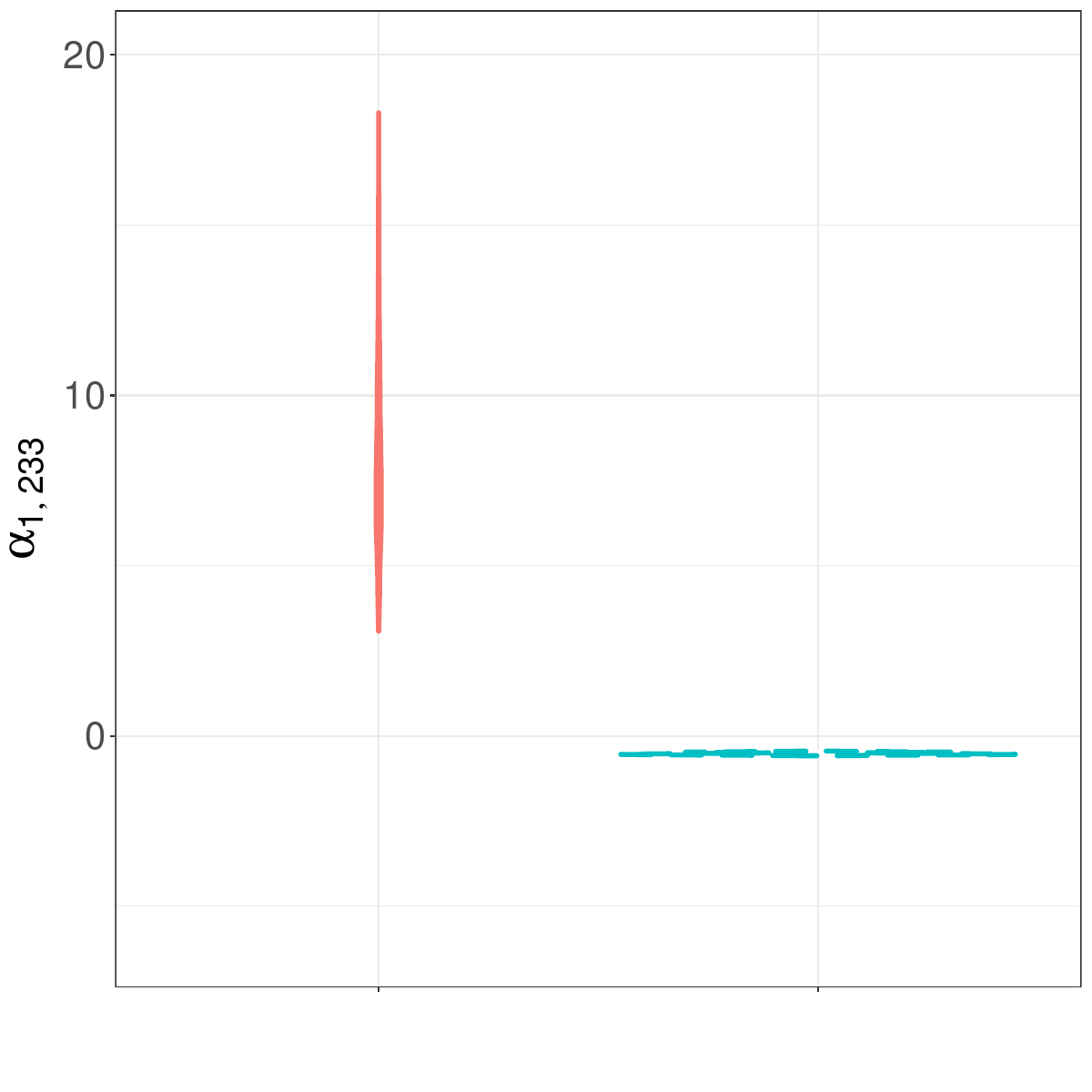}
        \caption{}\label{fig:stan_violin}
    \end{subfigure}
    \begin{subfigure}[t]{0.4\textwidth}
        \includegraphics[width = \textwidth]{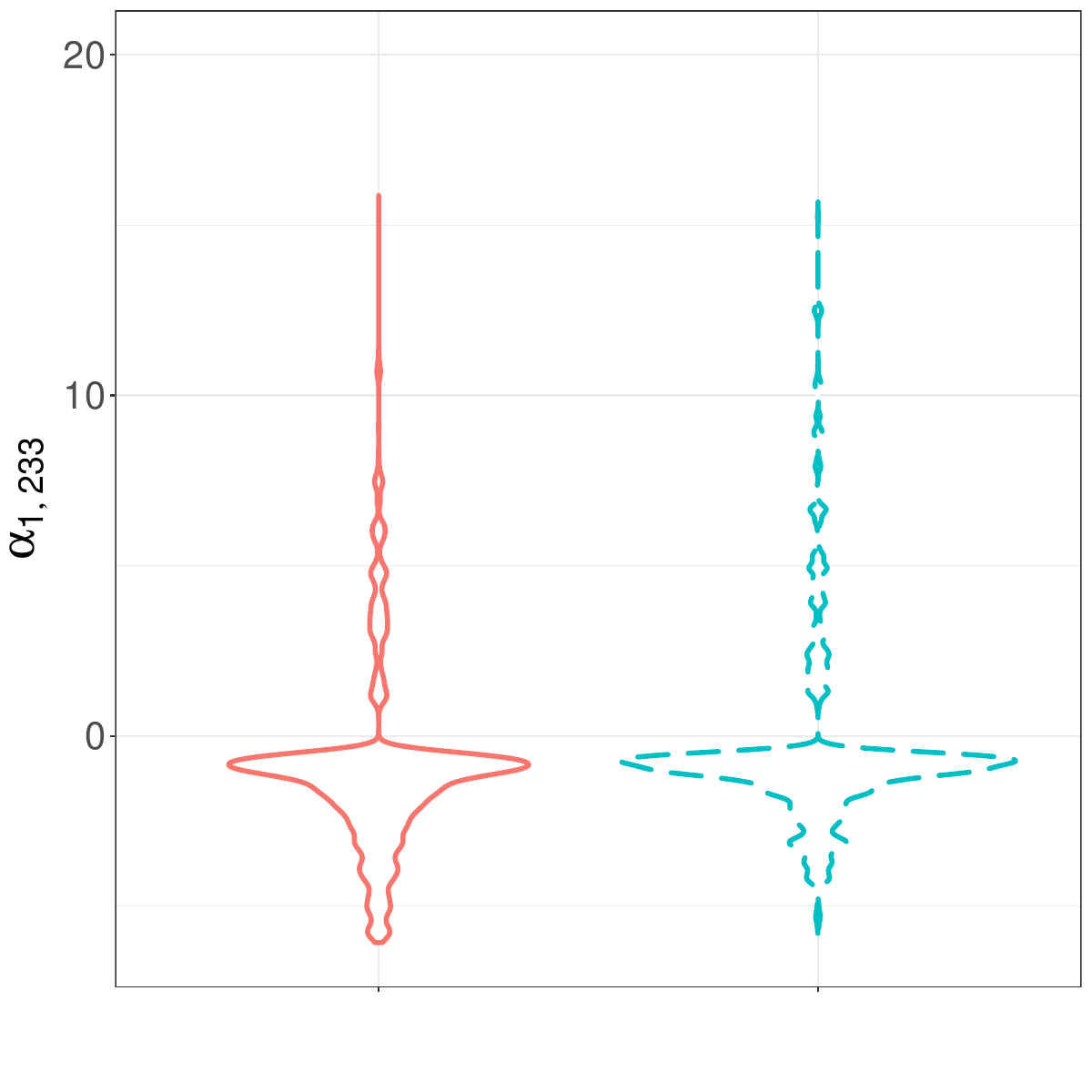}
        \caption{}\label{fig:gibbs_violin}
    \end{subfigure}
    \caption{Trace plots representative $\alpha_{j,1}$ parameter from two chains generated using the Stan and Gibbs implementation of the logistic unfolding model from Section \ref{sec:model_formulation}.  The only difference between the two chains is the initial values of 10\% of the $\bfalpha_j$s, which are placed on opposite orthants.  The data being fitted is the same voting data for the 116\textsuperscript{th} U.S.\ House of Representatives discussed in Section \ref{sec:empirical}. The dot in each graph indicates the initial value of the chain.}
    \label{fig:combined_example}
\end{figure*}

Before describing the computational algorithm we propose, we discuss in more detail the computational challenges associated with the proposed logistic unfolding model.  First, we discuss the challenges associated with trying to employ off-the-shelf software such as \texttt{RStan} to fit the model.  Note that the prior in \eqref{eq:alphadeltaprior} has support over two orthants that connect through a single point (the origin).  This makes it  impossible for generic Hamiltonian Monte Carlo algorithms to move between orthants, as paths connecting points on opposite orthants have zero probability of being proposed.  This is demonstrated empirically in Figures \ref{fig:stan_traceplot} and \ref{fig:stan_violin}, which show trace and violin plots for one representative $\alpha_{j,1}$ parameter for two chains generated using an \texttt{RStan} implementation of the logistic unfolding model applied to voting data from the 116\textsuperscript{th} U.S.\ House of Representatives (see Section \ref{sec:empirical}).  The two chains differ in the initial values of 10\% of the $\bfalpha_j$s, which are placed on opposite orthants.  The code and data required to reproduce these graphs is available at \url{https://github.com/rayleigh/logit_unfolding_model/}.  From the graphs, it seems that \texttt{RStan} is not only unable to move  between orthants, but it seems to fail to properly explore the posterior distribution within each orthant.  On the other hand, Figures \ref{fig:gibbs_traceplot} and \ref{fig:gibbs_violin} show  results for the algorithm we discuss in Section \ref{sec:gibbs}.  This approach does allow for the chains to cross over and to converge to the same posterior distribution (which, as the figures show, can be multimodal in some instances) no matter which orthant the initial value of the chain is in.

We consider now the computational challenges associated with more complex models that build on the  model introduced in Section \ref{sec:model_formulation}.  For the sake of concreteness, we briefly consider an extension inspired by \cite{lofland2017assessing}, in which votes are broken down into two (known) voting domains (e.g.,  procedural vs.\ substantive votes, or economic vs.\ social policy votes).  In such situations, legislators might exhibit different preferences across domains, and a key substantive question is identifying which ones among them actually do. This question can be addressed with a slight extension of the model in which each legislator is (in principle) allowed two ideal different ideal points (one for each group), say, $\beta_{i,1}$ and $\beta_{i,2}$.  The pair is then modeled jointly using a spike-and-slab type prior such that
$$
(\beta_{i,1}, \beta_{i,2})' \sim \begin{cases}
    \textrm{N}(\beta_{i,1} \mid 0, 1) \textrm{N}(\beta_{i,2} \mid 0, 1) & \xi_i=0 , \\
    \textrm{N}(\beta_{i,1} \mid 0, 1) \delta_{\beta_{i,1}} (\beta_{i,2}) & \xi_i = 1 ,
\end{cases}
$$
where $ \textrm{N}( \cdot \mid 0, 1)$ denotes a standard normal distribution and  $\delta_{a} ( \cdot )$ denotes a degenerate distribution placing all its mass at $a$.  Note that $\xi_i = 1$ corresponds to $\beta_{i,1} = \beta_{i,2}$, in which case legislator $i$ exhibits the same preferences on both domains.  The model is completed with an appropriate prior on the vector $(\xi_1, \ldots, \xi_I)$.  Unless there is a way to analytically integrate out the vector $(\beta_{i,1}, \beta_{i,2})$, computation for a model like this requires the use of Reversible Jump Markov chain Monte Carlo algorithms that are difficult to design and implement and, more importantly, tend to mix very poorly.  One key advantage of the approach we discuss in this paper is that such analytical integration is often possible.  While we do not pursue this kind of extensions further in this paper, the need to accommodate them in the future was one of the factor that led to us to consider the approaches discussed here.

\subsection{Approximating the Gumbel distribution}\label{sec:Gumbleapprox}

The key to developing our computational algorithm is to replace the Gumbel distribution for the random shocks $\epsilon_{i, j, 1}$, $\epsilon_{i, j, 2}$ and $\epsilon_{i, j, 3}$ in \eqref{eq:utility_def} with a finite mixture of Gaussian distributions:
\begin{align}
    g_K(\cdot) = \sum_{k = 1}^K \pi_{k} \textrm{N}(\cdot \mid m_{k}, s^2_k).    
    \label{eq:approx_mixture}
\end{align} 
Here, the number of components $K$, the weights $\pi_1, \ldots, \pi_K$, the means $m_1, \ldots, m_K$, and the variances $s_1^2, \ldots, s_K^2$ are selected so that $g_K$ closely matches the standard Gumbel distribution.  Once determined, these values will be treated as known for the purpose of the algorithm described in  Section \ref{sec:gibbs}.
\begin{figure*}[!t]
    \centering
    \begin{subfigure}[t]{0.45\textwidth}
        \includegraphics[width = \textwidth]{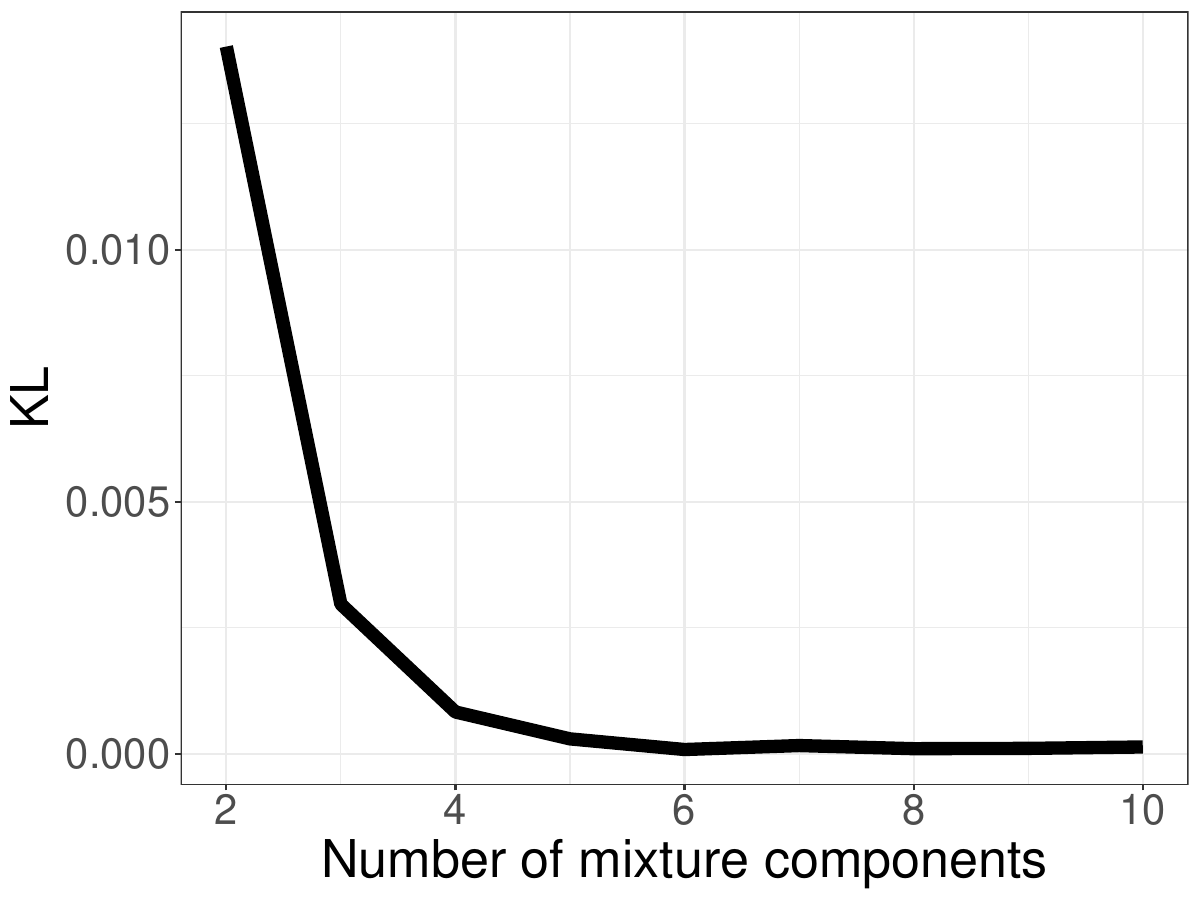}
        \caption{The KL divergence as a function of $K$}\label{fig:kl_plots_a}
    \end{subfigure}
    \begin{subfigure}[t]{0.45\textwidth}
        \includegraphics[width = \textwidth]{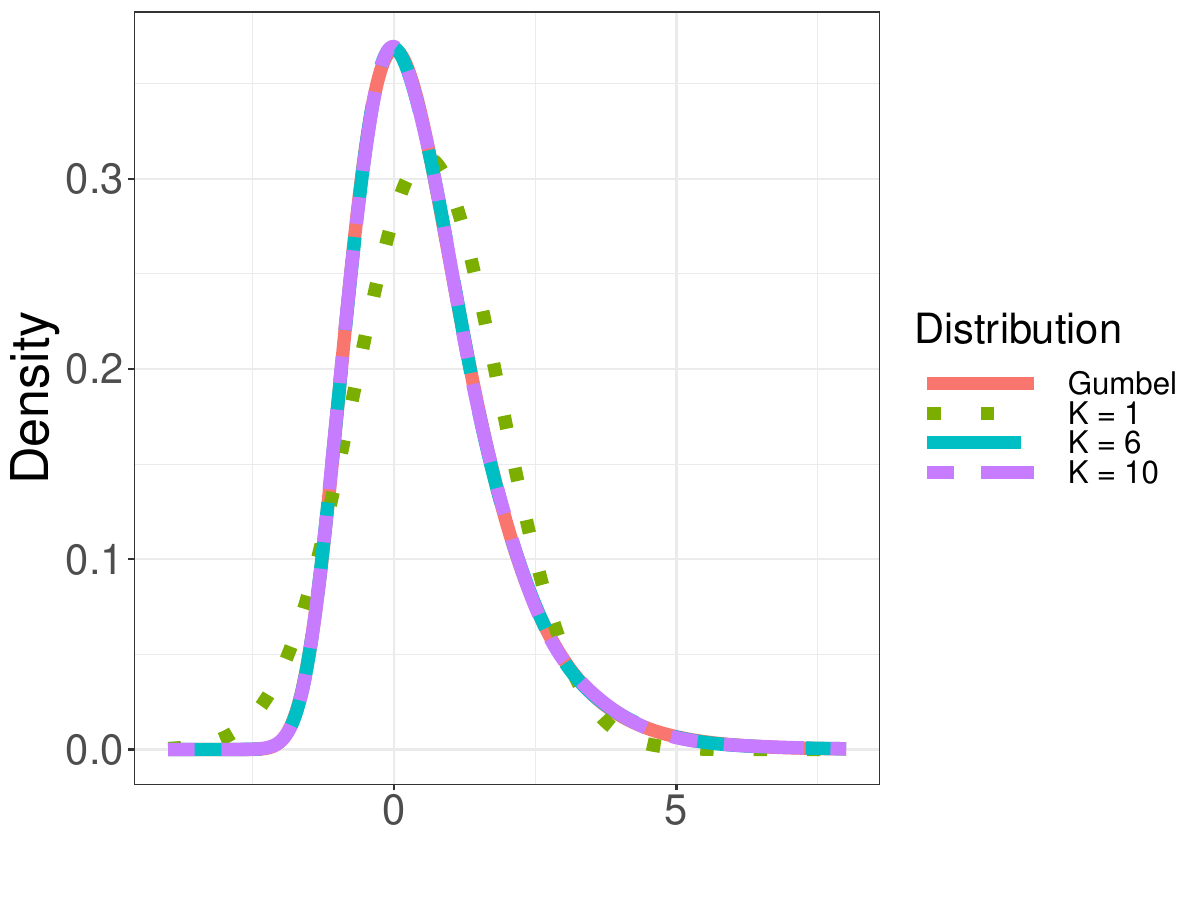}
        \caption{Densities $g$ and $g_K$ for various values of $K$}\label{fig:kl_plots_b}
    \end{subfigure}
    \caption{Plots showing the KL divergence between the Gumbel distribution and a mixture of normal distributions on the left and the density plot with different number of Gumbel distributions.}
    \label{fig:kl_plots}
\end{figure*}

The idea of replacing an intractable distribution with a mixture of Gaussians for the purpose of deriving MCMC algorithms for Bayesian model estimation has a long history in the literature.  Various approaches to identify the approximation have been proposed.  For example,  \cite{KimEtAlStochasticVolatilityLikelihood1998} and \cite{OmoriEtAlStochasticVolatilityLeverage2007} select the parameters of the mixture by matching moments.  Instead, in this paper, we follow \cite{Fruhwirth-SchnatterFruhwirthAuxiliaryMixtureSampling2007} and \cite{Fruhwirth-SchnatterFruhwirthBayesianInferenceMultinomial2012} and focus on minimizing the Kullback-Lieber (KL) divergence:
\begin{align}\label{eq:KLdiv}
d_{KL}(g, g_K) &= \int_{\mathbbm{R}} \log\left( \frac{g(\epsilon)}{g_K(\epsilon)} \right) g(\epsilon) \dd \epsilon,
\end{align}
where $g(\epsilon) = \exp\left\{ - \exp\left( -x \right) \right\}$ denotes the standard Gumbel distribution. More specifically, for a given $K$, $\pi_1, \ldots, \pi_K$, $m_1, \ldots, m_K$ and $s_1^2, \ldots, s_K^2$ are selected by solving
\begin{multline}
\label{eq:optimalparKL}
(\pi_1, \ldots, \pi_K, m_1, \ldots, m_K, s_1^2, \ldots, s_K^2) =\\ \argmax d_{KL}(g, g_K) .
\end{multline}
Then, because \eqref{eq:KLdiv} is strictly decreasing in $K$, the number of components is selected to be large enough to ensure that the marginal decrease in the KL distance is negligible.

While there is no closed form solution for the integral in \eqref{eq:KLdiv} or the optimization problem in \eqref{eq:optimalparKL}, a solution can be obtained numerically using standard algorithms.  To speed up convergence, the solutions obtained for a given value $K$ are used to generate the initial values for the parameters for $K+1$.  Figure \ref{fig:kl_plots_a} shows the value of the minimum KL divergence for various values of $K$.  We can see that the value tends to stabilize around $K=6$.  Tables \ref{table:gumbel_approx_6} and \ref{table:gumbel_approx_10} show the optimal values of $\pi_1, \ldots, \pi_K$, $m_1, \ldots, m_K$ and $s_1^2, \ldots, s_K^2$ for $K=6$ and $K=10$, respectively.  Finally, Figure \ref{fig:kl_plots_b} compares the density of $g_1$, $g_{6}$ and $g_{10}$ against that of the standard Gumbel.

\begin{table*}[!ht] \centering 
\begin{tabular}{@{\extracolsep{5pt}} ccccccc} 
\\[-1.8ex]\hline 
\textbf{Parameter} & \textbf{1} & \textbf{2} & \textbf{3} & \textbf{4} & \textbf{5} & \textbf{6}\\
\hline \\[-1.8ex] 
$m_k$ & 0.455 & -0.354 & 1.497 & 2.275 & -1.016 & 4.270 \\ 
$s_k$ & 0.649 & 0.516 & 0.768 & 1.297 & 0.397 & 1.948 \\ 
$\pi_k$ & 0.365 & 0.279 & 0.160 & 0.123 & 0.061 & 0.012\\ 
\hline \\[-1.8ex] 
\end{tabular} 
  \caption{Table showing the mean, standard deviation, and mixing probability for the mixture of six normal distributions approximating the Gumbel distribution arranged based on the mixing probability.} 
  \label{table:gumbel_approx_6} 
\end{table*} 

\begin{table*}[!htbp] \centering 
\begin{tabular}{@{\extracolsep{5pt}} ccccccccccc} 
\\[-1.8ex]\hline 
\textbf{Parameter} & \textbf{1} & \textbf{2} & \textbf{3} & \textbf{4} & \textbf{5} & \textbf{6} & \textbf{7} & \textbf{8} & \textbf{9} & \textbf{10}\\
\hline \\[-1.8ex] 
$m_k$ & -0.117 & 2.062 & 1.310 & -0.896 & 0.679 & 0.885 & -0.243 & 0.551 & 1.565 & 4.087 \\ 
$s_k$ & 0.529 & 1.265 & 0.733 & 0.427 & 0.448 & 0.678 & 0.456 & 0.603 & 0.693 & 1.914 \\ 
$\pi_k$ & 0.307 & 0.156 & 0.123 & 0.116 & 0.090 & 0.073 & 0.058 & 0.035 & 0.024 & 0.016 \\ 
\hline \\[-1.8ex] 
\end{tabular} 
\caption{Table showing the mean, standard deviation, and mixing probability for the mixture of ten normal distributions approximating the Gumbel distribution arranged based on the mixing probability.} 
\label{table:gumbel_approx_10} 
\end{table*}

\subsection{Gibbs sampler}\label{sec:gibbs}

We now discuss the details of the Gibbs sampler we propose to fit the logistic model from Section \ref{sec:model_formulation}.  The algorithm relies on three data augmentation tricks, which enable  us to rely heavily on Gibbs steps and to design transition kernels that allow us to address the computational challenges discussed in Section \ref{sec:compchallenges}.

The first augmentation trick is reminiscent of that described in \cite{albert1993bayesian} coupled with the approximation discussed in Section \ref{sec:Gumbleapprox}.  It involves the introduction of vectors of latent variables, $\bfy_{i,j}^{*} = (y^{*}_{i, j, 1}, y^{*}_{i, j, 2}, y^{*}_{i, j, 3})'$ for every $i=1\ldots,I$ and $j=1,\ldots,J$, such that 
\begin{align*}
    y^{*}_{i, j, 1} & = - \alpha_{j,1}(\beta_{i} - \delta_{j,1}) + e_{i, j, 1} \\
    y^{*}_{i, j, 2} & =  e_{i, j, 2} \\
    y^{*}_{i, j, 3} & = - \alpha_{j,2}(\beta_{i} - \delta_{j,2}) + e_{i, j, 3}
\end{align*}
where $e_{i, j, 1}, e_{i, j, 2}, e_{i, j, 3}$ denote the mixture approximations to $\epsilon_{i, j, 1}, \epsilon_{i, j, 2}, \epsilon_{i, j, 3}$ and are independently distributed according to  \eqref{eq:approx_mixture}. The definition of these latent variables is linked to the structure of the utility functions $U_{N+}, U_{Y}, U_{N-}$ introduced in Section \ref{sec:model_formulation}. Indeed, we can re-express $y_{i, j, 1}^*$ and $y_{i, j, 3}^*$ in terms of $\beta, \psi_{j, 1}, \psi_{j, 2},$ and $\psi_{j, 3}$ so that $y_{i, j, 1}^* = -(\beta_i - \psi_{j, 1})^2 + (\beta_i - \psi_{j, 2})^2 + e_{i, j, 1}$ and $y_{i, j, 3}^* = -(\beta_i - \psi_{j, 3})^2 + (\beta_i - \psi_{j, 2})^2 + e_{i, j, 3}.$ Thus, $y_{i, j, 1}^* < y_{i, j, 2}^*$ is equivalent to $U_{N-} < U_{Y}$ and $y_{i, j, 3}^* < y_{i, j, 2}^*$ is equivalent to $U_{N+} < U_{Y}$ and, under this augmentation, $y_{i,j} = 1$ if and only if $y^{*}_{i, j, 2} > y^{*}_{i, j, 1}$ and $y^{*}_{i, j, 2} > y^{*}_{i, j, 3}$.

The second augmentation relies on the standard latent variable representation for Gaussian mixtures.  More specifically, for every $i=1,\ldots, I$ and $j=1,\ldots, J$, let $\lambda_{i, j, \ell} \in \{1, 2, \ldots, K\}$ for $\ell \in \{1, 2, 3\}$ so that
\begin{align*}
    \lambda_{i, j, \ell} &\sim \textrm{Cat}(\pi_1, \ldots, \pi_K)\\
    e_{i, j, \ell} \mid \lambda_{i, j, \ell} &\sim \textrm{N}(\cdot \mid m_{\lambda_{i, j, \ell}},  s^2_{\lambda_{i, j, \ell}}).
\end{align*}
Clearly, integrating (summing) over all possible values of $\lambda_{i,j, \ell}$ gives back the mixture distribution for $e_{i,j, \ell}$ in \eqref{eq:approx_mixture}. Now, 
conditioning on these latent variables, the joint distribution of the latent utilities follows a trivariate normal distribution
\begin{multline*}
    \bfy_{i,j}^{*}
    \mid
    \alpha_{j,1}, \alpha_{j,2}, \delta_{j,1}, \delta_{j,2}, \beta_i,\\ \lambda_{i, j, 1},\lambda_{i, j, 2}, \lambda_{i, j, 3}
    \sim \textrm{N}(\cdot \mid \v{\mu^*}_{i, j}, \v{\Sigma}^*_{i, j}),
\end{multline*}
where 
\begin{align}\label{eq:utilitypar}
    \v{\mu^*}_{i, j} &= \begin{pmatrix}
        m_{\lambda_{i, j , 1}} -\alpha_{j,1}(\beta_i - \delta_{j,1})\\
        m_{\lambda_{i, j , 2}}\\
        m_{\lambda_{i, j , 3}} -\alpha_{j,2}(\beta_i - \delta_{j,2})\\
    \end{pmatrix} ,\\ 
    \v{\Sigma}^*_{i, j} &=
    \begin{pmatrix}
        s^2_{\lambda_{i, j , 1}} & 0 & 0\\
        0 &  s^2_{\lambda_{i, j , 2}} & 0\\
        0 & 0 &  s^2_{\lambda_{i, j , 3}}\\
    \end{pmatrix}.
    \nonumber
\end{align}

Taken together, these first two augmentations ensure that most of the full conditional distributions belong to standard families from which direct sampling is possible.  Furthermore, the conditional Gaussian distributions for the latent utilities imply that it will often be possible to integrate out $\beta_1, \ldots, \beta_I$ (recall the discussion in the last paragraph of Section \ref{sec:compchallenges}).

The third data augmentation relies on a similar trick to break down the mixture prior on $\bfalpha_j$ and $\bfdelta_j$ into the two fully connected regions.  In particular, for $j=1,\ldots,J$, we let $z_{j} = 1$ if and only if $\alpha_{j,1} > 0$ and $\alpha_{j,2} < 0$, and $z_{j} = -1$ otherwise.  Then, $\Pr(z_{i,j} = 1) = \Pr(z_{i,j} = -1) = 1/2$, and 
\begin{align*}
    p(& \bfalpha_j, \bfdelta_j \mid z_j) =\\ 
    & \begin{cases}
    \frac{1}{16 \pi^2 \omega^2 \kappa^2} \exp\left\{ -\frac{1}{2} \left(\frac{1}{\omega^2 }\bfalpha_j'\bfalpha_j  + \right.\right.\\
    \quad \left. \left. \frac{1}{\kappa^2}(\bfdelta_j - \v{\vartheta})'(\bfdelta_j - \v{\vartheta})\right)\right\}\\ 
    \quad \mathbbm{1}(\alpha_{j,1}>0, \alpha_{j,2}<0) , & z_{i,j} = 1 , \\
    \frac{1}{16 \pi^2 \omega^2 \kappa^2} \exp\left\{ -\frac{1}{2} \left(\frac{1}{\omega^2 }\bfalpha_j'\bfalpha_j  + \right.\right.\\
    \left. \left.
    \quad \frac{1}{\kappa^2}(\bfdelta_j + \v{\vartheta})'(\bfdelta_j + \v{\vartheta})\right)\right\}\\ 
    \quad \mathbbm{1}(\alpha_{j,1}<0, \alpha_{j,2}>0) , & z_{i,j} = -1 .
    \end{cases}
\end{align*}

This third augmentation is particular important to ensure that the algorithm can fully explore the posterior distribution of $\bfalpha_j$ and, in particular, that it can move between the quadrant where $\alpha_{j,1} >0$ and $\alpha_{j,2}<0$, and that where $\alpha_{j,1} <0$ and $\alpha_{j,2}>0$.  To accomplish this, we rely on three types of moves: (a) a Gibbs sampling move in which we simulate $\bfalpha_j$ and $z_j$ together by first sampling from $z_j$ after integrating out $\bfalpha_j$ from the full joint posterior, (b) a Metropolis-Hastings step in which the proposal is generated by flipping the signs of $\bfalpha_j$ and $\bfdelta_j$, and (c) another Metropolis-Hasting step in which a flip is proposed and the associated value of $\bfdelta_j$ is proposed from the prior conditioned on the proposed quadrant. 
Moves (b) and (c) are Metropolis-Hastings-within-Gibbs steps.  They lead to an overall Harris-recurrent Markov chain because the prior probability of $\alpha_{j,1}, \alpha_{j, 2}, \delta_{j, 1}, \delta_{j, 2}$ is greater than zero for the entire support and there is positive probability of changing the values of $\alpha_{j,1}$, $\alpha_{j, 2}$, $\delta_{j, 1}$ and $\delta_{j, 2}$ (see Theorem 12 in \citealp{roberts2006harris}).  Move (b) is expected to be most effective for items for which $\bfalpha_j$ is close to $\mathbf{0}$ (e.g., quasi-unanimous votes in our political science application).  The number of $(\bfalpha_j, \bfdelta_j)$ pairs in this category is typically relatively low, but when the move is useful, the acceptance probability tends to be very high.  Hence, we propose to perform move (b) relatively infrequently.  On the other hand, move (c) is most useful when the posterior distribution of $(\alpha_j, \delta_j)$ is clearly bimodal.  Move (c) tends to be useful for a larger number of $(\bfalpha_j, \bfdelta_j)$ pairs, but also have more moderate acceptance probabilities.  Hence, we propose to perform it relatively more frequently.

With these general considerations in mind, we proceed now to provide a detailed description of the kernels we use in our proposed MCMC algorithm:

\begin{enumerate}
    \item For $i = 1, \ldots, I$ and $j = 1, \ldots, J$, sample $\lambda_{i, j , 1}$, $\lambda_{i, j , 2}$ and $\lambda_{i, j , 3}$ from their full conditional distributions
    \begin{align*}
    p(&\lambda_{i, j , 1} = k \mid \cdots) =\\
    & \frac{\pi_k \textrm{N}(y^*_{i, j , 1} \mid m_k - \alpha_{j, 1}(\beta_i - \delta_{j, 1}), s^2_k)}{\sum_{k' = 1}^K \pi_{k'} \textrm{N}(y^*_{i, j , 1} \mid m_{k'} - \alpha_{j, 1}(\beta_i - \delta_{j, 1}), s^2_{k'})},\\
    p(&\lambda_{i, j , 2} = k \mid \cdots) =\\ & \frac{\pi_k\textrm{N}(y^*_{i, j , 2} \mid m_{k}, s^2_{k'})}{\sum_{k' = 1}^K \pi_{k'} \textrm{N}(y^*_{i, j , 2} \mid m_{k'}, s^2_{k'})},\\
    p(& \lambda_{i, j , 3} = k \mid \cdots) =\\ & \frac{\pi_k \textrm{N}(y^*_{i, j , 3} \mid m_k - \alpha_{j, 2}(\beta_i - \delta_{j, 2}), s^2_k)}{\sum_{k' = 1}^K \pi_{k'} \textrm{N}(y^*_{i, j , 3} \mid m_{k'} - \alpha_{j, 2}(\beta_i - \delta_{j, 2}), s^2_{k'})}.
    \end{align*}

    \item For $i = 1, \ldots, I$ and $j = 1, \ldots, J$, draw $y^*_{i, j , 1}$, $y^*_{i, j , 2}$ and $y^*_{i, j , 3}$ from the following full conditional distribution:
        \begin{align*}
        y^*_{i, j, 1} &\mid \cdots \sim\\ 
        & \begin{cases}
            \textrm{N}(\cdot \mid \mu^*_{i, j, 1}, s^2_{\lambda_{i, j, 1}})\\
            \quad\mathbbm{1}(-\infty,  y^*_{i, j, 2}) & y_{i, j} = 1,\\
            \textrm{N}(\cdot \mid \mu^*_{i, j, 1}, s^2_{\lambda_{i, j, 1}})\\
            \quad \mathbbm{1}(y^*_{i, j, 2}, \infty) & y_{i, j} = 0, y^*_{i, j, 3} < y^*_{i, j, 2},\\
            \textrm{N}(\cdot \mid \mu^*_{i, j, 1}, s^2_{\lambda_{i, j, 1}}) & \textrm{o.w.}\\
        \end{cases}\\
        y^*_{i, j, 2} &\mid \cdots \sim\\
        &\begin{cases}
            \textrm{N}(\cdot \mid \mu^*_{i, j, 2}, s^2_{\lambda_{i, j, 2}})\\ 
            \quad \mathbbm{1}(\max(y^*_{i, j, 1}, y^*_{i, j, 3}), \infty) & y_{i, j} = 1,\\
            \textrm{N}(\cdot \mid \mu^*_{i, j, 2}, s^2_{\lambda_{i, j, 2}})\\ 
            \quad \mathbbm{1}(-\infty, \max(y^*_{i, j, 1}, y^*_{i, j, 3})) & y_{i, j} = 0.\\
        \end{cases}\\
        y^*_{i, j, 3} &\mid \cdots \sim\\ 
        & \begin{cases}
            \textrm{N}(\cdot \mid \mu^*_{i, j, 3}, s^2_{\lambda_{i, j, 3}})\\ 
            \quad\mathbbm{1}(-\infty,  y^*_{i, j, 2}) & y_{i, j} = 1,\\
            \textrm{N}(\cdot \mid \mu^*_{i, j, 3}, s^2_{\lambda_{i, j, 3}})\\ 
            \quad\mathbbm{1}(y^*_{i, j, 2}, \infty) & y_{i, j} = 0, y^*_{i, j, 1} < y^*_{i, j, 2},\\
            \textrm{N}(\cdot \mid \mu^*_{i, j, 3}, s^2_{\lambda_{i, j, 3}}) & \textrm{o.w.}\\
        \end{cases}
    \end{align*}
    where $\bfmu^{*}_{i,j}$ was defined in  \eqref{eq:utilitypar}.

    \item For $i = 1, 2, \ldots, I$, sample $\beta_i \mid \cdots \sim \textrm{N}(\mu_{\beta_i}, \sigma^2_{\beta_i})$, where
    \begin{align*}
        \sigma^2_{\beta_i} &= \left(1 + \sum_{j = 1}^{J} \frac{1}{s^2_{\lambda_{i, j, 1}}s^2_{\lambda_{i, j, 3}}}\bfalpha_j' \bfalpha_j\right)^{-1},\\
        \mu_{\beta_i} &= -\sigma^2_{\beta_i}\left(\sum_{j = 1}^J \frac{1}{s^2_{\lambda_{i, j, 1}}s^2_{\lambda_{i, j, 3}}}\bfalpha_j' \right. \\ 
        & \qquad\left.\left(
        \begin{pmatrix}
            y^{*}_{i, j, 1}\\ y^{*}_{i, j, 3}
        \end{pmatrix} - \bfD_{\bfalpha_j}\bfdelta_j -
        \begin{pmatrix}
            m_{\lambda_{i, j, 1}}\\
            m_{\lambda_{i, j, 3}}
        \end{pmatrix}\right)\right),
    \end{align*}
    and $\bfD_{\bfalpha_j} = \diag\{ \alpha_{j,1}, \alpha_{j,2}\}$.
    \item For $j = 1, 2, \ldots, J$, sample $\bfalpha_j, z_j$ according to its full conditional distributions in two steps. Note that for each $j$, we only include the $i \in \{1, 2, \ldots, I\}$ such that $y_{i, j} \in \{0, 1\}$.
    \begin{enumerate}
        \item Draw $z_j$ from its full conditional distribution:
        \begin{align*}
          p(z_j &\mid \cdots) \propto\\
          & \begin{cases}
              \phi_2 \left(\bfdelta_j \mid \v{\vartheta}, \kappa^2 \bfI_{2}\right)\\ \quad\left\{ 1 - \Phi\left(  -\frac{(\bfmu_{\bfalpha_j})_1}{\left(\bfSigma_{\bfalpha_j}\right)_{1, 1}}  \right) \right\}\\ 
              \quad \Phi\left(-\frac{(\bfmu_{\bfalpha_j})_2}{\left(\bfSigma_{\bfalpha_j}\right)_{2, 2} }\right)
              & z_j = 1, \\
              \phi_2 \left(\bfdelta_j \mid -\v{\vartheta}, \kappa^2 \bfI_{2}\right)\\ 
              \quad \Phi\left(  -\frac{(\bfmu_{\bfalpha_j})_1}{\left(\bfSigma_{\bfalpha_j}\right)_{1, 1}}  \right)\\  
              \quad \left\{ 1 -\Phi\left(-\frac{(\bfmu_{\bfalpha_j})_2}{\left(\bfSigma_{\bfalpha_j}\right)_{2, 2} }\right) \right\}
              & z_j = -1,
          \end{cases}
        \end{align*}
    where $\phi_p(\cdot \mid \bfmu, \bfSigma)$ denotes the density of the $p$-variate normal distribution with mean $\bfmu$ and variance $\bfSigma$, $\Phi(\cdot)$ denotes the cumulative distribution function of the standard univariate normal distribution, $\bfI_{p}$ is the $p \times p$ identity matrix, and $\bfmu_{\bfalpha_j}$ and $\bfSigma_{\bfalpha_j}$ are such that
    \begin{align}
    &\bfSigma_{\bfalpha_j} = \nonumber\\ 
    & \left(\sum_{i = 1}^I \bfD_{\beta_i , \bfdelta_j} 
    \begin{pmatrix}
            s^2_{\lambda_{i, j, 1}} & 0\\
            0 & s^2_{\lambda_{i, j, 3}} 
    \end{pmatrix}^{-1} \right. \label{eq:alpha_post_cov}\\
    & \left. 
     \qquad\qquad  \bfD_{\beta_i , \bfdelta_j} + \frac{1}{\omega^2}\bfI_{2}\right)^{-1}, \nonumber
    \\
    & \bfmu_{\bfalpha_j} = - \bfSigma_{\bfalpha_j}  \nonumber\\  
    & \left(\sum_{i = 1}^I 
    \begin{pmatrix}
            s^2_{\lambda_{i, j, 1}} & 0\\
            0 & s^2_{\lambda_{i, j, 3}} 
    \end{pmatrix}^{-1}
    \bfD_{\beta_i , \bfdelta_j} \right.     \label{eq:alpha_post_mu}\\
    & \left. \quad \left(
    \begin{pmatrix}
        y^{*}_{i, j, 1}\\
        y^{*}_{i, j, 3}
    \end{pmatrix}
    - 
    \begin{pmatrix}
        m_{\lambda_{i, j, 1}}\\
        m_{\lambda_{i, j, 3}}
    \end{pmatrix}\right)\right).
    \nonumber
    \end{align}
    Here, $\bfD_{\beta_i , \bfdelta_j} = \beta_i \bfI_{2} - \diag\{\delta_{j, 1},  \delta_{j, 2}\}$.
    
    \item Sample $\bfalpha_j$ according to its full conditional distribution, a truncated bivariate normal distribution: 
    \begin{align*}
        \bfalpha_j \mid \cdots &\sim 
        \begin{cases}
            \textrm{N}\left(\cdot \mid \v{\mu}_{\bfalpha_j}, \bfSigma_{\bfalpha_j}\right)\\
            \quad \mathbbm{1}(\alpha_{j,1}>0, \alpha_{j,2}<0) & z_j = 1\\
            \textrm{N}\left(\cdot \mid \v{\mu_{\bfalpha_j}}, \bfSigma_{\bfalpha_j}\right)\\
            \quad \mathbbm{1}(\alpha_{j,1}<0, \alpha_{j,2}>0) & z_j = -1,
        \end{cases}
    \end{align*}
    where $\v{\mu}_{\bfalpha_j}$ and $\bfSigma_{\bfalpha_j}$ are given in \eqref{eq:alpha_post_cov} and \eqref{eq:alpha_post_mu}. 
    \end{enumerate}
    
    \item For $j = 1, 2, \ldots, J$, sample $\bfdelta_j$ from its full conditional distribution, $\textrm{N}(\bfmu_{\bfdelta_j}, \bfSigma_{\bfdelta_j})$, where
    \begin{align*}
        & \bfSigma_{\bfdelta_j} =\\ 
        & \left(\sum_{i = 1}^I \bfD_{\bfalpha_j} 
        \begin{pmatrix}
            s^2_{\lambda_{i, j, 1}} & 0\\
            0 & s^2_{\lambda_{i, j, 3}} 
        \end{pmatrix}^{-1}
        \bfD_{\bfalpha_j} + \frac{1}{\kappa^2}\bfI_{2}\right)^{-1},\\ 
        & \v{\mu}_{\bfdelta_j} =\\ 
        & \quad\bfSigma_{\bfdelta_j} \left(\sum_{i = 1}^I \begin{pmatrix}
            s^2_{\lambda_{i, j, 1}} & 0\\
            0 & s^2_{\lambda_{i, j, 3}} 
        \end{pmatrix}^{-1} \right. \\ 
        & \quad \left. \bfD_{\bfalpha_j} \left(
        \begin{pmatrix}
            y^{*}_{i, j, 1}\\ y^{*}_{i, j, 3}
        \end{pmatrix} + \bfalpha_j\beta_i - \begin{pmatrix}
            m_{\lambda_{i, j, 1}}\\
            m_{\lambda_{i, j, 3}}
        \end{pmatrix}\right) + z_j \frac{\v{\vartheta}}{\kappa^2}\right).
\end{align*}
\item For every fifth iteration, propose $\bfalpha_j', \bfdelta_j', z'_j$ for $j = 1, 2, \ldots, J$ in the following way.
\begin{itemize}
    \item With probability 0.1, set $z'_j = -z_j$, $\bfalpha_j' = -\bfalpha_j$ and $\bfdelta_j' = -\bfdelta_j$.

    \item With probability 0.9, let $z'_j = -z_j$ and draw proposals 
    \begin{align*}
        \bfalpha_j' &\sim \begin{cases}
            \textrm{N}(\cdot \mid \v{0}, \omega^2 \bfI_{2})\\
            \quad \mathbbm{1}(\alpha'_{j, 1} > 0, \alpha'_{j, 2} < 0) & z'_j = 1,\\
            \textrm{N}(\cdot \mid \v{0}, \omega^2 \bfI_{2})\\
            \quad \mathbbm{1}(\alpha'_{j, 1} < 0, \alpha'_{j, 2} > 0) & z'_j = -1,
        \end{cases}\\ 
        \bfdelta_j' &\sim \textrm{N}(\cdot \mid z'_j \v{\vartheta}, \kappa^2 \bfI_{2}).
    \end{align*}
\end{itemize}
Accept the proposed $\bfalpha_j', \bfdelta_j', z'_j$ with probability:
    \begin{multline*}
        \prod_{i = 1}^I \left(\frac{p(y_{i, j} = 1 \mid \beta_i, \bfalpha_j', \bfdelta_j')}{p(y_{i, j} = 1 \mid \beta_i, \bfalpha_j, \bfdelta_j)}\right)^{y_{i, j}}\\ 
        \left(\frac{1 - p(y_{i, j} = 1 \mid \beta_i, \bfalpha_j', \bfdelta_j')}{1 - p(y_{i, j} = 1 \mid \beta_i, \bfalpha_j, \bfdelta_j)}\right)^{1 - y_{i, j}},
    \end{multline*}
where $p(y_{i, j} = 1 \mid \beta_i, \bfalpha_j, \bfdelta_j)$ was defined in \eqref{eq:logisticBGUM}.
\end{enumerate}

The derivation of the acceptance probability for step 6 is provided in Section 2 of the Supplementary Material. Code implementing this algorithm is also available at \url{https://github.com/rayleigh/logit_unfolding_model/}.
\section{Empirical evaluation}\label{sec:empirical}

As we foreshadowed in Sections \ref{sec:intro} and \ref{sec:model_formulation}, discrete choice models are widely used in political science to estimate the  preferences of members of deliberative bodies based on their votes \citep{poole1985spatial,
JackmanMultidimensionalAnalysisRoll2001}. 
In this context, they serve to operationalize the notion of legislator's \text{ideology}.  In this Section, we use the sampler described in Section \ref{sec:computation} to fit the logistic unfolding model to four datasets of votes from the U.S.\ House of Representatives (corresponding to the 103\textsuperscript{rd}, 107\textsuperscript{th}, 110\textsuperscript{th} and 116\textsuperscript{th} Houses) and compare the results against those generated by the original probit unfolding model of \cite{LeiRodriguezNovelClassUnfolding2023} and the traditional IDEAL apporoach described in \cite{JackmanMultidimensionalAnalysisRoll2001}.  We focus on these four datasets because they encompass a broad spectrum of voting patterns (e.g., see \citealp{LeiRodriguezNovelClassUnfolding2023}).  The raw voting data is available at \url{https://voteview.com/data}.  As is common practice (e.g., see \citealp{MartinQuinnParkMCMCpack}) unanimous votes were removed from the dataset as they do not provide any information about the legislator's relative preferences.  In addition, the very small number  of legislators who missed more than 40\% of the non-unanimous votes were also removed from the dataset.  We follow most of the literature in treating any remaining missing votes as missing completely at random.  While there is some evidence that legislators might use abstentions and absences strategically (e.g., see \citealp{rodriguez2015measuring} and references therein), the number of actual missing votes in the U.S.\ House of Representative is small overall (usually, somewhere between 3\% and 5\% of all votes).  Hence, we expect this assumption to have very little impact on our results.

We use the algorithm from Section \ref{sec:gibbs} with $K=6$ mixture components to approximate the Gumbel distribution to fit the logistic unfolding model, and a slight variant of it (where the mixture is replaced by a single normal distribution) to fit the probit unfolding model of \cite{LeiRodriguezNovelClassUnfolding2023}.  IDEAL is fitted using the implementation in the \texttt{R} package \texttt{MCMCpack} \cite{MartinQuinnParkMCMCpack}.  Results for the logistic version of the unfolding model are based on 20,000 samples obtained after burning the first 500,000 samples and thinning the next 1,000,0000 samples every 50 iterations.  Results for the probit version are also based on 20,000 samples, obtained in this case after burning the first 200,000 samples of the chain and thinning the next 200,000 samples every 10 iterations.  Finally, the results for IDEAL are based on 20,000 samples obtained after burning the first 10,000 samples of the chain and with no thinning applied.  As the numbers above suggest, the algorithm for the logistic version of the unfolding model tends to mix more slowly than the probit version, and both mix substantially more slowly than the algorithm for IDEAL.  This is likely driven by the fact that the algorithm for the logit unfolding model relies on a large number of auxiliary variables.

\begin{figure*}[!ht]
    \centering
    \includegraphics[width = 0.45\textwidth]{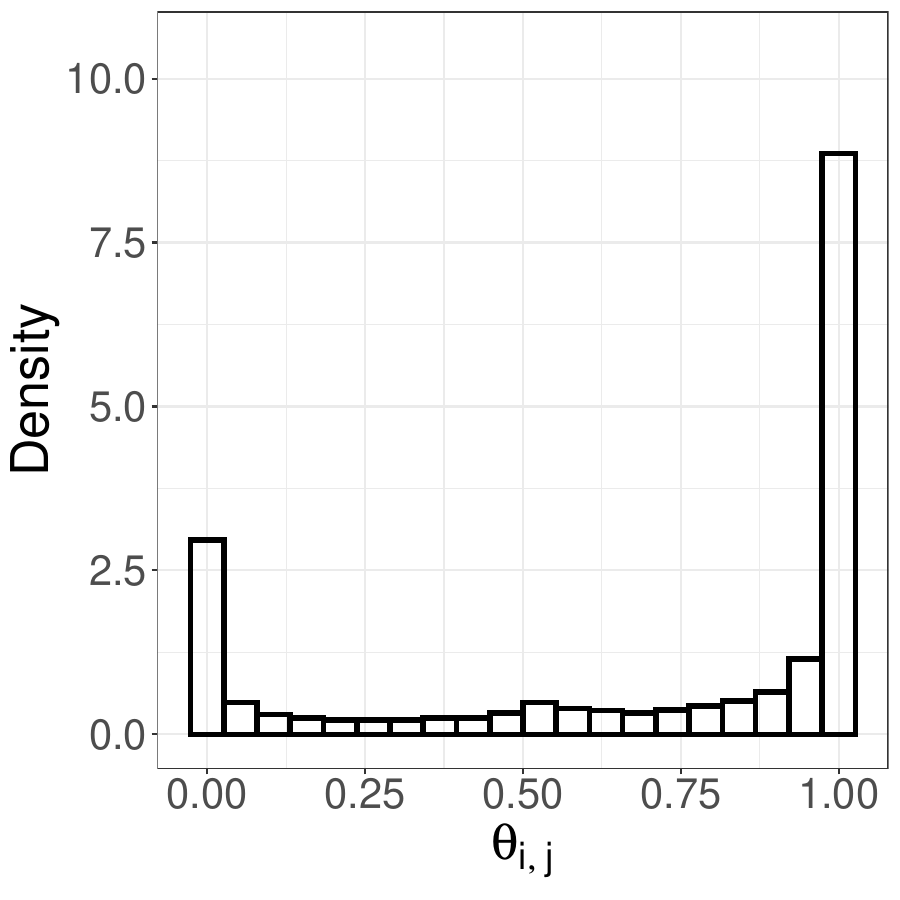}
    \caption{Histogram based on 10,000 samples from the implied prior distribution on $\theta_{i,j}$ under the logistic unfolding model with $\bfmu = (-2,10)'$, $\omega^2 = 25$ and $\kappa^2 = 10$.}
    \label{fig:prior_theta_plots}
\end{figure*}

\subsection{Prior selection}

To guide the selection of the hyperparameters $\bfmu$, $\omega$ and $\kappa$, we examine the implied prior distribution on $\theta_{i,j}$, the  probability of an affirmative vote,
\begin{multline*}
\theta_{i,j}  =\\  \frac{1}{1 + \exp\left\{-\alpha_{j,1}(\beta_i - \delta_{j,1})\right\} + \exp\left\{-\alpha_{j,2}(\beta_i - \delta_{j,2})\right\}}    
\end{multline*}
\begin{table*}[!t] \centering 
\begin{tabular}{ccccc} \hline 
\textbf{House} & \textbf{WAIC}(LUM,IDEAL) & \textbf{WAIC}(LUM,PUM) & $\rho$(LUM, IDEAL) & $\rho$(LUM, PUM)\\
\hline
$103$ & $2752.388$ & $359.697$ & $0.994$ $(0.993, 0.994)$ & $0.998$ $(0.997, 0.998)$\\ 
$107$ & $3866.190$ & $404.722$ & $0.959$ $(0.956, 0.963)$ & $0.993$ $(0.992, 0.995)$\\ 
$110$  & $3748.697$ & $1176.145$ & $0.984$ $(0.982, 0.986)$ & $0.996$ $(0.996, 0.997)$\\ 
$116$ & $2675.145$ & $349.273$ & $0.960$ $(0.956, 0.964)$ & $0.990$ $(0.988, 0.992)$\\
\hline
\end{tabular} 
\vspace{1mm}
  \caption{Difference in WAIC scores between the logit unfolding model and IDEAL ($WAIC$(LUM) - $WAIC$(IDEAL)) and the logit unfolding model and the probit unfolding model ($WAIC$(LUM) - $WAIC$(PUM)), and posterior summaries of the Spearman correlation of legislator's ranks between IDEAL, the probit unfolding models, and the logit unfolding models.} 
  \label{table:model_comp_stats} 
\end{table*} 
In the context of applications to political science, it is common to assume that the the preferences of the legislators are quite clear on most votes (e.g., see 
\citealp{yu2021spatial},
\citealp{LeiRodriguezNovelClassUnfolding2023},
\citealp{spirling2010identifying}, and
\citealp{paganin2021computational}).  Furthermore, since the majority party controls the agenda of the House, most votes are expected to pass and the probability of a positive vote can be expected to be slightly larger than the probability of a negative one.  These observations lead us to prefer roughly U-shaped priors that place a slightly higher probability to values close to 1.  An example of such a prior, which corresponds to $\bfmu = (-2,10)'$, $\omega^2 = 25$ and $\kappa^2 = 10$, is presented in Figure \ref{fig:prior_theta_plots}.  Sensitivity analysis conducted as part of this research, along with those in \cite{LeiRodriguezNovelClassUnfolding2023}, suggest that the results are robust to alternative hyperparameter specifications that lead to priors on $\theta_{i,j}$ similar to those in Figure \ref{fig:prior_theta_plots}.

\begin{figure*}[!t]
    \centering
    \begin{subfigure}[t]{0.45\textwidth}
        \includegraphics[width = \textwidth]{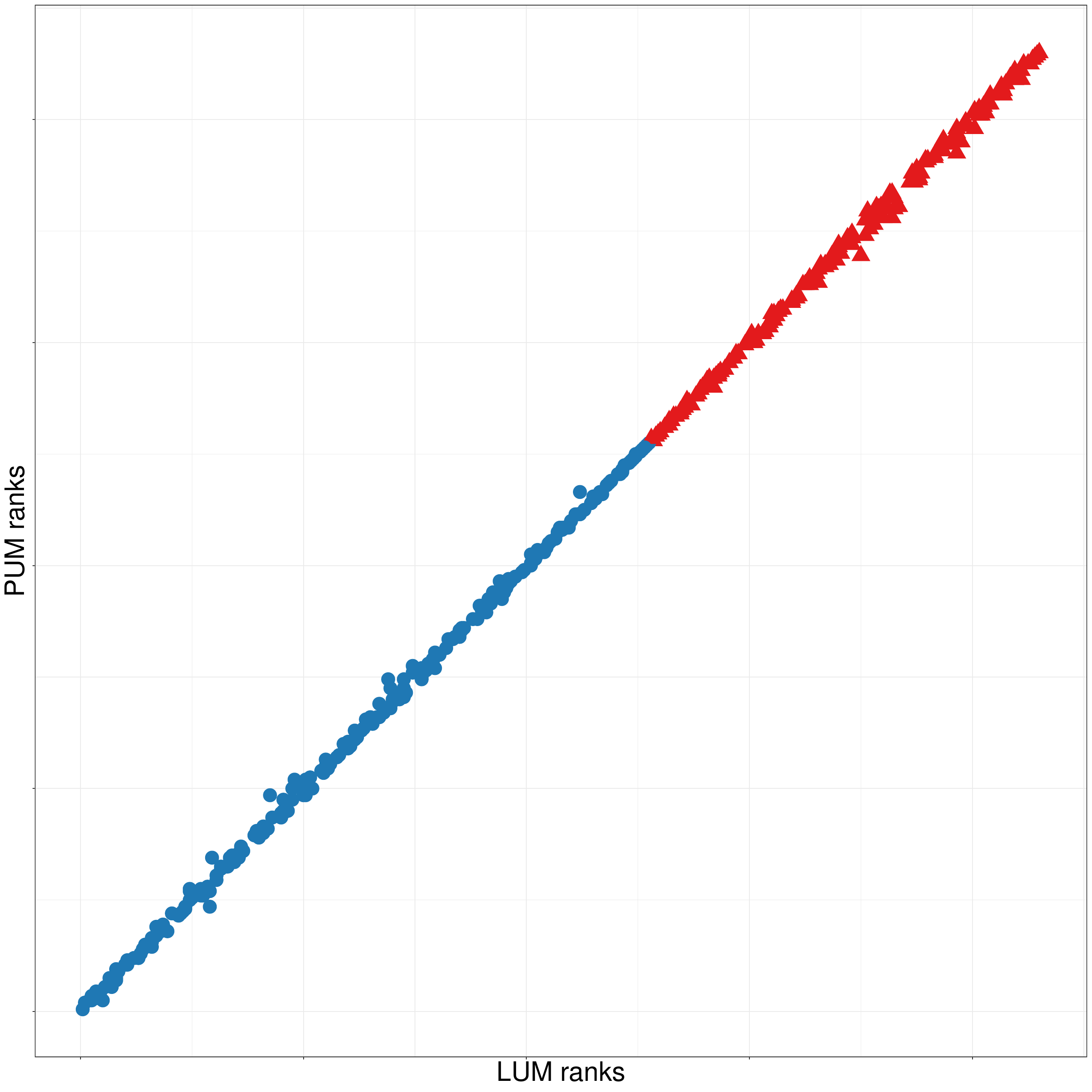}
        \caption{103rd}
    \end{subfigure}
        \begin{subfigure}[t]{0.45\textwidth}
        \includegraphics[width = \textwidth]{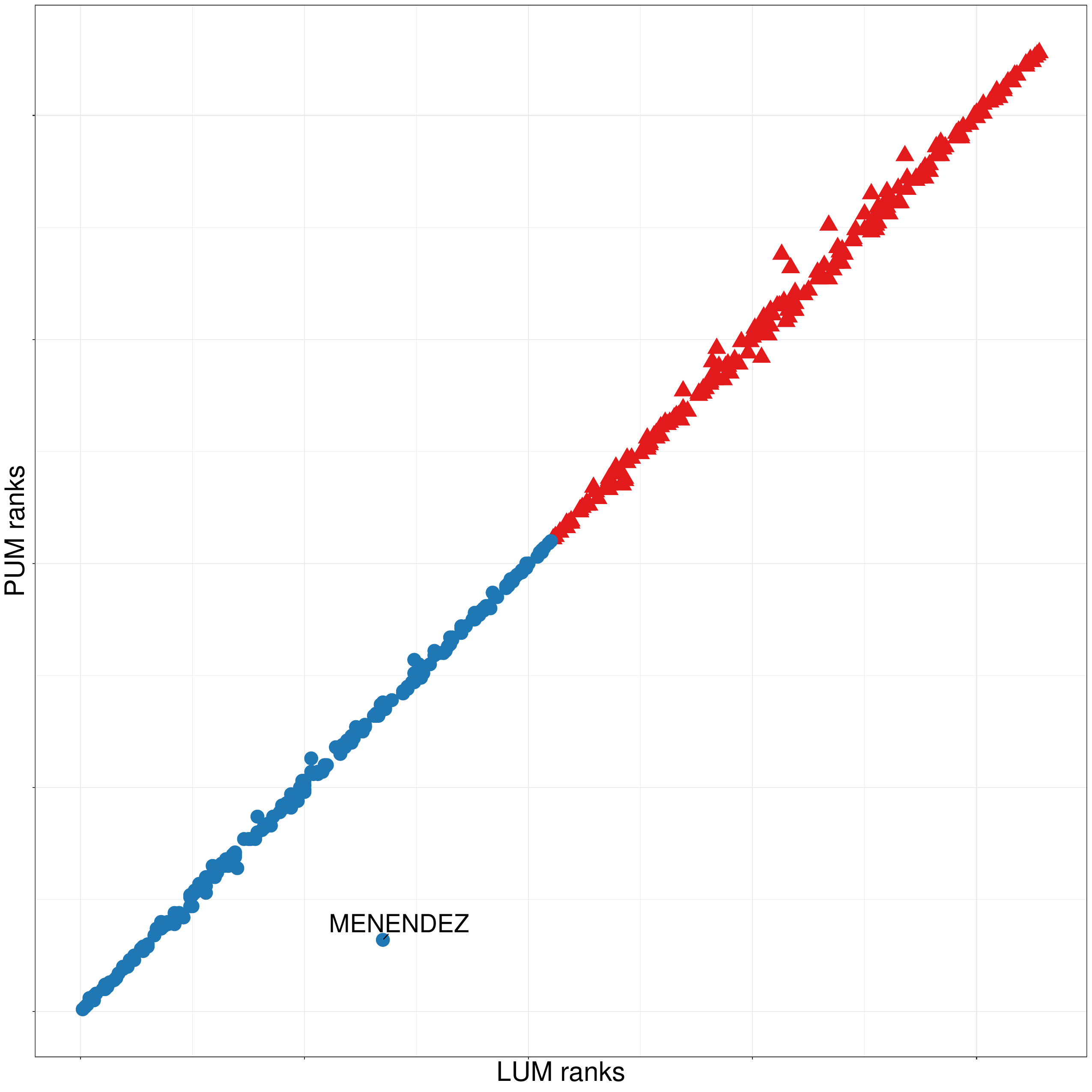}
        \caption{107th}
    \end{subfigure}\\
    \begin{subfigure}[t]{0.45\textwidth}
        \includegraphics[width = \textwidth]{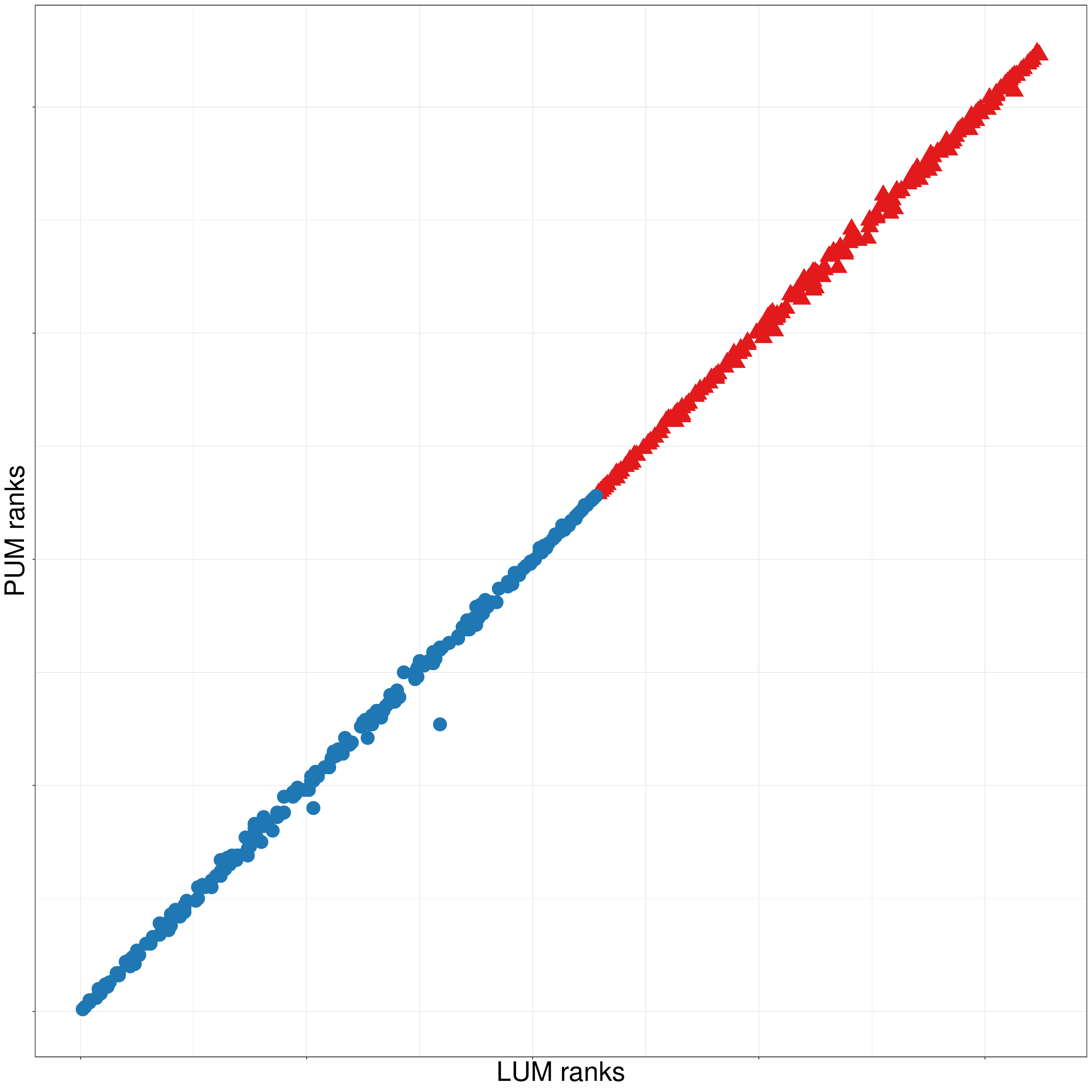}
        \caption{110th}
    \end{subfigure}
    \begin{subfigure}[t]{0.45\textwidth}
        \includegraphics[width = \textwidth]{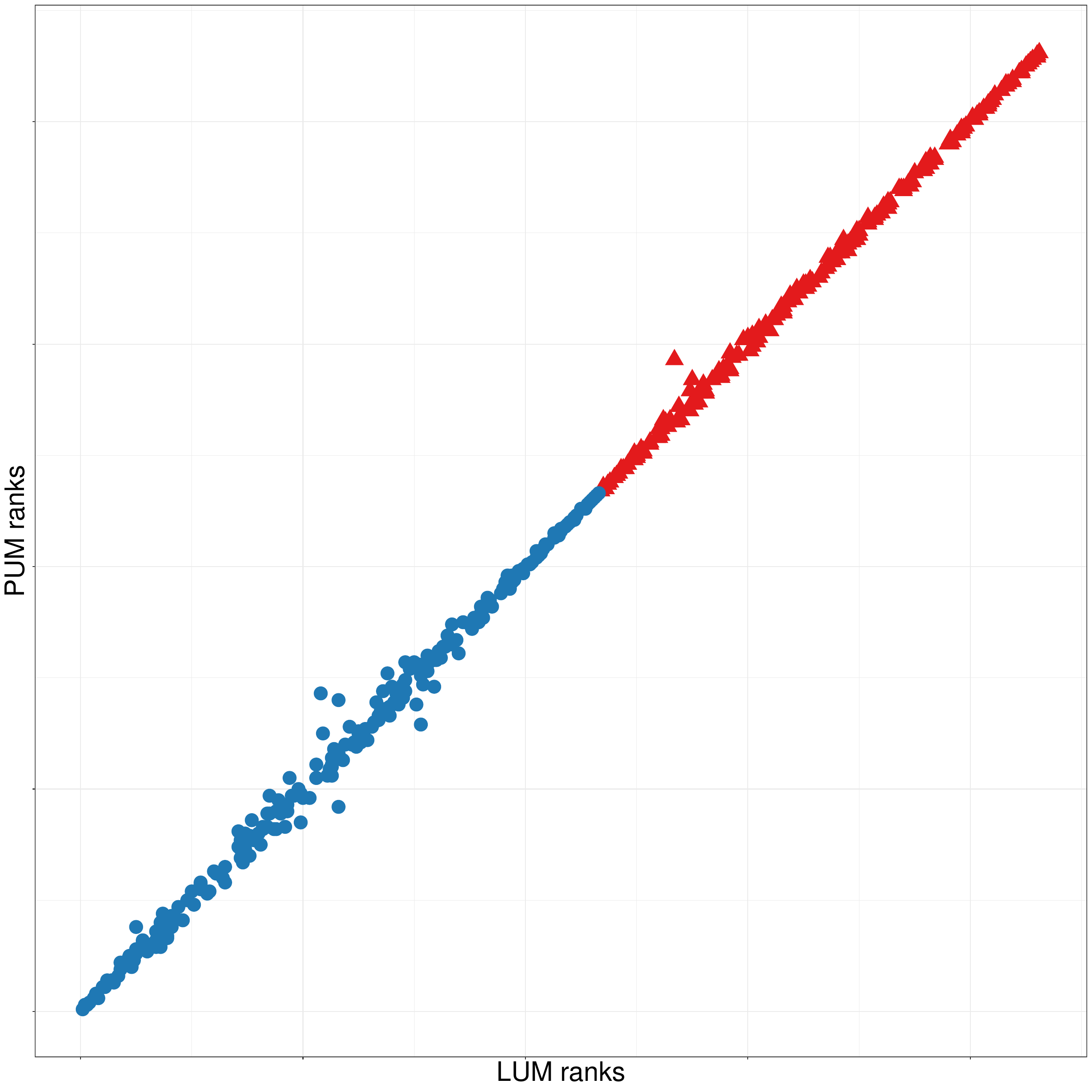}
        \caption{116th}
    \end{subfigure}
    \caption{Comparisons between the posterior mean rank of legislators under the logistic and probit unfolding models for the four U.S.\ House of Represenatives under consideration.}
    \label{fig:house_rankcomp_LUM_PUM}
\end{figure*}

\subsection{Model performance}

Table \ref{table:model_comp_stats} presents the difference in Watanabe-Akaike Information Criterion scores (WAIC, \citealp{watanabe2010asymptotic, watanabe2013widely, gelman2014understanding}), as well as the posterior mean and 90\% credible intervals for the Spearman correlation between the logistic unfolding model (LUM) and both the probit unfolding model (PUM) and IDEAL.
For each House, the WAIC score under model $m$ is computed as:
\begin{multline}\label{eq:lWAIC}
    WAIC(m) =\\ 
    \sum_{i=1}^{I} \log\left( \textrm{E}_{\textrm{post}} \left\{ \prod_{j = 1}^J 
     \theta_{i,j}(m)^{y_{i,j}} \right. \right. \\
     \left. \left. \left[1 -\theta_{i,j}(m)\right]^{1 - y_{i,j}} \right\} \right) -\\
    \sum_{i=1}^{I} \textrm{var}_{\textrm{post}}\left\{ 
     \left[\sum_{j = 1}^{J} y_{i,j} \log \theta_{i,j}(m) + \right. \right.\\ 
     \left. \left. (1-y_{i,j}) \log(1-\theta_{i,j}(m)) \right] \right\} ,
\end{multline}
Here, $\theta_{i, j}(m)$ represents the probability that legislator $i$ votes yes on issue $j$ under model $m$.  The WAIC is similar to the Akaike Information Criterion and Bayesian Information Criterion in that it attempts to balance model fit with model complexity.  A larger WAIC score indicates a better complexity-adjusted fit to the data. On the other hand, $\rho_i(m,m')$, the Spearman correlation of the ranks estimated from models $m$ and $m'$, is computed using the well-known formula 
$$
\rho_i(m,m') = 1 - \frac{6}{n(n^2-1)}\sum_{i=1}^{I} \left(r_i^{(m)} - r_i^{(m')}\right)^2 ,
$$ 
where $r_i^{(m)}$ indicates the ideological rank of legislator $i$ under model $m$.  A value of 1 for $\rho_i(m,m')$ indicates that both models yield exactly the same ranking for all legislators.

\begin{figure*}[!t]
    \centering
    \begin{subfigure}[t]{0.45\textwidth}
        \includegraphics[width = \textwidth]{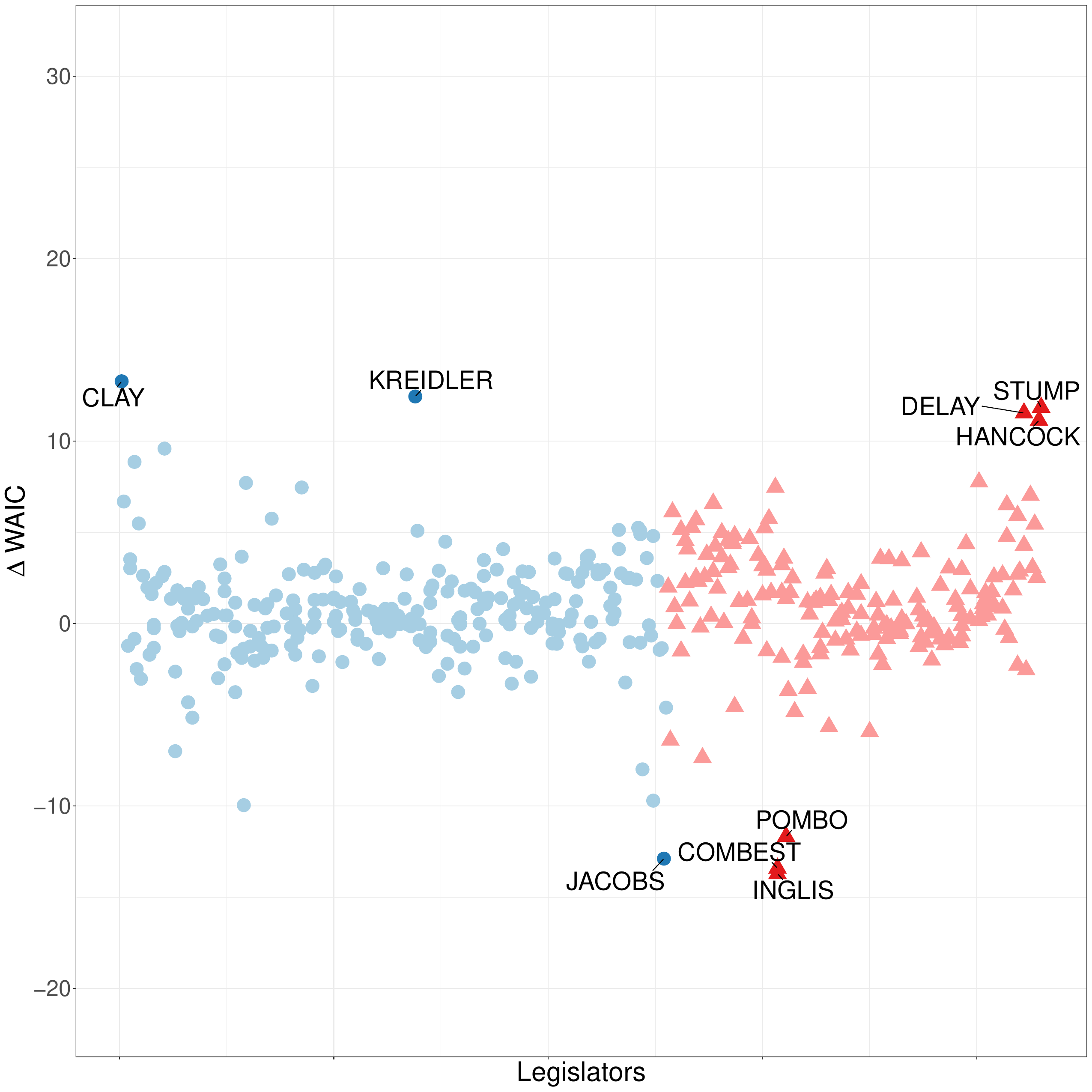}
        \caption{103\textsuperscript{th} House}
    \end{subfigure}
    \begin{subfigure}[t]{0.45\textwidth}
        \includegraphics[width = \textwidth]{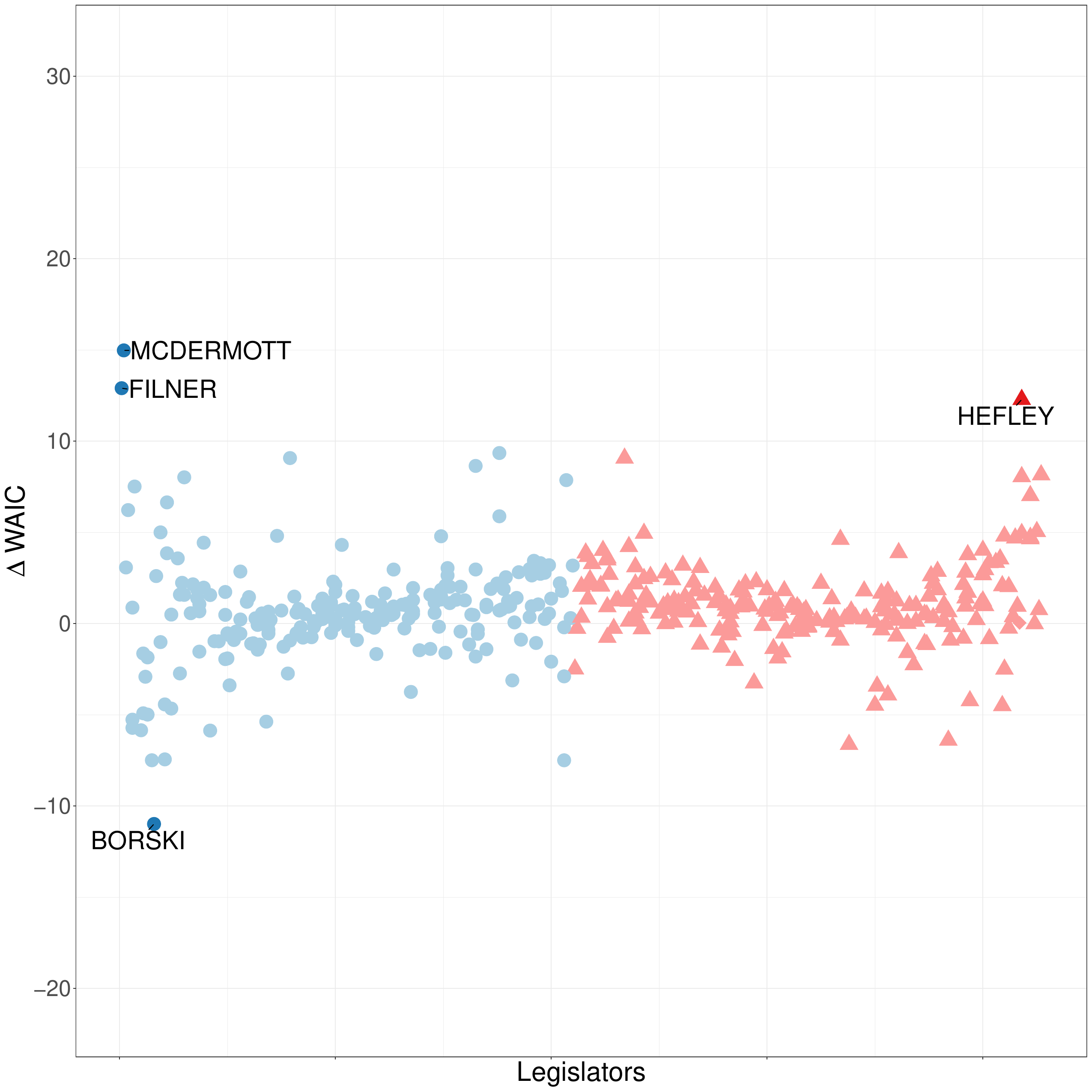}
        \caption{107\textsuperscript{th} House}
        \label{fig:logit_comp_waic_107}
    \end{subfigure}\\
    \begin{subfigure}[t]{0.45\textwidth}
        \includegraphics[width = \textwidth]{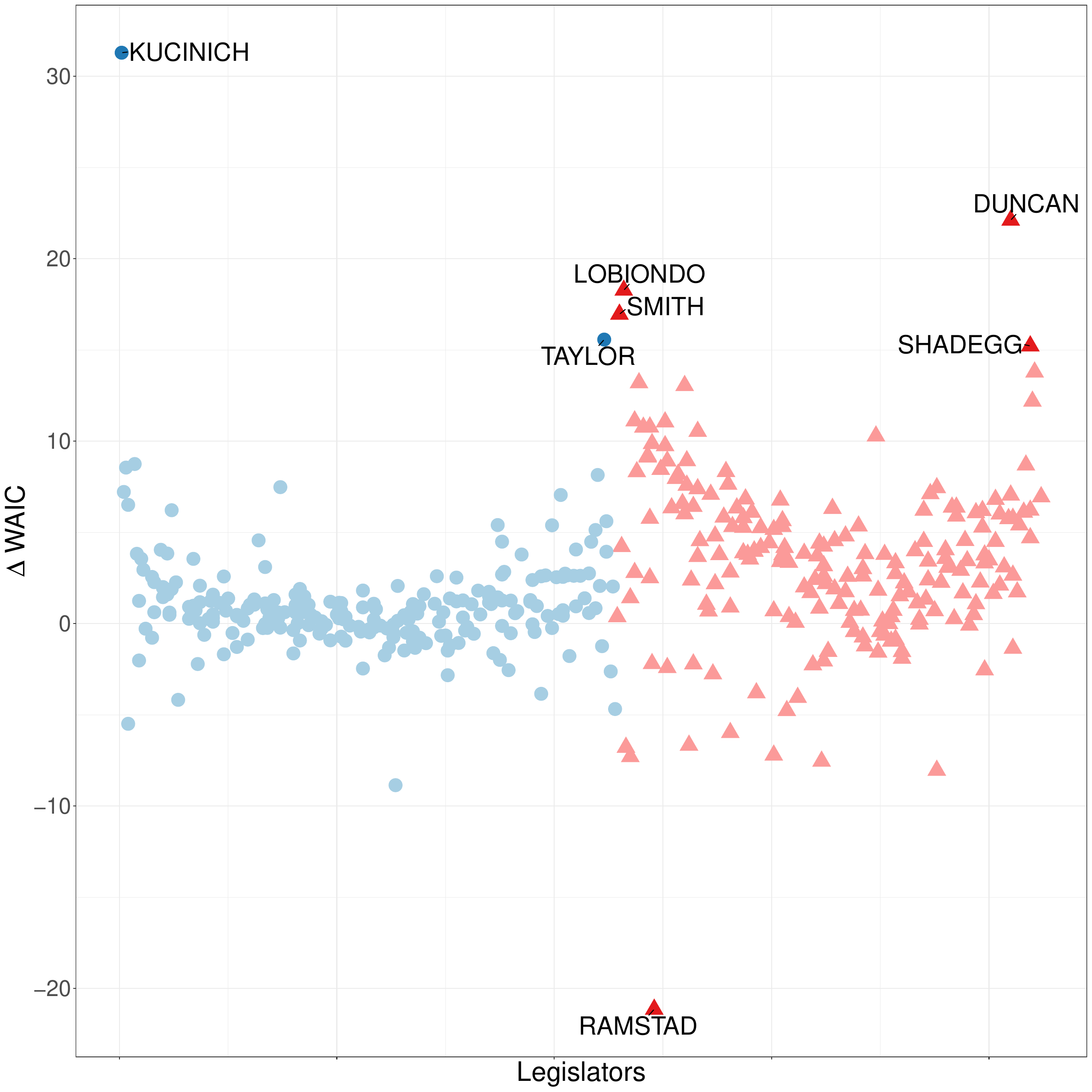}
        \caption{110\textsuperscript{th} House}
    \end{subfigure}
    \begin{subfigure}[t]{0.45\textwidth}
        \includegraphics[width = \textwidth]{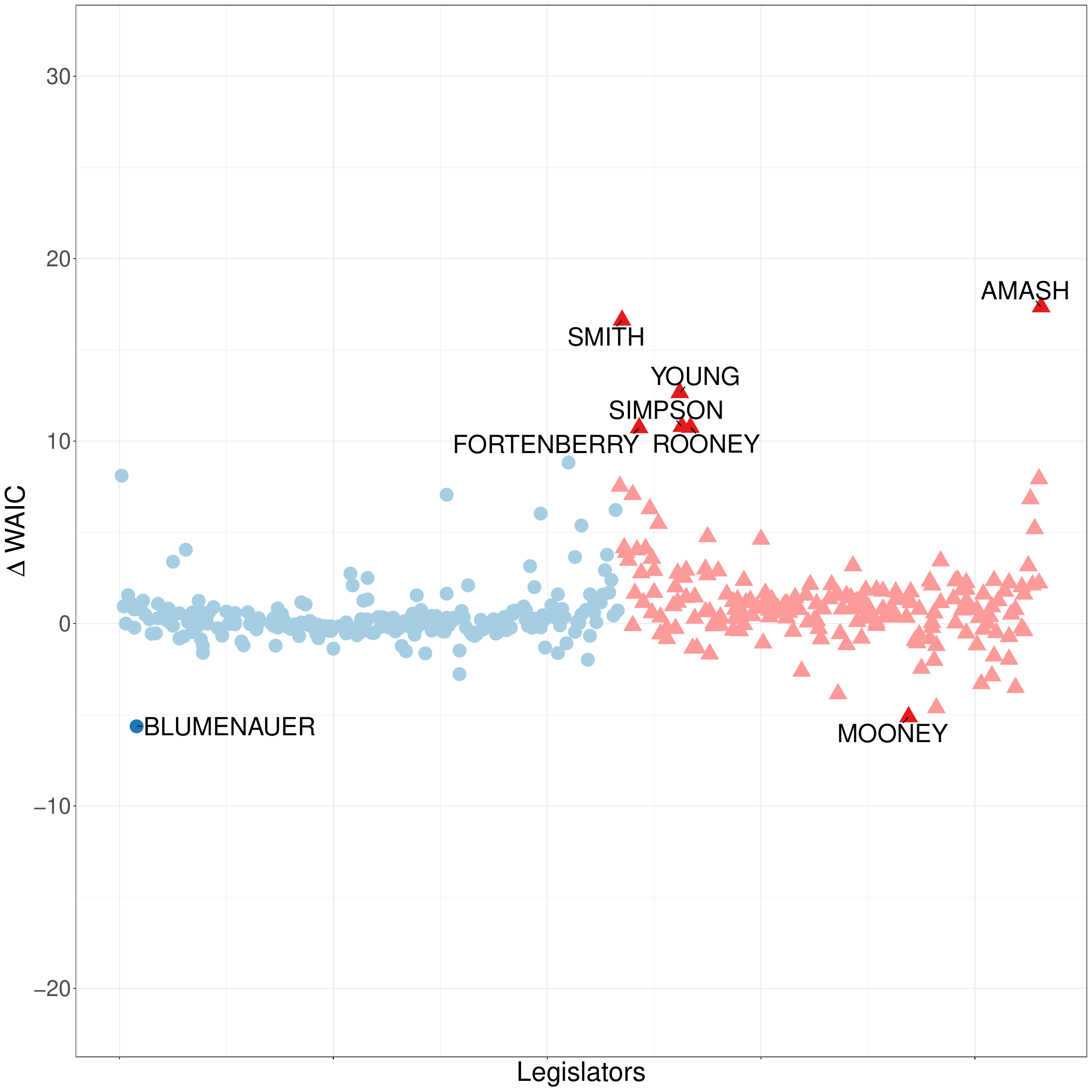}
        \caption{116\textsuperscript{th} House}
        \label{fig:logit_comp_waic_116}
    \end{subfigure}
    \caption{Breakdown of the difference in WAIC scores by legislator in each of the four Houses under consideration. Note that Representative Ron Paul from the 107th and 110th House and Representative Jim Cooper from the 110th House are not shown for visualization purposes.} \label{fig:house_WAICbreakdown}
\end{figure*}

The second and third columns of Table \ref{table:model_comp_stats} show that both PUM and LUM provide a better complexity-adjusted fit to the data than IDEAL.  They also provide strong evidence of the importance of allowing for heavy-tailed shocks in discrete choice models.  Indeed, in all four Houses we consider, the WAIC scores for LUM are higher than those for PUM, especially for the 110\textsuperscript{th} House.  On the other hand, the fourth and fifth columns of Table \ref{table:model_comp_stats} show that, in spite of these differences, the ranks recovered by all three models agree to a large degree. However, as would be expected, the ideological ranks of legislators recovered by LUM agree much more closely with those recovered by PUM than those recovered by IDEAL.

\begin{figure*}[!t]
    \centering
    \begin{subfigure}[t]{0.3\textwidth}
        \includegraphics[width = \textwidth]{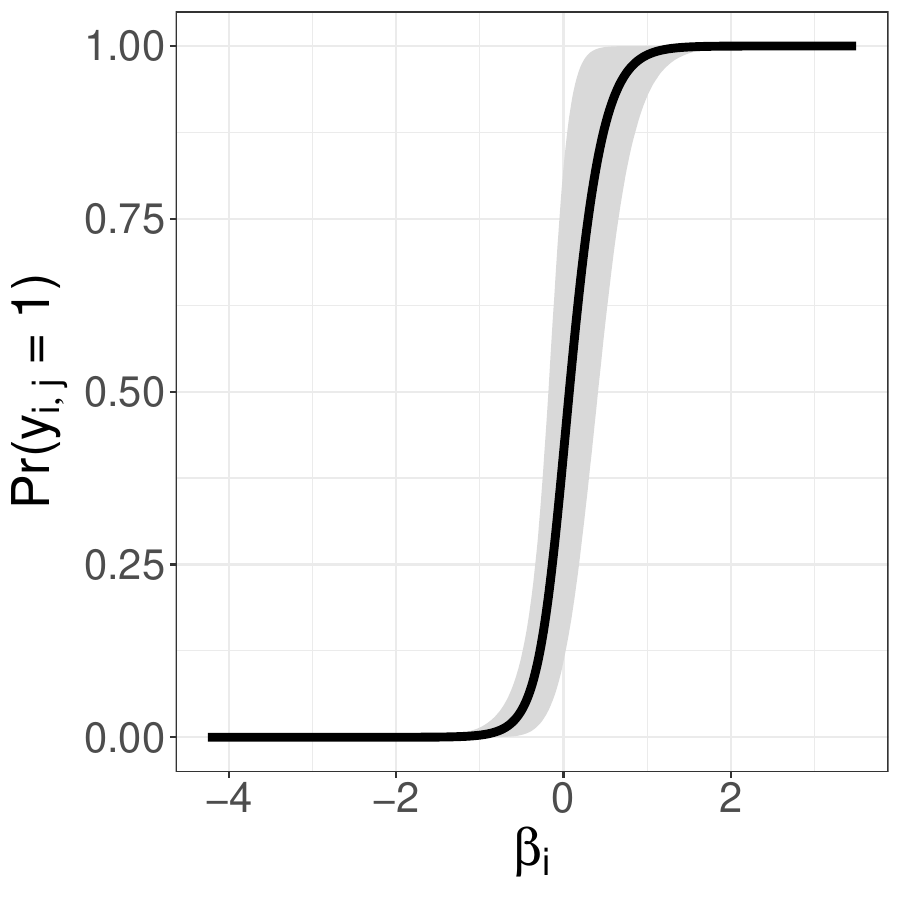}
        \caption{HR21 - LUM}
    \end{subfigure}
    \begin{subfigure}[t]{0.3\textwidth}
        \includegraphics[width = \textwidth]{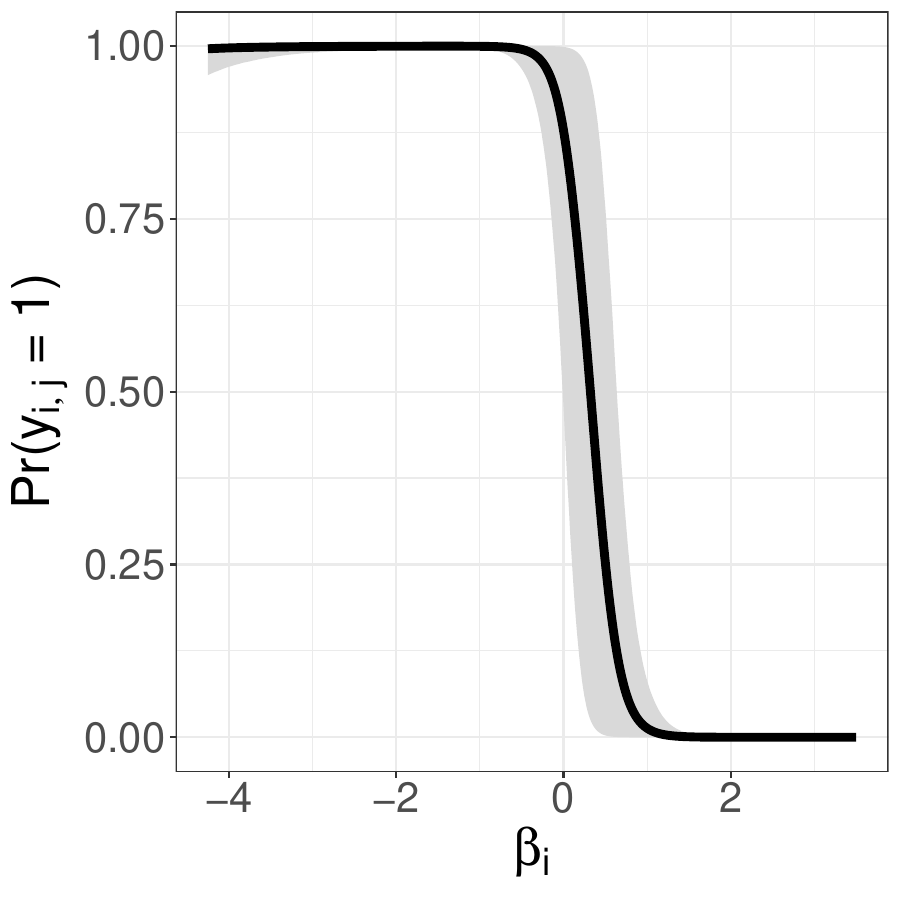}
        \caption{HRES5 - LUM}
    \end{subfigure}
    \begin{subfigure}[t]{0.3\textwidth}
        \includegraphics[width = \textwidth]{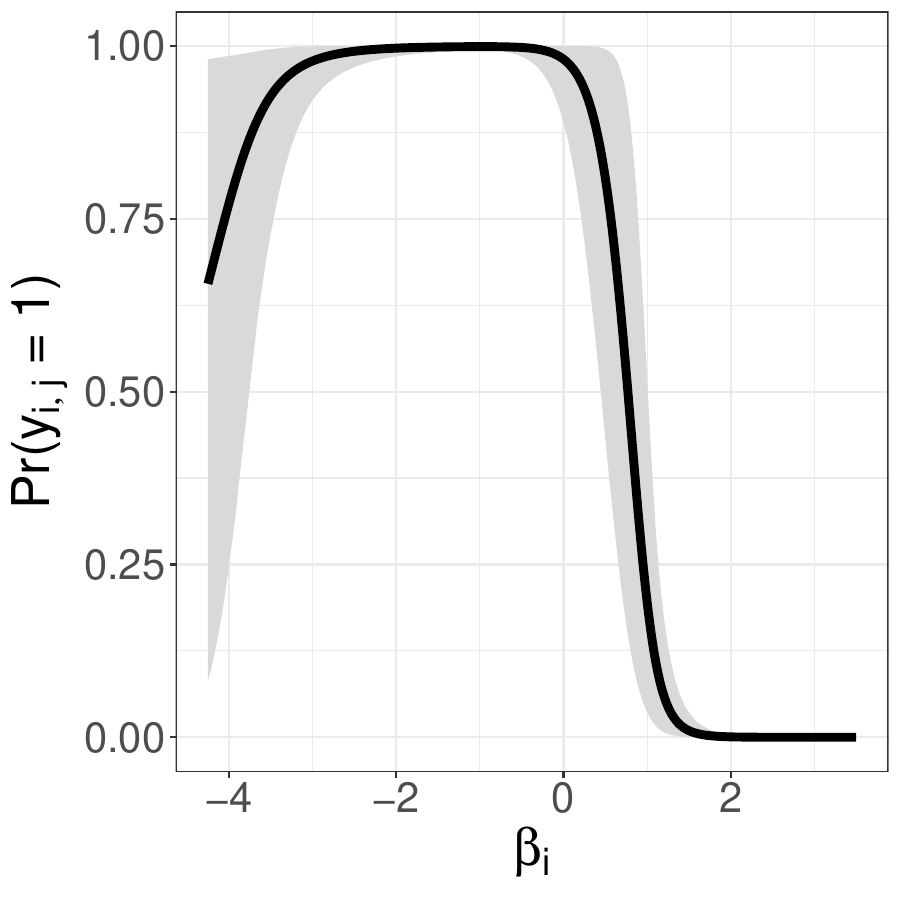}
        \caption{HRES6 - LUM}
    \end{subfigure}\\
    \begin{subfigure}[t]{0.3\textwidth}
        \includegraphics[width = \textwidth]{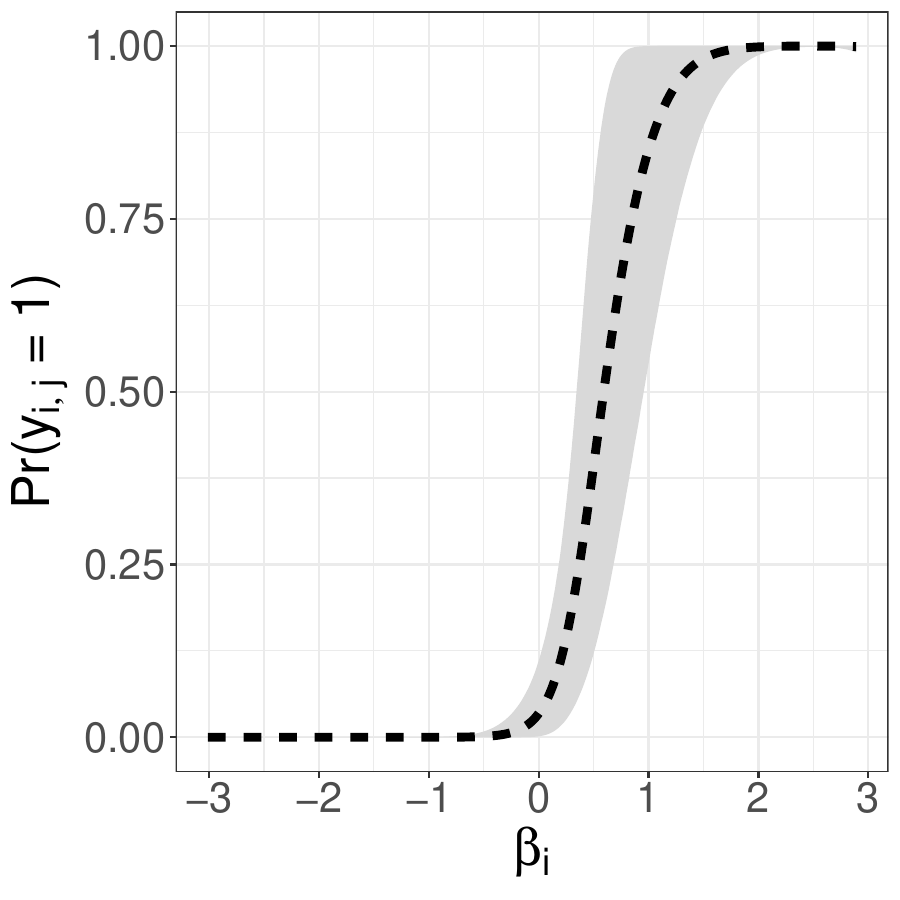}
        \caption{HR21 - PUM}
    \end{subfigure}
    \begin{subfigure}[t]{0.3\textwidth}
        \includegraphics[width = \textwidth]{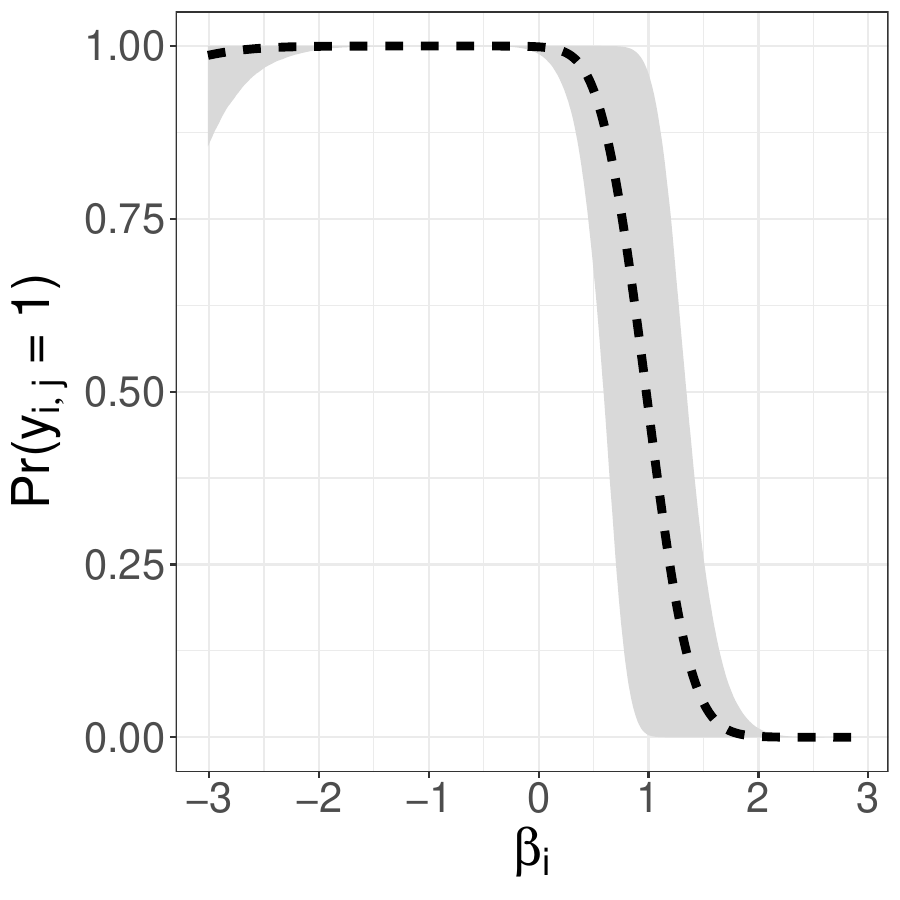}
        \caption{HRES5 - PUM}
    \end{subfigure}
    \begin{subfigure}[t]{0.3\textwidth}
        \includegraphics[width = \textwidth]{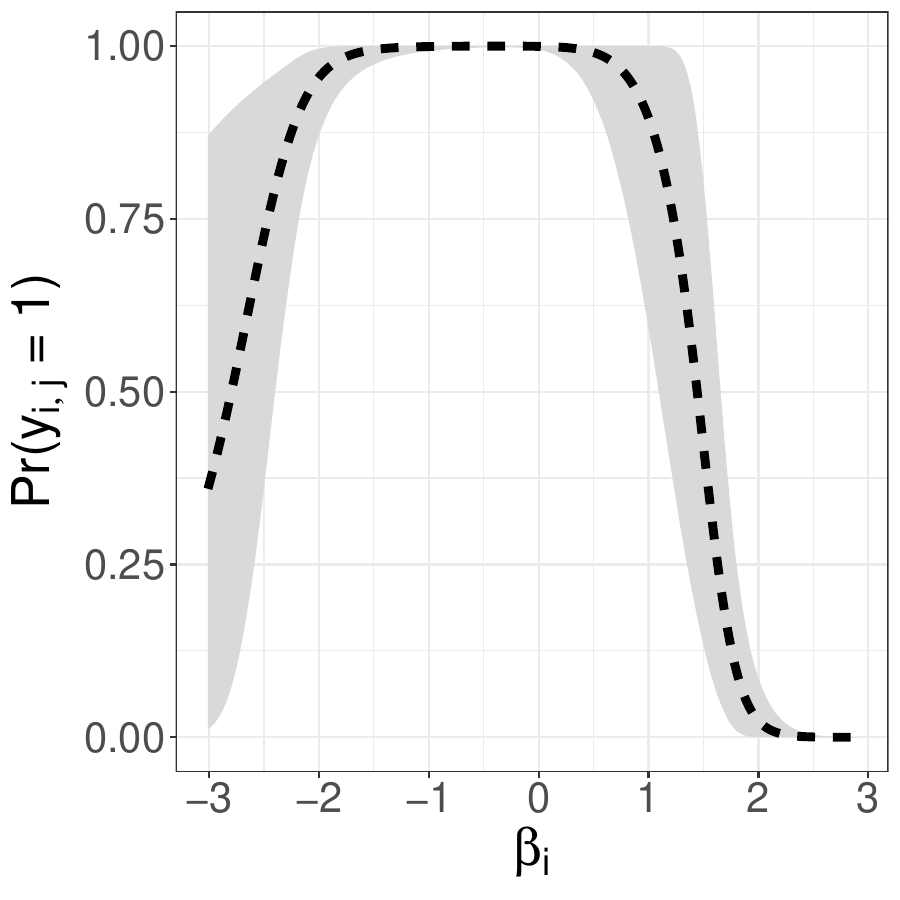}
        \caption{HRES6 - PUM}
    \end{subfigure}
     \caption{Posterior mean response curves and pointwise credible bands for the response function $\Pr(y_{i,j}=1 \mid \beta_i)$ under the logit (top row) and probit (bottom row) unfolding model for three votes in the 116\textsuperscript{th} House of representatives. The $x$ axis on each graph were selected based on the range of values that the ideal points take under each model.}
    \label{fig:model_response_curves}
\end{figure*}

To better understand the differences between PUM and LUM, we present in Figure \ref{fig:house_rankcomp_LUM_PUM} a detailed comparison of the posterior mean ranks of the legislators  recovered by both models.  In agreement with the Spearman correlations in Table \ref{table:model_comp_stats}, we see that the points mostly fall on a straight line.  The discrepancies, when they arise, seem to be focused on a very small number of legislators (e.g., Representative Menendez from New Jersey during the 107\textsuperscript{th} House).  Meanwhile, Figure \ref{fig:house_WAICbreakdown} presents a legislator-level breakdown of the difference in WAIC between LUM and PUM. We can see that, generally speaking, the differences are again dominated by a small number of legislators which, for the most part, tend to be either centrists or extremists.  This observation is consistent with the idea, also foreshadowed in the introduction, that  heavy-tailed structural shocks might provide an alternative explanation for some end-against-the-middle votes.

Finally, Figure \ref{fig:model_response_curves} presents point and intervals estimates for the response functions (the probability of an affirmative vote as a function of the legislator's ideal point, recall Equation \eqref{eq:logisticBGUM}) for three votes taken during the 116\textsuperscript{th} House.  These votes are representative of three scenarios:  two highly partisan votes (one in which the underlying measure is favored by Republicans and one in which the underlying measure is favored by Democrats), and one ``end-against-the-middle'' vote.  Note that, in order to facilitate comparisons, the $x$ axis of the graphs was set based on the range of the ideal points estimated by each of the two models.  We see that, overall, the shapes are very similar.  Interestingly, the uncertainty associated with the curves under LUM seems to be somewhat lower than the uncertainty under PUM.  The estimate under LUM also seems to avoid the small dip in the response function for HRES5 that PUM shows for extreme Democrats.  We see this as another sign that LUM fits the model slightly better than PUM:  the outcome of HRES5 was strictly along partisan lines, so the small dip in the PUM curve would seem to be an artifact of the model (driven by the fact that traditional item response theory models are obtained in the limit).  We also see a small difference in the behavior of the two estimates for extreme Democrats for HRES6, with the curve estimated by PUM dipping somewhat faster than the curve for LUM.  Unlike HRES5, the outcome of HRES6 is a canonical example of an ``end-against-the-middle'' vote, with all Republicans and a small group of three Democrats voting against the measure. Two members, Representative Alexandria Ocasio-Cortez and Ro Khanna, are often considered to be on the extreme of their party. Again, the estimate generated by LUM seems more appropriate, as the low number of extreme Democrats voting against the measure hardly seems to justify the relatively low values of the response function under PUM at the lower end of the ideological scale.

\begin{figure*}[!t]
    \centering
    \begin{subfigure}[t]{\textwidth}
        \centering
        \includegraphics[width = 0.45\textwidth]{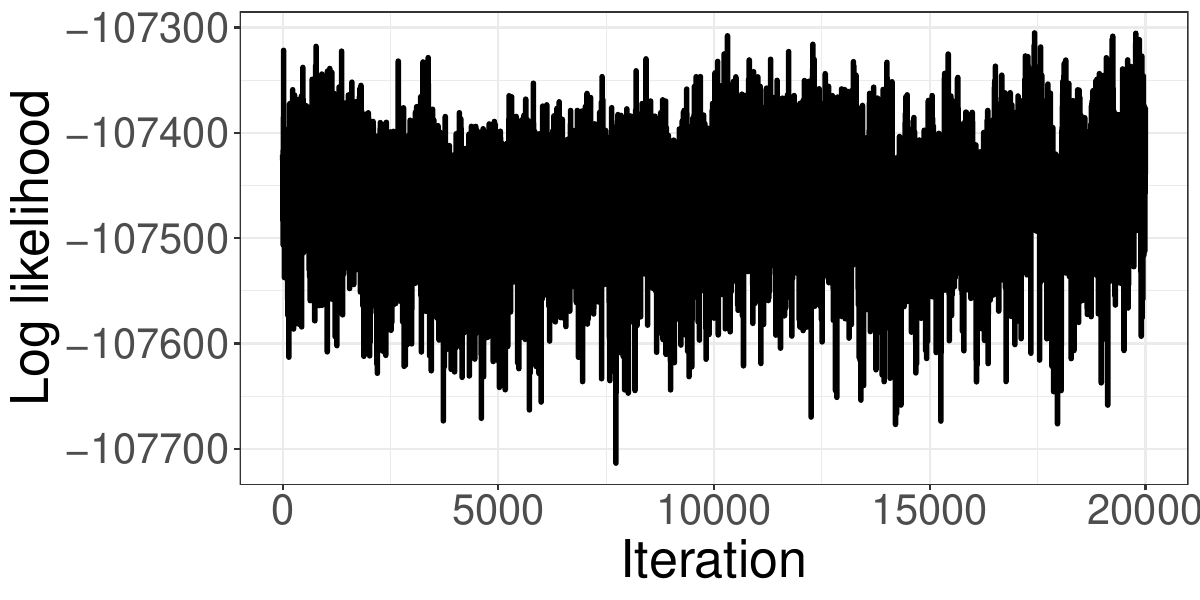}
        \includegraphics[width = 0.45\textwidth]{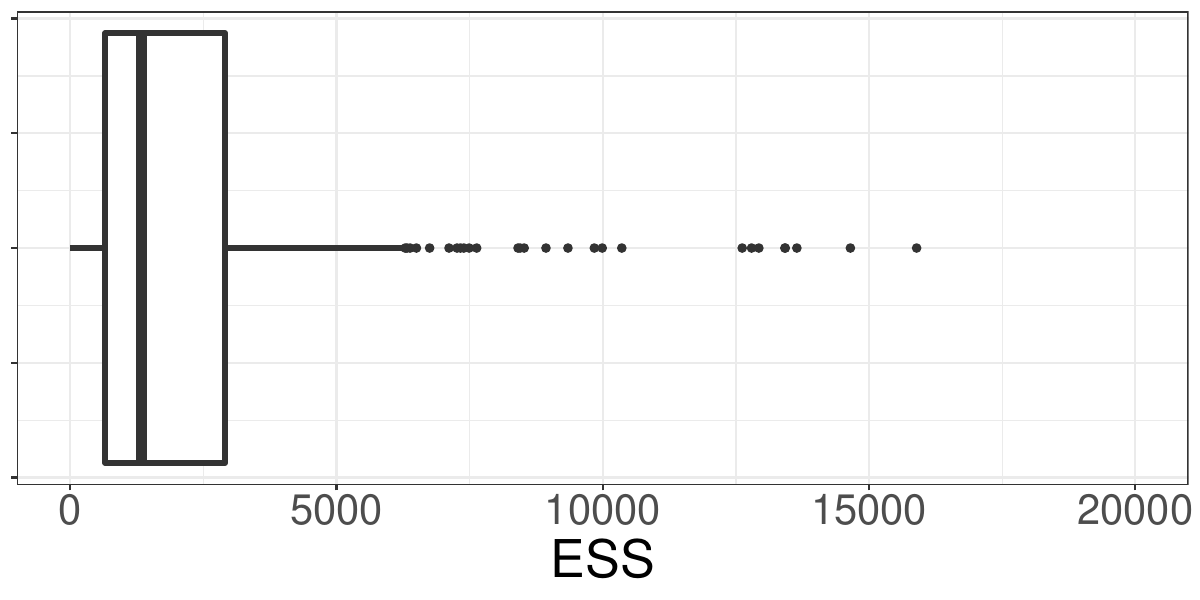}
        \caption{103\textsuperscript{th} House}
    \end{subfigure}
    \begin{subfigure}[t]{\textwidth}
        \centering
        \includegraphics[width = 0.45\textwidth]{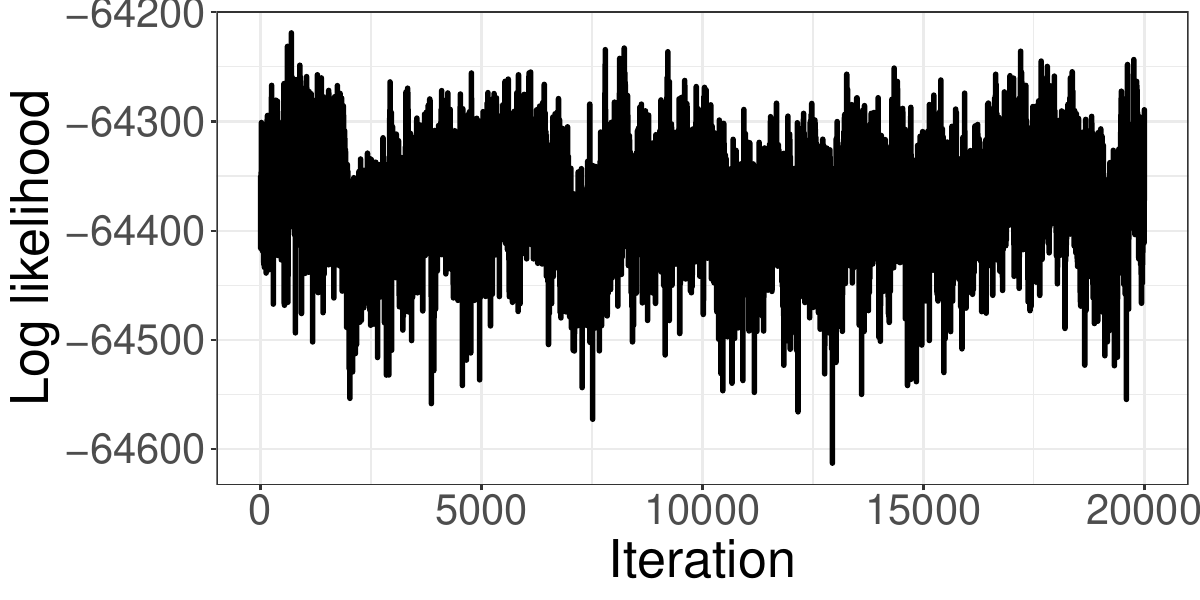}
        \includegraphics[width = 0.45\textwidth]{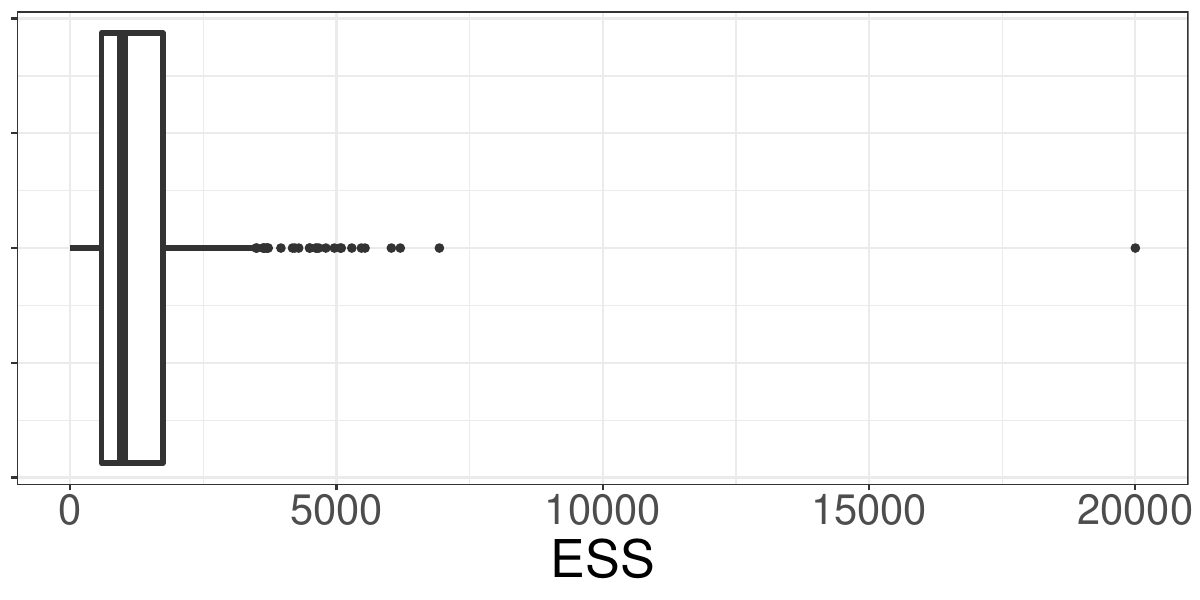}
        \caption{107\textsuperscript{th} House}
    \end{subfigure}
    \begin{subfigure}[t]{\textwidth}
        \centering
        \includegraphics[width = 0.45\textwidth]{img/trace_plots/house_107/house_107_logit_ll_traceplot.pdf}
        \includegraphics[width = 0.45\textwidth]{img/trace_plots/house_107/house_107_logit_rank_ess_plot.pdf}
        \caption{110\textsuperscript{th} House}
    \end{subfigure}
    \begin{subfigure}[t]{\textwidth}
        \centering
        \includegraphics[width = 0.45\textwidth]{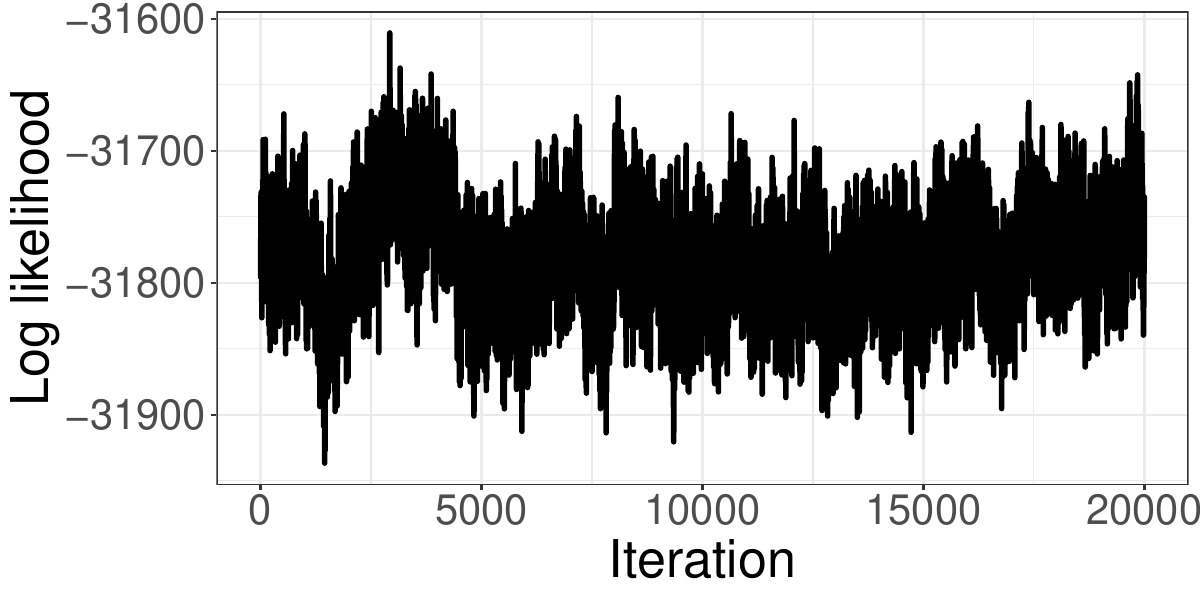}
        \includegraphics[width = 0.45\textwidth]{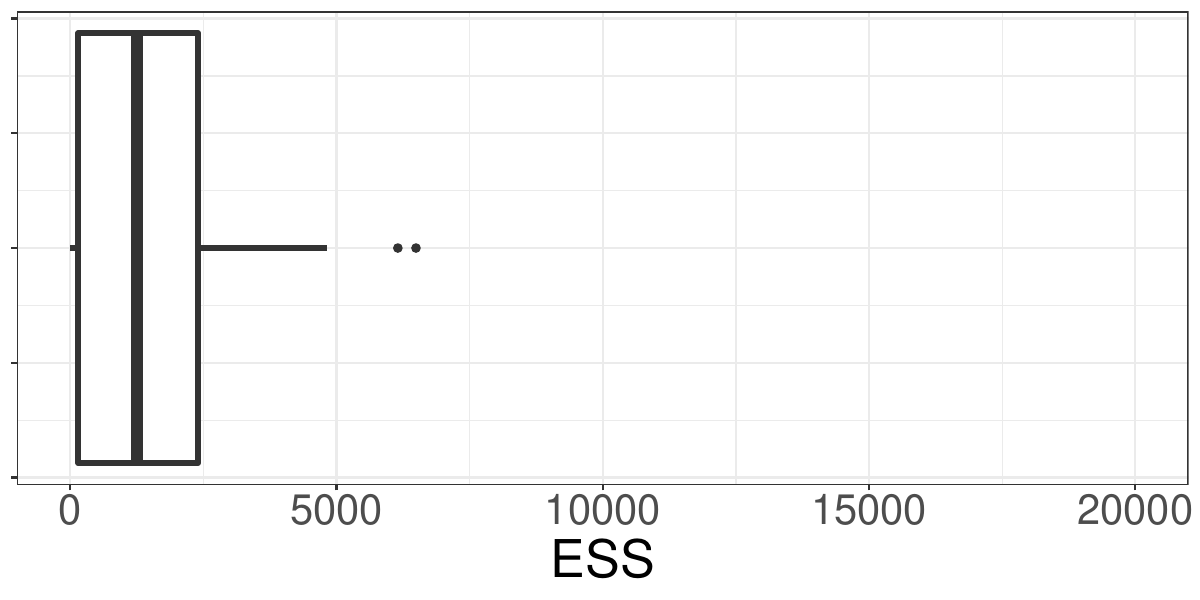}
        \caption{116\textsuperscript{th} House}
    \end{subfigure}
    \caption{Trace plot of the logistic unfolding model's log likelihood (left column) and the effective sample size for the ideological ranks of the legislators (right column) for each of the four Houses under consideration.}
    \label{fig:logliktrace_and_ESS}
\end{figure*}

\subsection{Algorithmic performance}

Figure \ref{fig:logliktrace_and_ESS} shows trace plots of the logistic unfolding model's log likelihood (left column) and boxplots of the effective sample size (ESS) for the ideological ranks of the legislators for each of the four Houses under consideration in one run of the algorithm. For this second summary, we focus on the ideological ranks because they are strongly identified in the model (as opposed to the underlying ideal points from which they are derived, which are only weakly identified).  
The trace plots suggest that, in all cases, the algorithm has converged and is able to fully explore the parameter space.  To verify this observation, we used \texttt{RStan}'s built-in functions to compute the Gelman-Rubin $\hat{R}$ statistic \citep{GelmanRubinInferenceIterativeSimulation1992} for each of the four chains.  The resulting values, 1.028, 1.005, 1.055, and 1.002 for the 103\textsuperscript{rd}, 107\textsuperscript{th}, 110\textsuperscript{th}, and 116\textsuperscript{th} Houses, respectively, are all well below the customary threshold of 1.10, again suggesting that the algorithm has converged in all cases.
From the ESS boxplots, we can see that the median ESS is around 1,000-1,500 in all four datasets, although the quartiles of the distribution change substantially depending on the dataset. Interestingly, we see a couple of examples of parameters with an ESS very close to the maximum value of 20,000.  We also see some examples of parameters with very low ESS, although we caution that this is in part an artifact of the fact that we focus our analysis of the rankings:  the ideological position of centrists and extreme legislators are discrete (and dependent) quantities that often have low uncertainty and change very little from one iteration to next, even if the underlying estimates of their ideal points are changing.  This might result in a low ESS even when the algorithm is mixing properly. 

\begin{figure*}[!t]
    \centering
    \begin{subfigure}[t]{0.45\textwidth}
        \includegraphics[width = \textwidth]{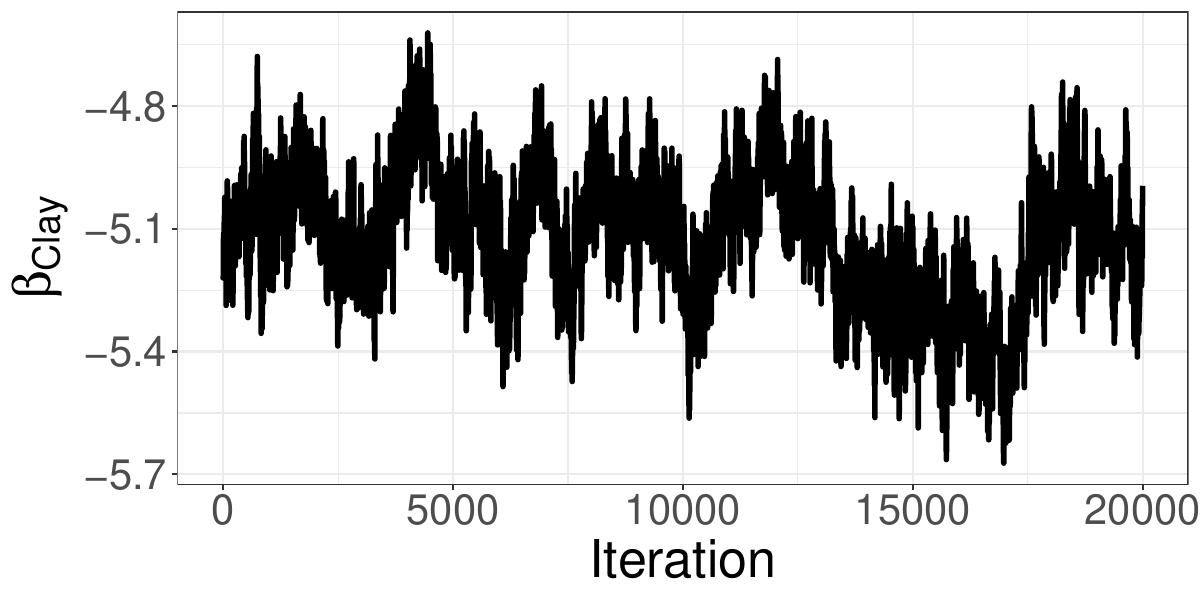}
        \caption{Representative William Clay Sr}
    \end{subfigure}
    \begin{subfigure}[t]{0.45\textwidth}
        \includegraphics[width = \textwidth]{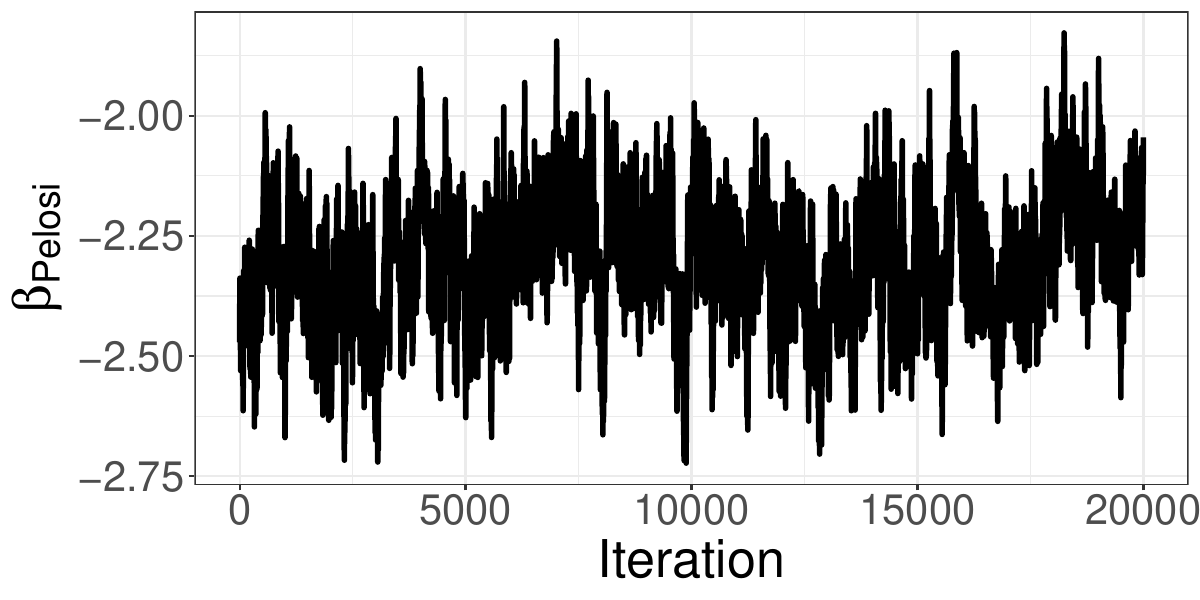}
        \caption{Representative Nancy Pelosi}
    \end{subfigure}\\
    \begin{subfigure}[t]{0.45\textwidth}
        \includegraphics[width = \textwidth]{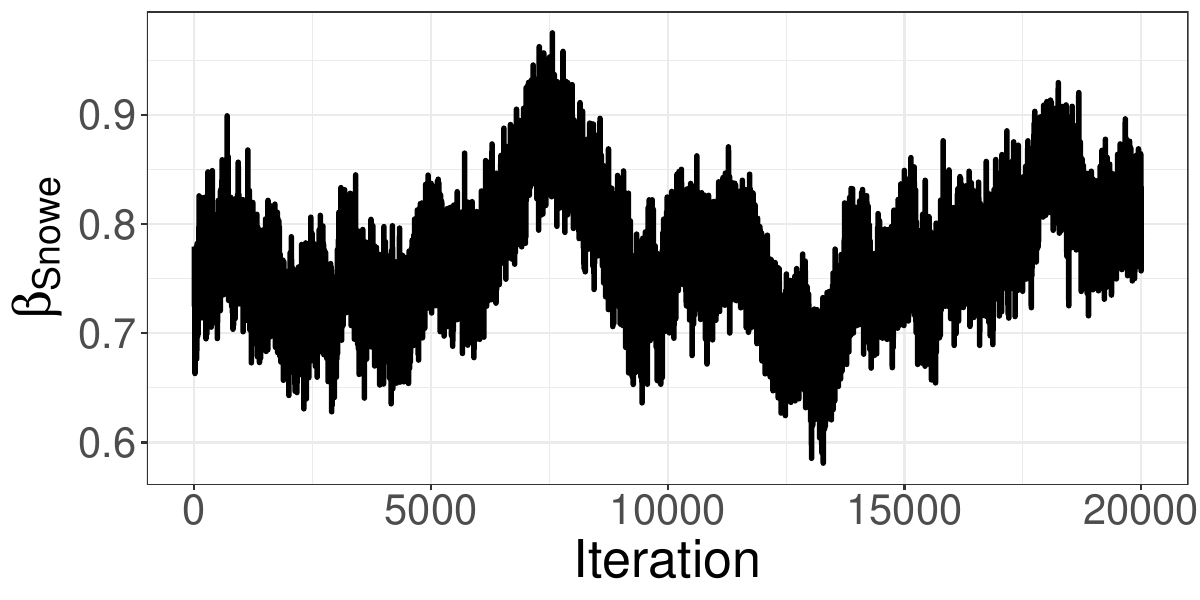}
    \caption{Representative Olympia Snowe}
    \end{subfigure}
    \begin{subfigure}[t]{0.45\textwidth}
        \includegraphics[width = \textwidth]{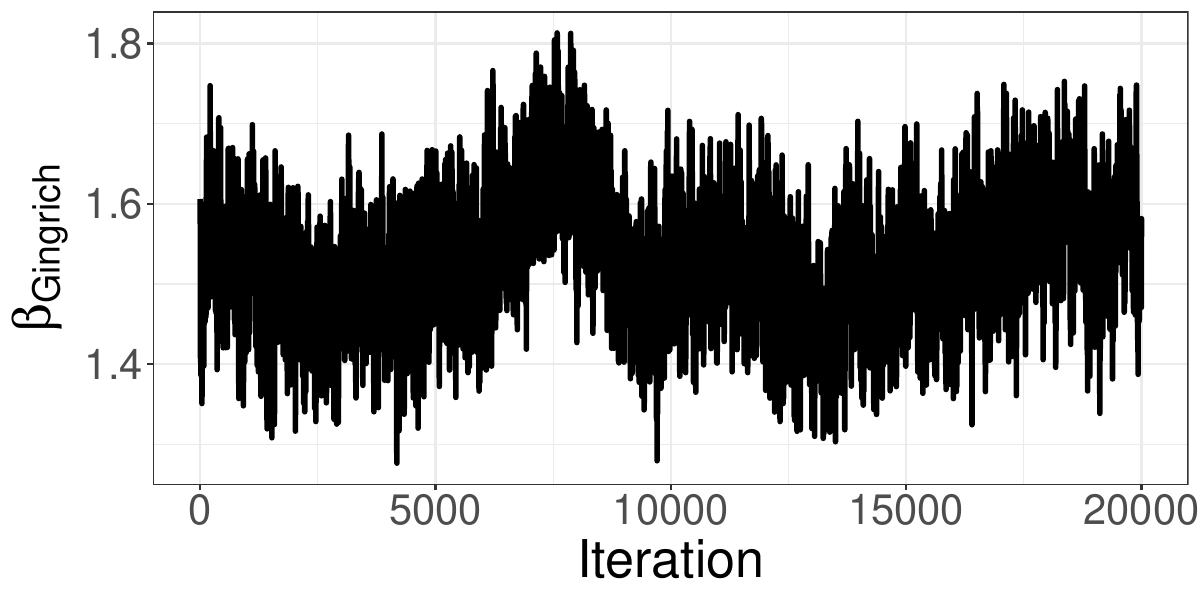}
    \caption{Representative Newt Gingrich}
    \end{subfigure}
    \caption{Trace plots of the ideological preference for various Representatives in the 103\textsuperscript{rd} House.}\label{fig:house_103_leg_traceplots}
\end{figure*}

\begin{figure*}[!t]
    \centering
    \begin{subfigure}[t]{0.45\textwidth}
        \includegraphics[width = \textwidth]{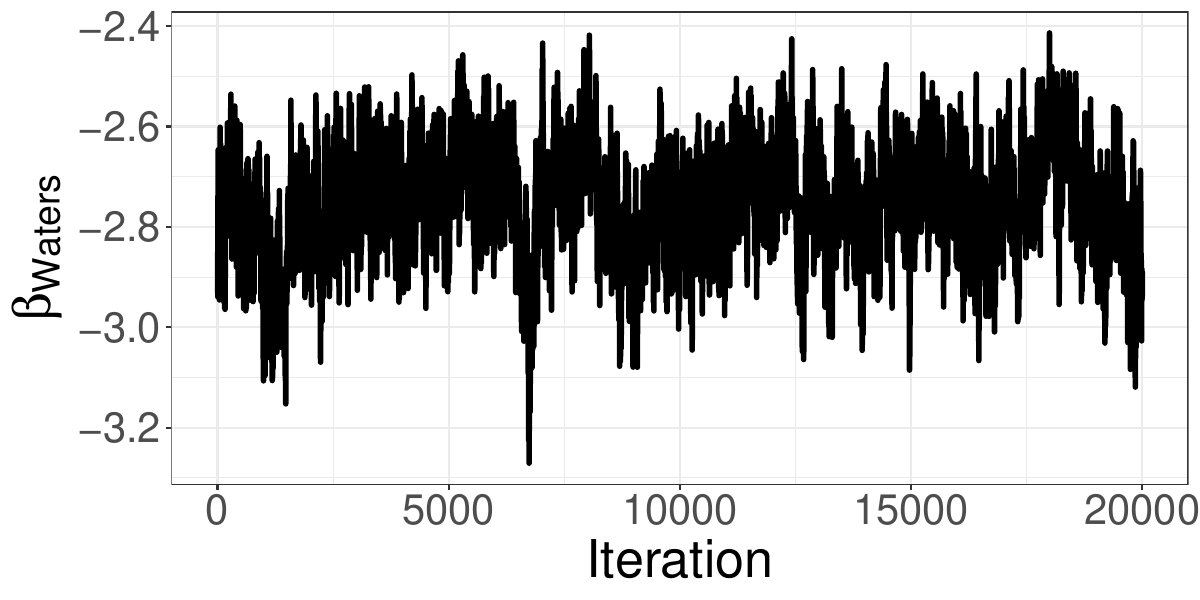}
    \caption{Representative Maxine Waters}
    \end{subfigure}
    \begin{subfigure}[t]{0.45\textwidth}
        \includegraphics[width = \textwidth]{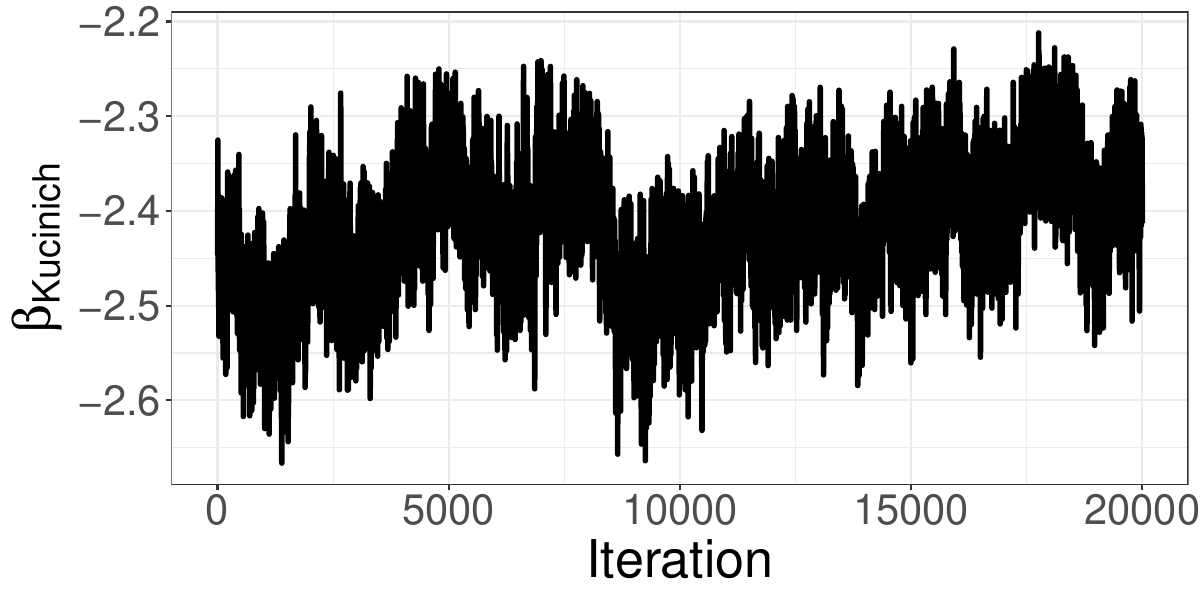}
        \caption{Representative Dennis Kucinich}
    \end{subfigure}\\
    \begin{subfigure}[t]{0.45\textwidth}
        \includegraphics[width = \textwidth]{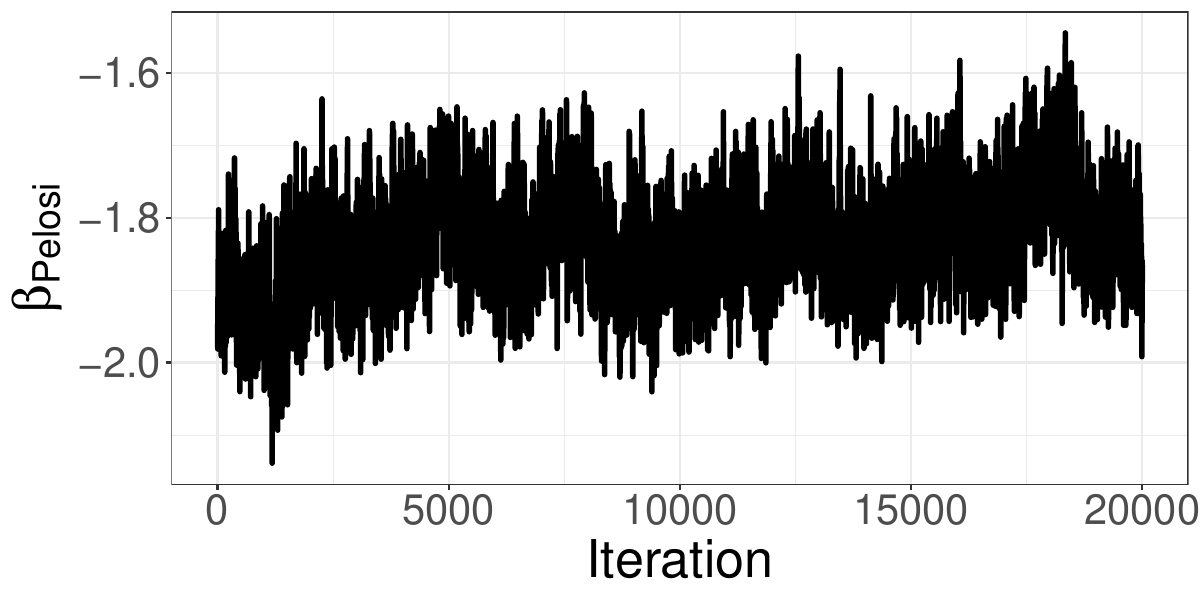}
        \caption{Representative Nancy Pelosi}
    \end{subfigure}
    \begin{subfigure}[t]{0.45\textwidth}
        \includegraphics[width = \textwidth]{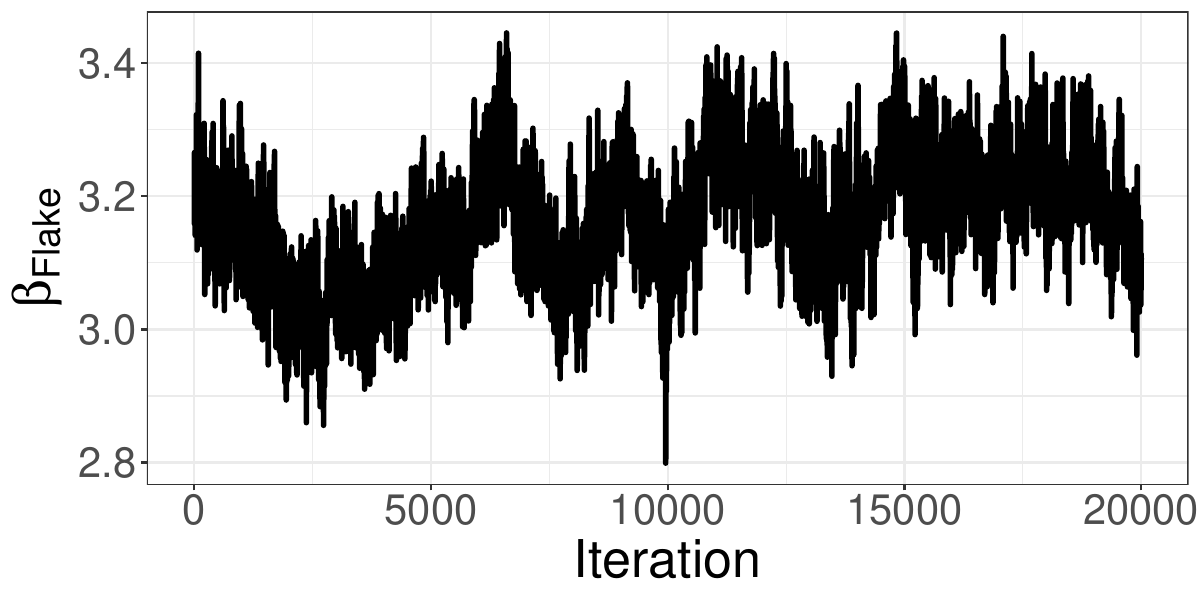}
    \caption{Representative Jeff Flake}
    \end{subfigure}
    \caption{Trace plots of the ideological preference for various Representatives in the 107\textsuperscript{rd} House.}\label{fig:house_107_leg_traceplots}
\end{figure*}

\begin{figure*}[!t]
    \centering
    \begin{subfigure}[t]{0.45\textwidth}
        \includegraphics[width = \textwidth]{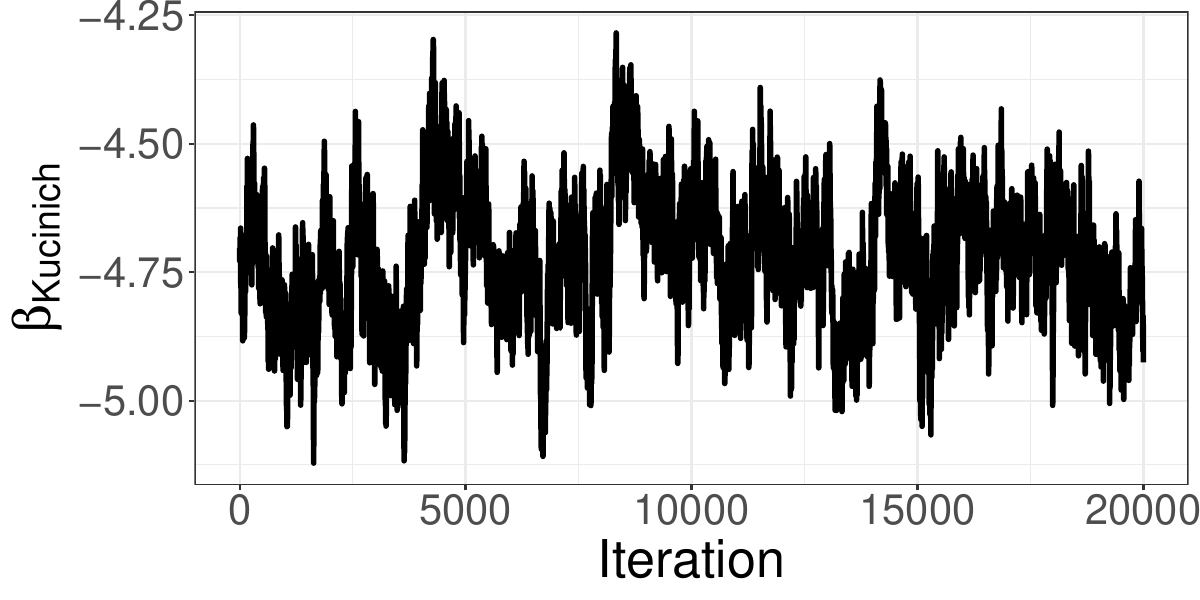}
        \caption{Representative Dennis Kucinich}
    \end{subfigure}
    \begin{subfigure}[t]{0.45\textwidth}
        \includegraphics[width = \textwidth]{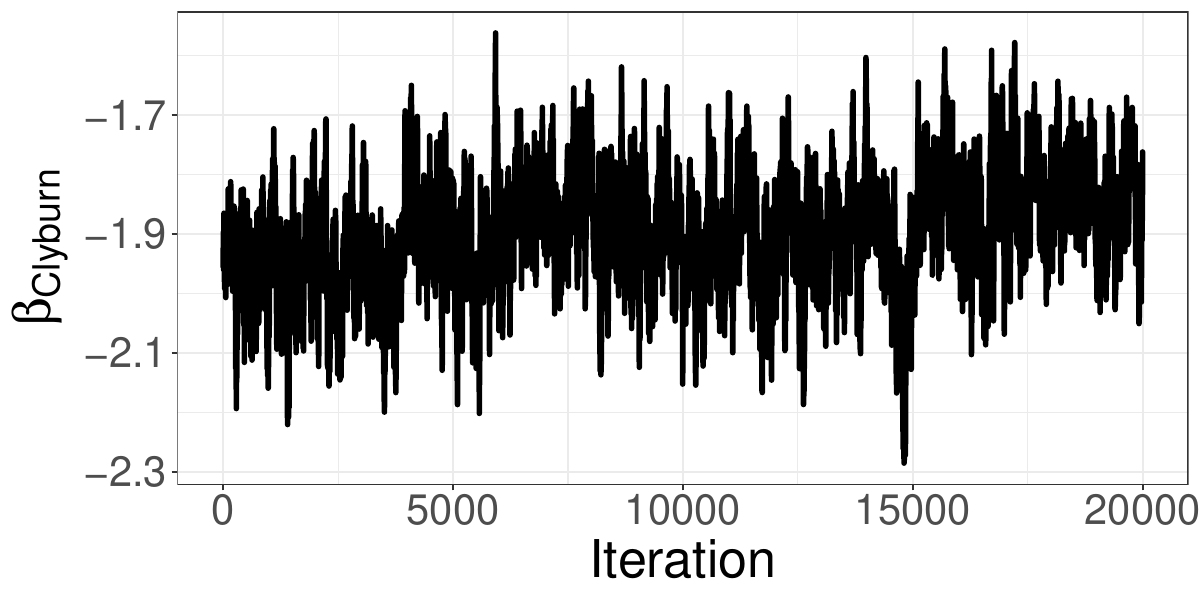}
        \caption{Representative Jim Clyburn}
    \end{subfigure}\\
    \begin{subfigure}[t]{0.45\textwidth}
        \includegraphics[width = \textwidth]{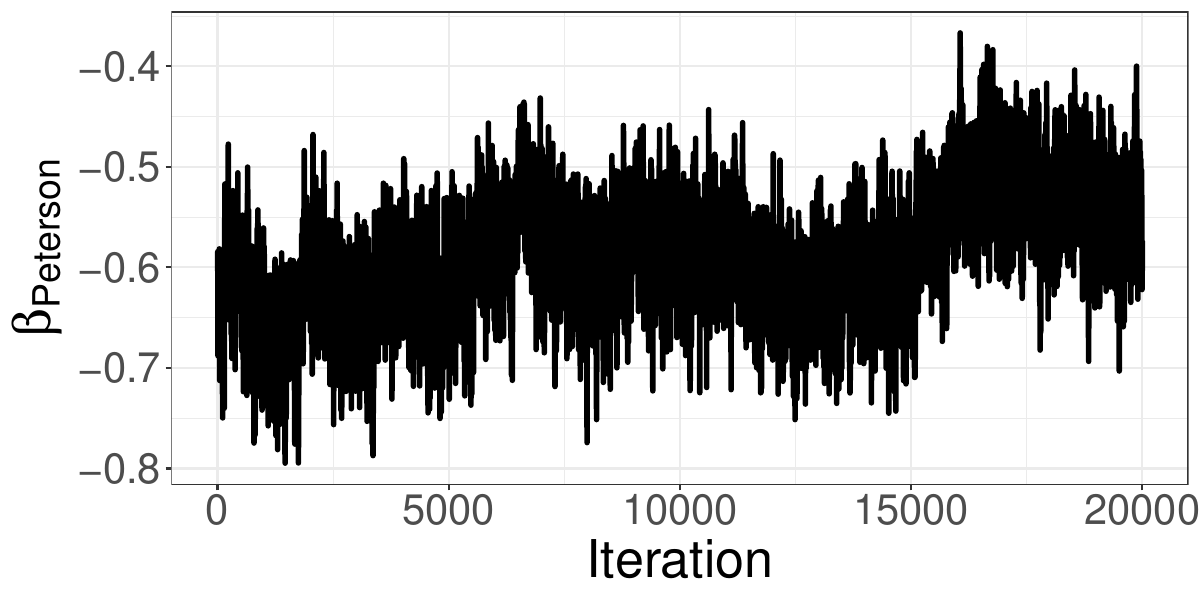}
    \caption{Representative Colin Peterson}
    \end{subfigure}
    \begin{subfigure}[t]{0.45\textwidth}
        \includegraphics[width = \textwidth]{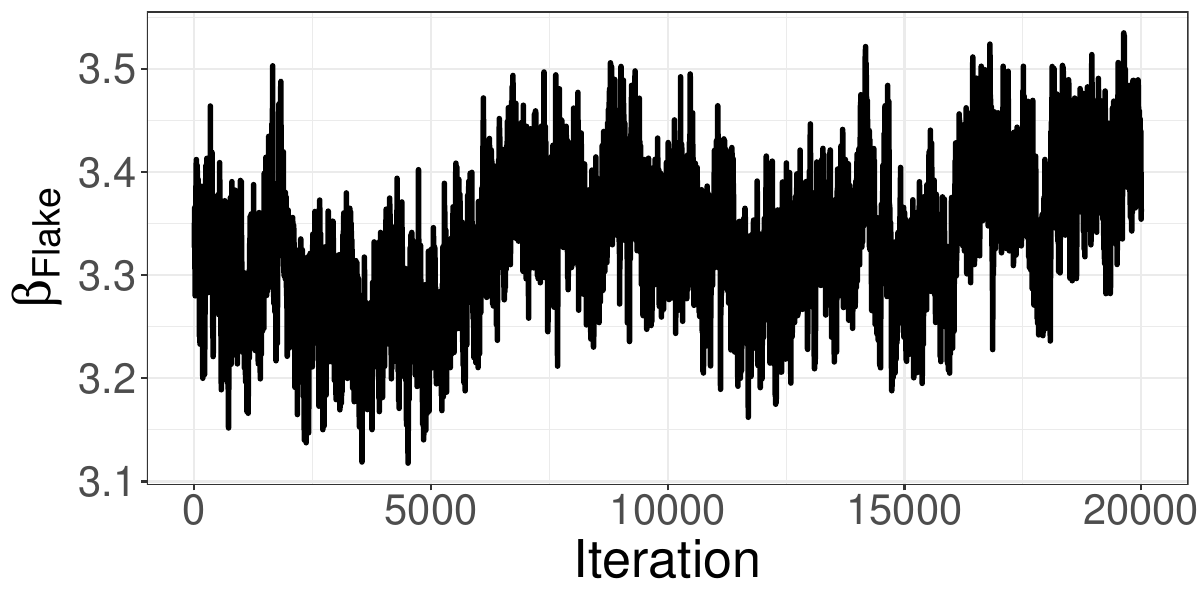}
    \caption{Representative Jeff Flake}
    \end{subfigure}
    \caption{Trace plots of the ideological preference for various Representatives in the 110\textsuperscript{rd} House.}\label{fig:house_110_leg_traceplots}
\end{figure*}

\begin{figure*}[!t]
    \centering
    \begin{subfigure}[t]{0.45\textwidth}
        \includegraphics[width = \textwidth]{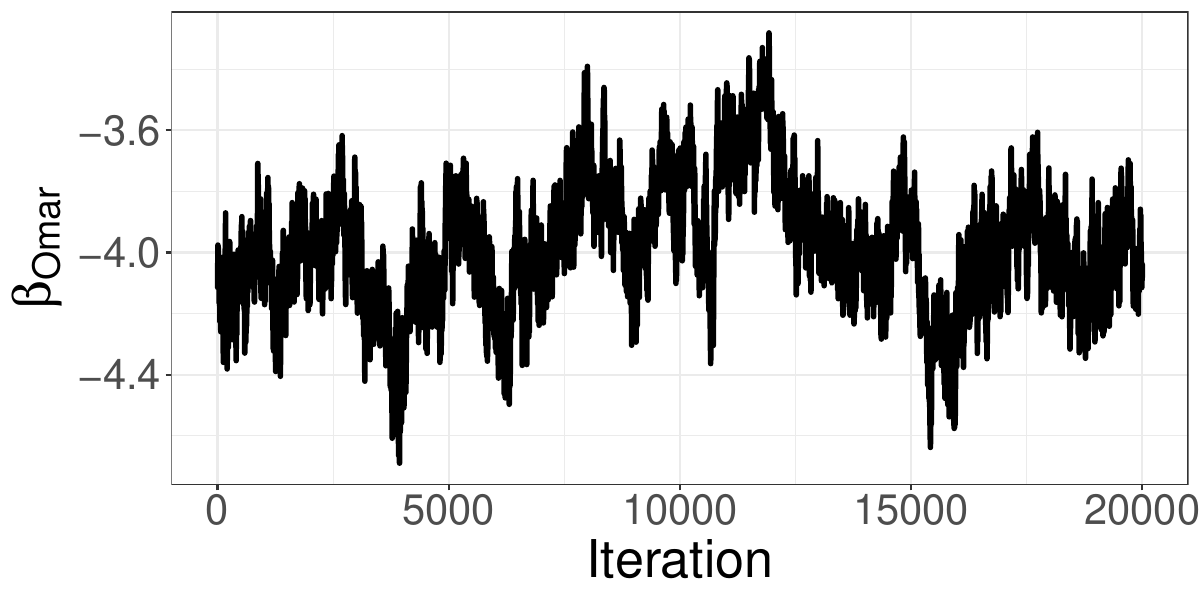}
    \caption{Representative Ilhan Omar}
    \end{subfigure}
    \begin{subfigure}[t]{0.45\textwidth}
        \includegraphics[width = \textwidth]{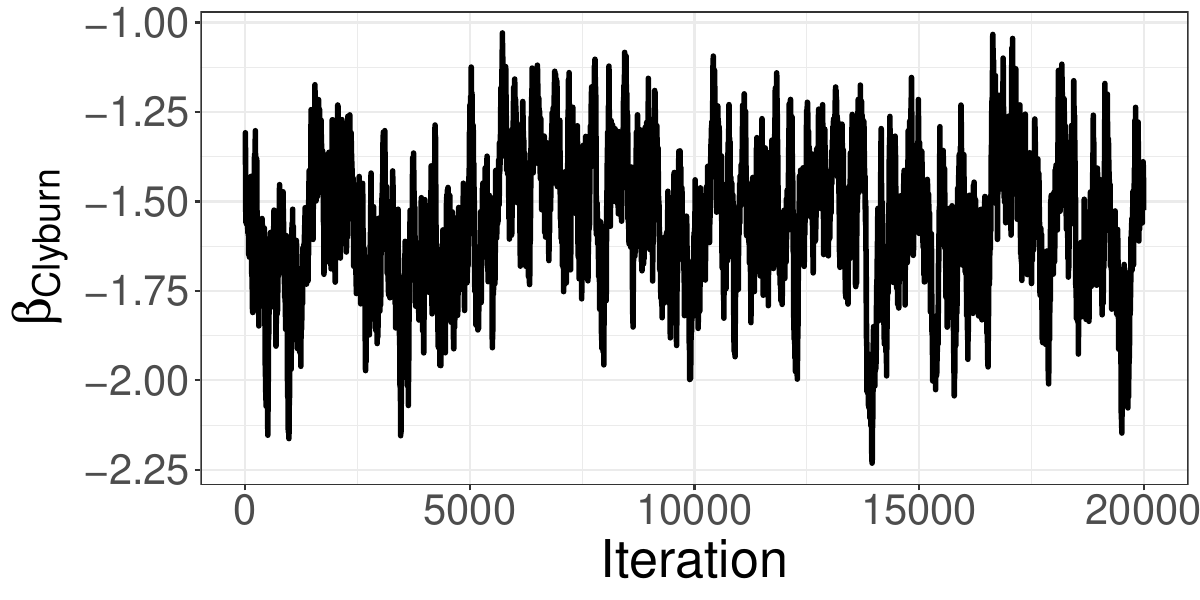}
        \caption{Representative Jim Clyburn}
    \end{subfigure} \\
    \begin{subfigure}[t]{0.45\textwidth}
        \includegraphics[width = \textwidth]{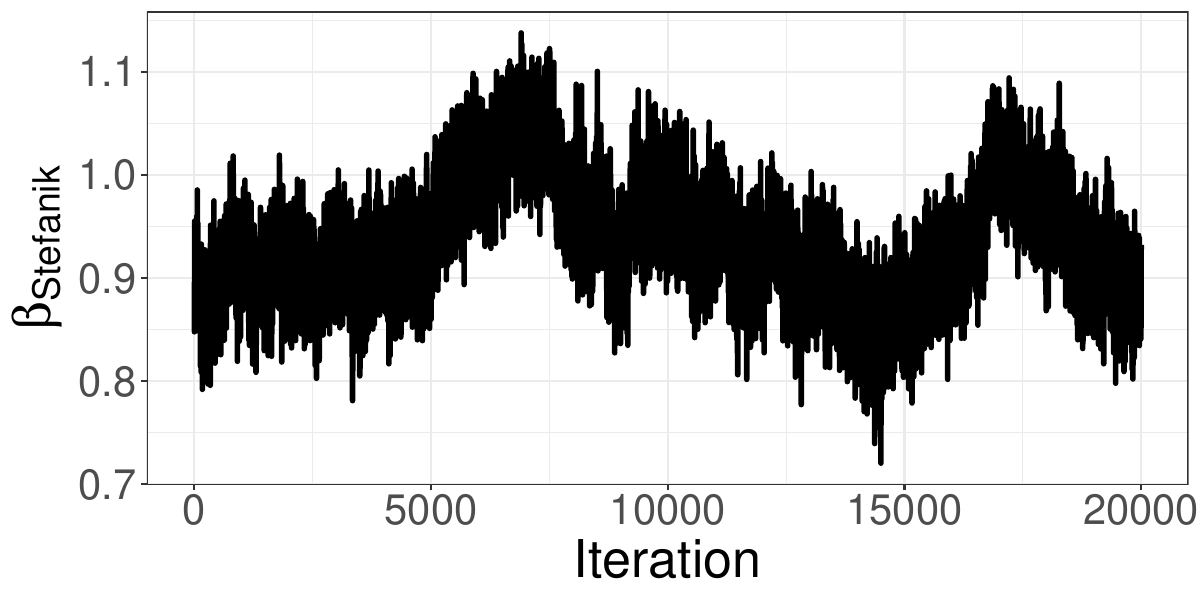}
    \caption{Representative Elise Stefanik}
    \end{subfigure}
    \begin{subfigure}[t]{0.45\textwidth}
        \includegraphics[width = \textwidth]{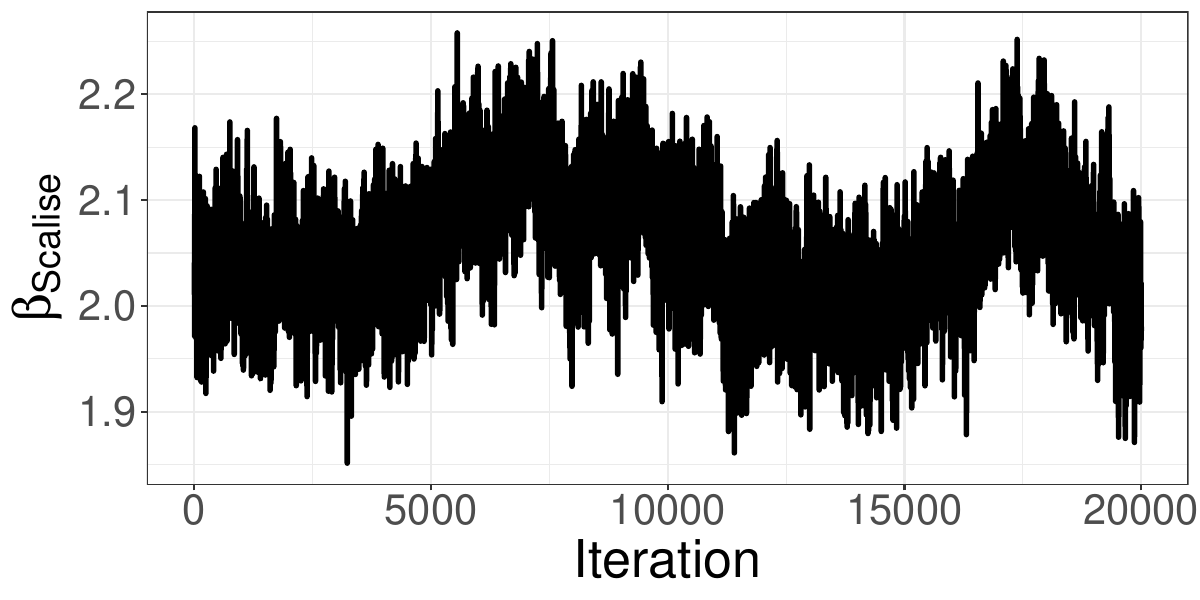}
    \caption{Representative Steve Scalise}
    \end{subfigure}
    \caption{Trace plots of the ideological preferences for various Representatives in the 116\textsuperscript{rd} House.}
    \label{fig:house_116_leg_traceplots}
\end{figure*}

\begin{figure*}[!t]
    \centering
    \begin{subfigure}[t]{0.45\textwidth}
        \includegraphics[width = \textwidth]{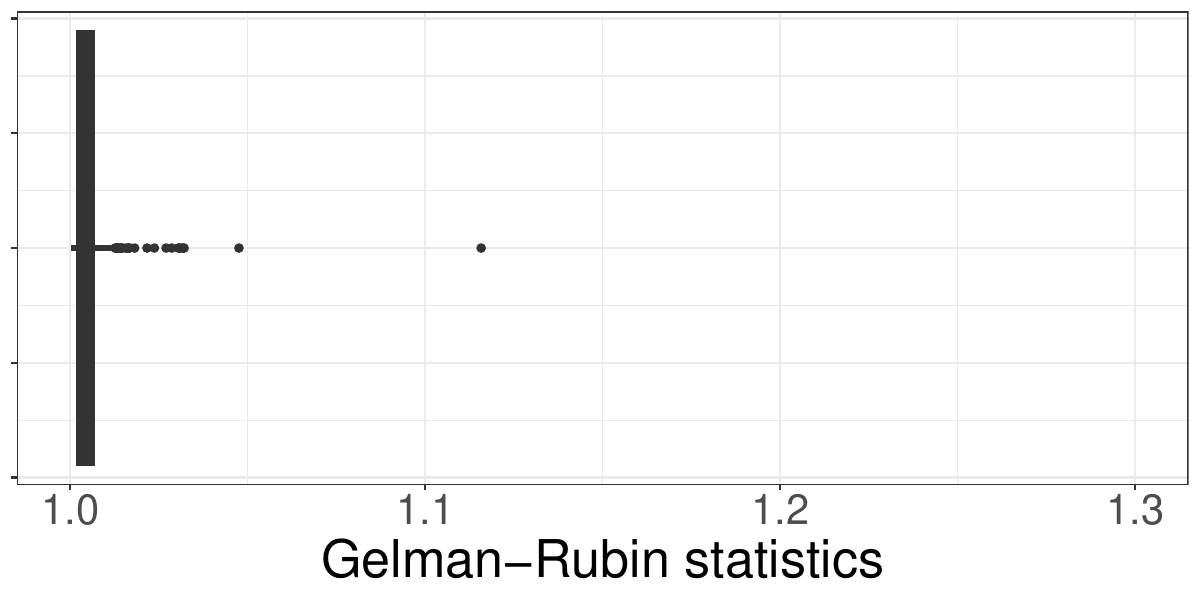}
    \caption{103\textsuperscript{rd} House}
    \end{subfigure}
    \begin{subfigure}[t]{0.45\textwidth}
        \includegraphics[width = \textwidth]{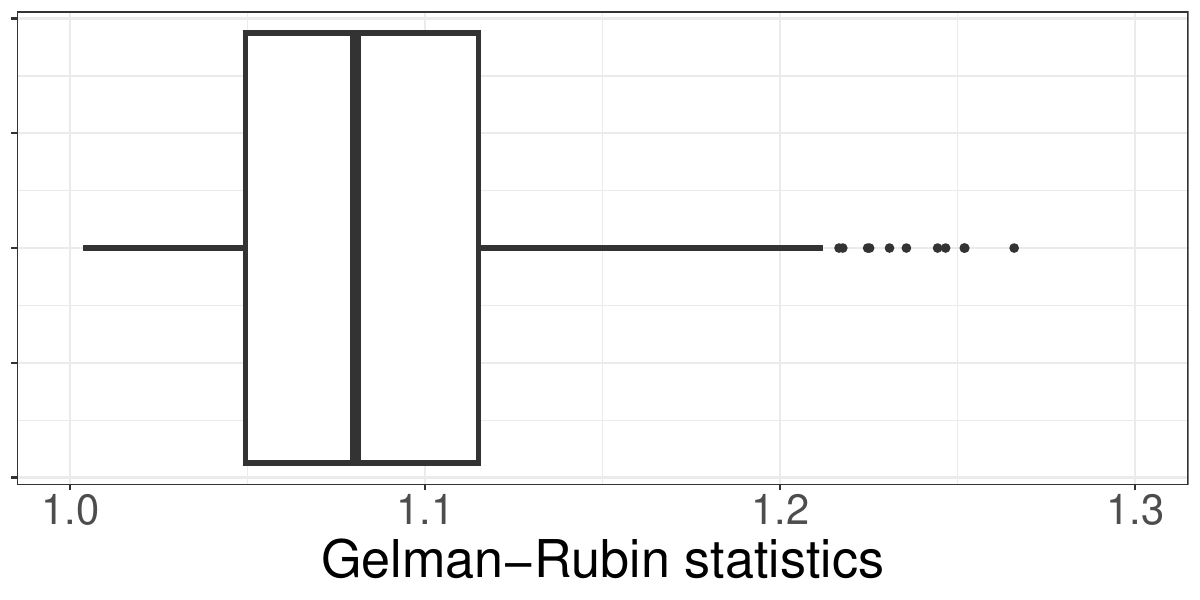}
        \caption{107\textsuperscript{th} House}
    \end{subfigure}
    \begin{subfigure}[t]{0.45\textwidth}
        \includegraphics[width = \textwidth]{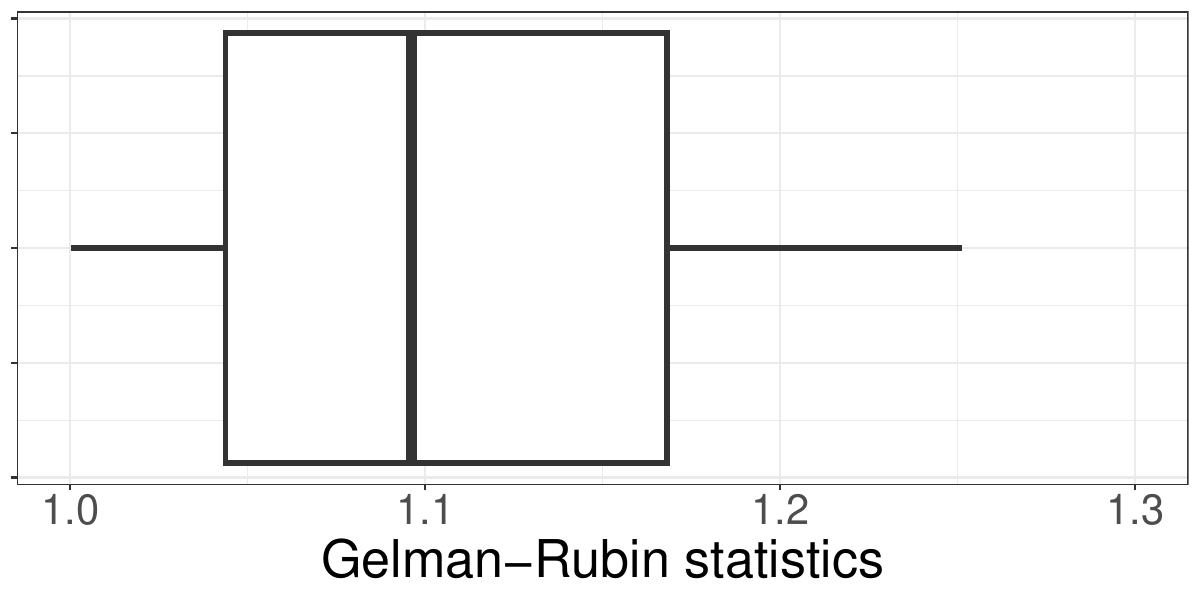}
    \caption{110\textsuperscript{th} House}
    \end{subfigure}
    \begin{subfigure}[t]{0.45\textwidth}
        \includegraphics[width = \textwidth]{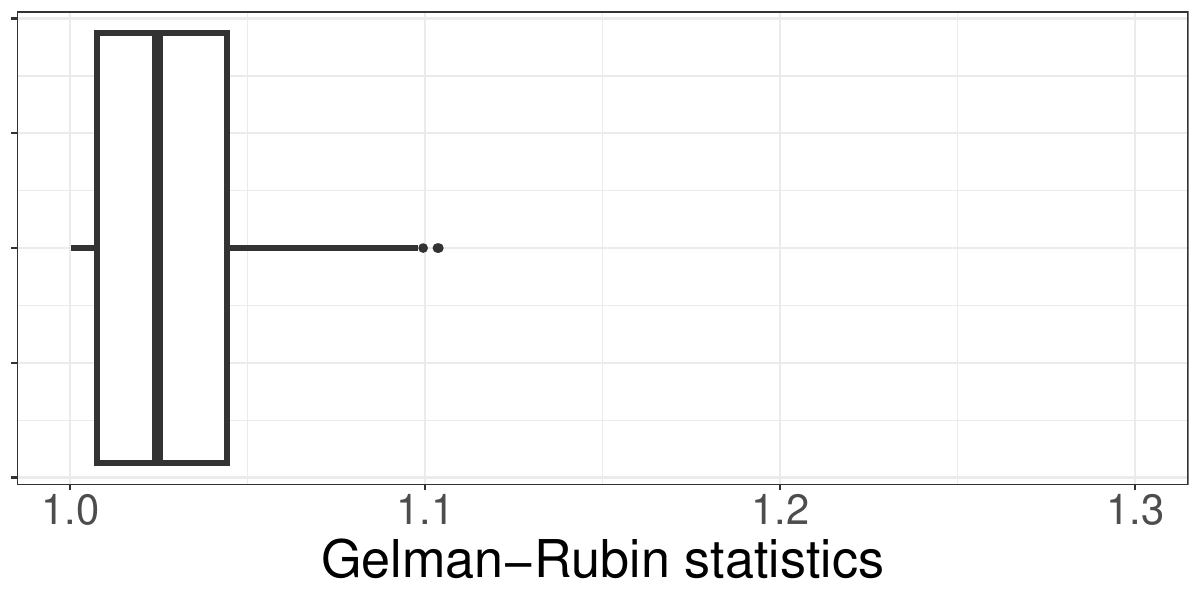}
        \caption{116\textsuperscript{th} House}
    \end{subfigure}
    \caption{Boxplots of the Gelman-Rubin statistics for the ideal points in various houses.}\label{fig:house_rhats_boxplots}
\end{figure*}

To explore these issues in more detail, we present in Figures \ref{fig:house_103_leg_traceplots} through \ref{fig:house_116_leg_traceplots} trace plots for the ideal points of various legislators in each of the four congresses under consideration. These include examples of both centrists and extreme legislators in both parties, as well as mainstream ones.  All trace plots show relatively high autocorrelation, but they also suggest that the algorithm is able to fully explore the support of the corresponding posterior distributions. To further explore the convergence of the algorithm, we present in Figure \ref{fig:house_rhats_boxplots} boxplots of the Gelman-Rubin $\hat{R}$ statistic for the ideal points on each House. We can see that the convergence rate of the algorithm varies substantially with the dataset.  For the 103\textsuperscript{rd} and 116\textsuperscript{th} Houses, virtually all of the $\beta_i$s have values of $\hat{R}$ that are below 1.1, suggesting that the algorithm has fully converged in both cases.  The picture is a bit more nuanced for the 107\textsuperscript{th} and the 110\textsuperscript{th} Houses.  In these two cases, somewhere between 50\% and 65\% of the estimated ideal points have $\hat{R}$ values that are under the customary 1.1 threshold.  But even then, over 90\% of them have $\hat{R}$ values below 1.2.  Furthermore, while these results for the 107\textsuperscript{th} and the 110\textsuperscript{th} Houses might appear disappointing at first sight, we believe that they should be interpreted carefully.  Indeed, it is important to recall that the value of the $\beta_i$'s are only weakly identified in our model.  In our experience, this kind of parameters often show much slower convergence and mixing than any strongly identifiable reparameterization.

\section{Conclusion}\label{sec:discussion}

This paper described a logistic version of the unfolding model introduced in \cite{LeiRodriguezNovelClassUnfolding2023} and discussed a computational algorithm that relies on a mixture approximation to the the Gumbel distribution to derive an easy-to-implement Gibbs sampler that outperforms off-the-shelf algorithms based on Hamiltonian Monte Carlo methods.  Our numerical experiments suggest that logistic models tend to provide better estimates and a better complexity-adjusted fit to the voting data from the U.S.\ House of Representatives than models based on a probit link. 

Moving forward, there are several interesting directions to explore. As we noted in Sections \ref{sec:intro} and \ref{sec:compchallenges}, one of the motivations to develop this particular kind of algorithm to fit logit unfolding models is the desire to explore extensions of the model that allow for the testing of specific hypotheses about the ideal points of legislators, along the lines considered in \cite{lofland2017assessing}, \cite{MoserEtAlMultipleIdealPoints2021} and \cite{lipman2023explaining}.  While developing such an extension is beyond of the scope of this paper, it is important to reemphasize that even if black-box samplers such as those implemented in \texttt{Stan} were to work for the basic version of the model, they would likely not be useful for models in which mixtures involving point mass priors are used and the associated parameters cannot be marginalized out.  A second type of extension enabled by the development of the algorithm in this paper is dynamic models that allow individual preferences to vary over time. A natural application of such a model is in the estimation of the ideology of justices in the U.S.\ Supreme Court (e.g., see \citealp{MartinQuinnDynamicIdealPoint2002a}). From a substantive point of view, the development of such a model would allow us to investigate whether heavy-tailed utility shocks are also prevalent in judicial settings.  Given the differences in both settings (unlike members of the House of Representatives, justices in the U.S.\ Supreme Court are appointed for life and are typically not subject to party or constituency pressure), discovering that heavy-tailed shocked are important in judicial settings would be a noteworthy finding.  Another direction is to develop a higher dimensional version of the logit unfolding model. As far as we know, all unfolding models available in the literature assume that the latent space is unidimensional.  While this is enough in many applications (including the study of the U.S.\ Congress we consider in this paper), such an assumption can be limiting in other contexts.  Extending our model in that direction would be relatively straightforward. Finally, one last direction to explore in the future is ordinal responses. While developing such model would require a careful choice of a ``reference'' category, the remaining ones could be represented by pairs of coordinates on the preference line.

\section*{Statements and Declarations}
\subsection{Conflict of interest/Competing interests}
The authors have no conflict of interest to report.

\newpage

\bibliography{my_refs,newrefs}

\end{document}


\title[Logit unfolding models]{Logit unfolding choice models for binary data - Supplementary Material}

\author*[1]{\fnm{Rayleigh} \sur{Lei}}\email{rlei13@uw.edu}

\author[1]{\fnm{Abel} \sur{Rodriguez}}\email{abelrow@uw.edu}

\affil*[1]{\orgdiv{Department of Statistics}, \orgname{University of Washington}, \orgaddress{\street{Padelford Hall}, \city{Seattle}, \postcode{98915}, \state{WA}, \country{United States}}}


\maketitle

\section{Model identifiability}

The likelihood function for our model is given by
\begin{multline*}
p(\mathbf{Y} \mid \beta_i, \alpha_{j,1}, \delta_{j,1}, \alpha_{j,2}, \delta_{j,2}) = \\
\prod_{i=1}^{I}\prod_{j=1}^{J} \left[ \theta_{i, j} (\beta_i, \alpha_{j,1}, \delta_{j,1}, \alpha_{j,2}, \delta_{j,2}) \right]^{y_{i,}}
\left[ 1 - \theta_{i, j} (\beta_i, \alpha_{j,1}, \delta_{j,1}, \alpha_{j,2}, \delta_{j,2}) \right]^{1 - y_{i,}} ,
\end{multline*}
where, per Equation (2) in the main manuscript,
\begin{multline*}
     \theta_{i, j} (\beta_i, \alpha_{j,1}, \delta_{j,1}, \alpha_{j,2}, \delta_{j,2}) =  \frac{1}{1 + \exp\left\{-\alpha_{j,1}(\beta_i - \delta_{j,1})\right\} + \exp\left\{-\alpha_{j,2}(\beta_i - \delta_{j,2})\right\}} .
\end{multline*}

Similarly, the posterior distribution is given by 
\begin{multline*}
p(\beta_i, \alpha_{j,1}, \delta_{j,1}, \alpha_{j,2}, \delta_{j,2} \mid \mathbf{Y}) \propto \\
p(\mathbf{Y} \mid \beta_i, \alpha_{j,1}, \delta_{j,1}, \alpha_{j,2}, \delta_{j,2})
\prod_{i=1}^{I} p(\beta_i)  \prod_{j=1}^{J}  p(\alpha_{j,1}, \alpha_{j,2}, \delta_{j,1}, \delta_{j,2}) .
\end{multline*}

Consider a reparameterization such that $\alpha'_{j, 1} = \frac{1}{a} \alpha_{j, 1}$, $\alpha'_{j, 2} = \frac{1}{a} \alpha_{j, 2}$, $\delta'_{j, 1} = a (\delta_{j, 1} + b)$, $\delta'_{j, 2} = a (\delta_{j, 2} + b)$ and $\beta'_i = a (\beta_i + b)$
for some real constants $a$ and $b$ with $a >0$. Note that
\begin{align*}
    \alpha'_{j, 1} (\beta'_i - \delta'_{j, 1}) &= \frac{1}{a}\alpha_{j, 1} \left\{ a (\beta_i + b) - a (\delta_{j, 1} + b) \right\}\\
    &= \alpha_{j, 1} \left\{ (\beta_i + b) - (\delta_{j, 1} + b) \right\} \\
    &= \alpha_{j, 1} (\beta_i - \delta_{j, 1}) ,
\end{align*}
and similarly,
\begin{align*}
    \alpha'_{j, 2} (\beta'_i - \delta'_{j, 2}) &= \frac{1}{a}\alpha_{j, 2} \left\{ a (\beta_i + b) - a (\delta_{j, 2} + b) \right\}\\
    &= \alpha_{j, 2} \left\{ (\beta_i + b) - (\delta_{j, 2} + b) \right\} \\
    &= \alpha_{j, 2} (\beta_i - \delta_{j, 2}).
\end{align*}
Hence,
\begin{align*}
p(\mathbf{Y} \mid \beta_i', \alpha_{j,1}', \delta_{j,1}', \alpha_{j,2}', \delta_{j,2}') &=
p(\mathbf{Y} \mid \beta_i, \alpha_{j,1}, \delta_{j,1}, \alpha_{j,2}, \delta_{j,2}),
\end{align*}
which shows that under the likelihood alone, the parameters are not identifiable under shifts and rescalings.  However, note that $p(\beta_i')$ corresponds to the density of a Gaussian distribution with mean $ab$ and variance $a^2$.  Hence, $p(\beta_i') \ne p(\beta_i)$ and therefore $p(\beta_i', \alpha_{j,1}', \delta_{j,1}', \alpha_{j,2}', \delta_{j,2}' \mid \mathbf{Y}) \ne p(\beta_i, \alpha_{j,1}, \delta_{j,1}, \alpha_{j,2}, \delta_{j,2} \mid \mathbf{Y})$.  This shows that the model is weakly identifiable under shifts and rescalings (i.e., identified only through the choice of prior distribution).

On the other hand, if we relax the restriction $a > 0$ and consider the case where $a = -1$ and $b=0$ (i.e., a reflection of the latent space), we do have $p(\beta_i') = p(\beta_i)$ and therefore $p(\beta_i', \alpha_{j,1}', \delta_{j,1}', \alpha_{j,2}', \delta_{j,2}' \mid \mathbf{Y}) = p(\beta_i, \alpha_{j,1}, \delta_{j,1}, \alpha_{j,2}, \delta_{j,2} \mid \mathbf{Y})$.  Hence, fixing the mean and variance of the latent traits does not address this other source of unidentifiability.  As discussed in the main manuscript, we resolve this issue by fixing the sign of the ideal point of a carefully selected legislator.

\section{Derivation of the acceptance probability for step 6 of the MCMC algorithm in Section 3}

Step 6 of our MCMC algorithm attempts to flip the sign of $z_j$ by randomly choosing between two possible proposals for the new values of $\bfalpha_j$, $\bfdelta_j$, and $z_j$. Here, we denote the new values as $\v{\alpha}'_j, \v{\delta}'_j$, and $z'_j$. The acceptance probability for both of these proposals happens to be the same, although the derivations are slightly different.

Generally speaking,  the acceptance probability for a generic proposal associated with a flip in the sign of $z_j$ is given by
\begin{align*}
        \prod_{i = 1}^I & \left(\frac{p(y_{i, j} = 1 \mid \beta_i, \bfalpha_j', \bfdelta_j')}{p(y_{i, j} = 1 \mid \beta_i, \bfalpha_j, \bfdelta_j)}\right)^{y_{i, j} } 
        \left(\frac{1 - p(y_{i, j} = 1 \mid \beta_i, \bfalpha_j', \bfdelta_j')}{1 - p(y_{i, j} = 1 \mid \beta_i, \bfalpha_j, \bfdelta_j)}\right)^{1 - y_{i, j}}\\
        & \qquad\qquad \frac{p(\v{\alpha}'_j, \v{\delta}'_j \mid z_j')}{p(\v{\alpha}_j, \v{\delta}_j \mid z_j )}
        \frac{q(\v{\alpha}_j, \v{\delta}_j \mid \v{\alpha}'_j, \v{\delta}'_j, z_j')}{q(\v{\alpha}'_j, \v{\delta}'_j \mid \v{\alpha}_j, \v{\delta}_j, z_j)}.
\end{align*}
where $p(y_{i, j} = 1 \mid \beta_i, \bfalpha_j, \bfdelta_j)$ is given in Equation (2) of the main manuscript, $p(\v{\alpha}_j, \v{\delta}_j \mid z_j )$ is the prior on $(\v{\alpha}_j, \v{\delta}_j)$ after conditioning on $z_j$ (i.e., the appropriate component of the mixture in Equation (3) of the main manuscript, see Section 3.3), and $q(\v{\alpha}'_j, \v{\delta}'_j \mid \v{\alpha}_j, \v{\delta}_j, z_j)$ is the proposal distribution for the new values of $\v{\alpha}_j$ and $\v{\delta}_j)$.

Consider first the case where $z_j' = -z_j$ and the proposal for $(\v{\alpha}_j, \v{\delta}_j, z_j)$ is given by $\bfalpha_j' = -\bfalpha_j$ and $\bfdelta_j' = -\bfdelta_j$.  Such a deterministic proposal can be justified as the limit of reflecting random walks where $\bfalpha_j' \mid \bfalpha_j, z_j \sim \textrm{N}( \cdot \mid - \bfalpha_j, \tau_1^2)$ and $\bfdelta_j' \mid \bfdelta_j, z_j \sim \textrm{N}( \cdot \mid - \bfdelta_j, \tau_2^2)$ when both $\tau_1, \tau_2 \to 0$.  In this case, we can see that
\begin{align*}
    \frac{q(\v{\alpha}_j, \v{\delta}_j \mid \v{\alpha}'_j, \v{\delta}'_j, z_j')} {q(\v{\alpha}'_j, \v{\delta}'_j \mid \v{\alpha}_j, \v{\delta}_j, z_j)} = 1,
    & &
    \frac{p(\v{\alpha}'_j, \v{\delta}'_j \mid z_j')}{p(\v{\alpha}_j, \v{\delta}_j \mid z_j )} = 1.
\end{align*}


Hence, the acceptance probability reduces to the ratio of likelihoods for the appropriate subset of observations.

Consider now the second proposal described in step 6, where $q(\v{\alpha}'_j, \v{\delta}'_j \mid \v{\alpha}_j, \v{\delta}_j, z_j) = p(\v{\alpha}'_j, \v{\delta}'_j \mid z_j')$.  Because of this, we have 
\begin{multline*}
\frac{q(\v{\alpha}_j, \v{\delta}_j \mid \v{\alpha}'_j, \v{\delta}'_j, z_j')} {q(\v{\alpha}'_j, \v{\delta}'_j \mid \v{\alpha}_j, \v{\delta}_j, z_j)} 
\frac{p(\v{\alpha}'_j, \v{\delta}'_j \mid z_j')}{p(\v{\alpha}_j, \v{\delta}_j \mid z_j )} =
\frac{q(\v{\alpha}_j, \v{\delta}_j \mid \v{\alpha}'_j, \v{\delta}'_j, z_j')}
{p(\v{\alpha}_j, \v{\delta}_j \mid z_j )}
\frac{p(\v{\alpha}'_j, \v{\delta}'_j \mid z_j')}{q(\v{\alpha}'_j, \v{\delta}'_j \mid \v{\alpha}_j, \v{\delta}_j, z_j)}
= 1    
\end{multline*}
and the acceptance probability again simplifies to the ratio of likelihoods.